\documentclass[12pt]{iopart}

\newcommand{\SU}[1]{\mathrm{SU}(#1)}
\newcommand{\Pauli}{\frac{\tau^{a}}{2}}
\newcommand{\UA}{\mathrm{U}(1)_{\mathrm{A}}}
\newcommand{\CS}{\SU{2}_{\mathrm{L}} \times \SU{2}_{\mathrm{R}}}
\newcommand{\cq}{m_{q\,\mathrm{cons.}}}
\newcommand{\qq}{{q\bar{q}}}
\usepackage{amsfonts}

\newcommand{\ket}[1]{|#1 \rangle}

\usepackage{graphicx}
\usepackage{subfigure}
\usepackage[utf8]{inputenc}
\usepackage{slashed}
\usepackage{cite}
\usepackage{nicefrac}
\usepackage{xcolor}

\begin{document}

\begin{flushright}
\small ADP-17-22/T1028
\end{flushright}
\vspace{-12pt}

\title[Centre vortex removal restores chiral symmetry]{Centre vortex removal restores chiral symmetry}

\author{Daniel Trewartha$^1$, Waseem Kamleh$^2$, Derek B Leinweber$^2$}

\address{$^1$ Thomas Jefferson National Accelerator Facility, 12000 Jefferson Avenue, Newport News, VA 23606, USA \newline$^2$ Centre for the Subatomic Structure of Matter(CSSM), Department of Physics, University of Adelaide 5005, Australia}

\begin{abstract}
The influence of centre vortices on dynamical chiral symmetry breaking is investigated through the light hadron spectrum on the lattice. Recent studies of the quark propagator and other quantities have provided evidence that centre vortices are the fundamental objects underpinning dynamical chiral symmetry breaking in $\SU{3}$ gauge theory. For the first time, we use the chiral overlap fermion action to study the low-lying hadron spectrum on lattice ensembles consisting of Monte Carlo, vortex-removed, and vortex-projected gauge fields. We find that gauge field configurations consisting solely of smoothed centre vortices are capable of reproducing all the salient features of the hadron spectrum, including dynamical chiral symmetry breaking. The hadron spectrum on vortex-removed fields shows clear signals of chiral symmetry restoration at light values of the bare quark mass, while at heavy masses the spectrum is consistent with a theory of weakly-interacting constituent quarks.
\end{abstract}
\submitto{\jpg}
\maketitle

\section{Introduction}

Dynamical chiral symmetry breaking is one of the signature features of quantum chromodynamics (QCD), along with the confinement of quarks inside hadrons. These phenomena appear to be emergent properties of QCD, and are generally accepted to originate from some topological feature of the non-trivial QCD vacuum. The centre vortex model~\cite{'tHooft:1977hy,'tHooft:1979uj,Cornwall:1979hz,Nielsen:1979xu,Ambjorn:1980ms,Vinciarelli:1978kp,Yoneya:1978dt,Mack:1978rq} is a well-known candidate for the origin of confinement, and has been extensively studied on the lattice in both $\SU{2}$ and $\SU{3}$ gauge theory~\cite{DelDebbio:1996mh,DelDebbio:1998uu,Engelhardt:1998wu,Langfeld:1997jx,Kovacs:1998xm,deForcrand:1999ms,Langfeld:2003ev}. The role of centre vortices on dynamical chiral symmetry breaking has also been examined on the lattice, where in $\SU{2}$ theory~\cite{Bowman:2008qd,deForcrand:1999ms,Engelhardt:2002qs,Bornyakov:2007fz,Hollwieser:2008tq,Hollwieser:2013xja,Hoellwieser:2014isa,Alexandrou:1999vx,Kovalenko:2005rz} it has been found that they are the fundamental long-range objects responsible. In $\SU{3}$, initial studies showed mixed results. While the Landau-gauge AsqTad quark propagator showed no role for centre vortices in $\SU{3}$ dynamical chiral symmetry breaking~\cite{Bowman:2010zr}, the opposite result was found in the low-lying hadron spectrum~\cite{OMalley:2011aa} using Wilson-type fermions.

Recently, the Landau-gauge quark propagator has been studied using the chirally-sensitive overlap fermion action~\cite{Trewartha:2015nna, Trewartha:2015ida}. There, it was found that removal of centre vortices from gauge field backgrounds results in the loss of dynamical chiral symmetry breaking.  Backgrounds consisting solely of centre vortices could reproduce dynamical chiral symmetry breaking after a small amount of gauge field smoothing. This is part of a consistent picture that has emerged from work by the CSSM lattice collaboration demonstrating that smoothed vortex-only gauge fields are able to reproduce a number of salient features of QCD~\cite{Kamleh:2017lij}.
In addition to macroscopic quantities such as the quark mass function and static quark potential, the microscopic structure of the gluon field has also been examined.  It was seen that after smoothing, a gauge field background consisting solely of centre vortices displays a structure of instanton-like objects similar in both size and density to those seen in untouched configurations after cooling, hence providing a mechanism for dynamical chiral symmetry breaking.

In this work we study the role of centre vortices in dynamical chiral symmetry breaking via the low-lying hadron spectrum. While this has been considered previously in Ref.~\cite{OMalley:2011aa}, here we offer several significant improvements. Firstly, the work of Refs.~\cite{Trewartha:2015nna, Trewartha:2015ida} has shown that the chiral nature of the overlap fermion action is vital to correctly discern the role of centre vortices in dynamical chiral symmetry breaking, and so it is used here. Additionally, one may be concerned that the procedure of removing centre vortices from gauge field configurations has changed the quark mass renormalization, and so using a Wilson-like action one may have difficulty matching bare quark masses across ensembles. Evidence of this was indeed seen in Ref.~\cite{OMalley:2011aa}. The overlap fermion action, thanks to its lattice-deformed chiral symmetry, does not suffer from additive mass renormalization, and so we may unambiguously compare ensembles with equivalent bare quark masses. In order to study dynamical chiral symmetry breaking, one naturally wishes to minimise the impact of the explicit chiral symmetry breaking induced by the bare quark mass. The overlap action enables us to consider very light masses, with a smallest value considered of $m_{q} = 13$  MeV.

\section{Simulation Details}

\begin{table}[!b]
\caption{A list of the meson and baryon interpolators considered herein.}
\smallskip
\centerline{\begin{tabular}{ ccc }
\hline 
\rule{0pt}{4ex} Meson & $\mathrm{I},\,\mathrm{J}^{\mathrm{PC}}$ & Operator \\
\hline
\hline
\rule{0pt}{3ex} $\pi$ & $1,0^{-+}$ & $\bar{q}\, \gamma_{5}\, \Pauli\, q$ \\
\rule{0pt}{3ex} $\rho$ & $1,1^{--}$ & $\bar{q}\, \gamma_{i}\, \Pauli\, q $ \\
\rule{0pt}{3ex} $a_{0}$ & $1,0^{++}$ & $\bar{q}\, \Pauli\, q $ \\
\rule{0pt}{3ex} $a_{1}$ & $1,1^{++}$ & $\bar{q}\, \gamma_{i}\, \gamma_{5}\, \Pauli\, q $ \\
\hline
\end{tabular}}
\bigskip
\centerline{\begin{tabular}{ ccc }
\hline
\rule{0pt}{4ex} Baryon & $\mathrm{I},\,\mathrm{J}^{\mathrm{P}}$ & Operator \\
\hline
\hline
\rule{0pt}{3ex} Nucleon &  $\frac{1}{2},\frac{1}{2}^{+}$ & $[u^{\mathrm{T}}\, \mathrm{C}\, \gamma_{5}\, d\,]\, u$\\
\rule{0pt}{3ex} $\Delta$ & $\frac{3}{2},\frac{3}{2}^{+}$ & $[u^{\mathrm{T}}\, \mathrm{C}\, \gamma_{i}\, u\,]\, u$\\
\hline
\end{tabular}}
\label{Tab:mesonsbaryons}
\end{table}

%Performing lattice calculations of key quantities on the three different ensembles and comparing the differences that emerge from the absence or presence of centre vortices has proven to be a rich source of information regarding the properties of pure Yang-Mills theory. Here we review the recent work that has been performed by the CSSM lattice collaboration~\cite{OMalley:2011aa,Trewartha:2015ida,Trewartha:2015nna}, and direct the reader therein for the details of the results presented here and a comprehensive list of references.

A centre vortex intersects with a two-dimensional region $A$ of the gauge manifold $U$ if the Wilson loop identified with the boundary has a non-trivial transformation property $U(\partial A) \rightarrow zU(\partial A),\ z \neq 1,$ under an element $Z = zI \in \mathbb{Z}_3$ of the centre group of SU(3), 
where $z \in \{ 1, e^{\pm2\pi i/3} \}$ is a cube root of unity. On the lattice we study centre vortices by seeking to decompose gauge links $U_\mu(x)$ in the form
\begin{equation}
U_{\mu}(x) = Z_{\mu}(x)\cdot R_{\mu}(x),
\end{equation}
in such a way that all vortex information is captured in the field of centre-projected elements $Z_{\mu}(x)$, with the remaining short-range fluctuations described by the vortex-removed field $R_{\mu}(x).$ By fixing to Maximal Centre Gauge and identifying $Z_{\mu}(x)$ as the projection of the gauge-fixed links to the nearest centre element, we produce configurations with vortices removed, as well as configurations consisting solely of vortex matter. Vortex matter is identified by searching for plaquettes with a nontrivial centre flux around the boundary. The procedure used is outlined in detail in Ref.~\cite{Trewartha:2015ida}. Throughout this work, we use three ensembles; an original, `untouched' ensemble (UT) of Monte Carlo gauge fields $U_\mu(x)$, an vortex-only ensemble (VO) consisting solely of centre projected elements $Z_\mu(x),$ and a vortex-removed ensemble (VR) of the remainder fields $R_\mu(x)$.

Results are calculated on $50$ pure gauge-field configurations using the L{\"u}scher-Weisz $\mathcal{O}(a^{2})$ mean-field improved action~\cite{Luscher:1984xn}, with a $20^{3} \times 40$ volume at a lattice spacing of $0.125 \, \mathrm{fm}$. We use the FLIC operator~\cite{Zanotti:2001yb,Kamleh:2004aw,Kamleh:2004xk,Kamleh:2001ff} as the overlap kernel, with negative Wilson mass $m_{w} = 1$. As per Refs.~\cite{Trewartha:2015nna, Trewartha:2015ida}, 10 sweeps of cooling are performed on the vortex-only ensemble in order to ensure the smoothness condition required for locality of the overlap operator.

The hadron interpolating fields we use here are listed in Table~\ref{Tab:mesonsbaryons}. Note that we consider only isovector mesons in order to avoid disconnected contributions.

The hadron spectrum is calculated with bare quark masses varying over a large range. The values of the overlap mass parameter $\mu$ and the corresponding bare quark mass are given in Table~\ref{Tab:hadmasses}. 100 iterations of Gauge-invariant Gaussian smearing~\cite{Gusken:1989qx,Burch:2004he} are performed at the fermion source and sink. Fixed boundary conditions are applied in the temporal direction, with our quark sources placed at $n_{t} = 10$ relative to the lattice length of 40. We have investigated the use of the variational method~\cite{Luscher:1990ck,Michael:1985ne}, and found no significant improvement in our ability to discern the ground state signals salient to our investigation. Hadron effective masses are extracted in the standard way, with uncertainties obtained via a second-order single-elimination jackknife analysis.

\begin{table}[t]
\caption{Values of the overlap mass parameter, $\mu$, considered, with corresponding bare quark masses in physical units, using $a=0.125$ fm.}
\smallskip
\centerline{\begin{tabular}{ cc }
\hline 
$\mu$ & $m_{q}$ (MeV) \\
\hline
\hline
0.004 & 13 \\
0.008 & 25 \\
0.012 & 38 \\
0.016 & 50 \\
0.032 & 101 \\
0.040 & 126 \\
0.048 & 151 \\
0.056 & 177 \\
\hline 
\end{tabular}}
\label{Tab:hadmasses}
\end{table}

\section{Vortex-Only Spectrum}
\label{sec:VOhadspec}

\begin{figure*}[p]
  \centering
  \includegraphics[width=0.4\columnwidth]{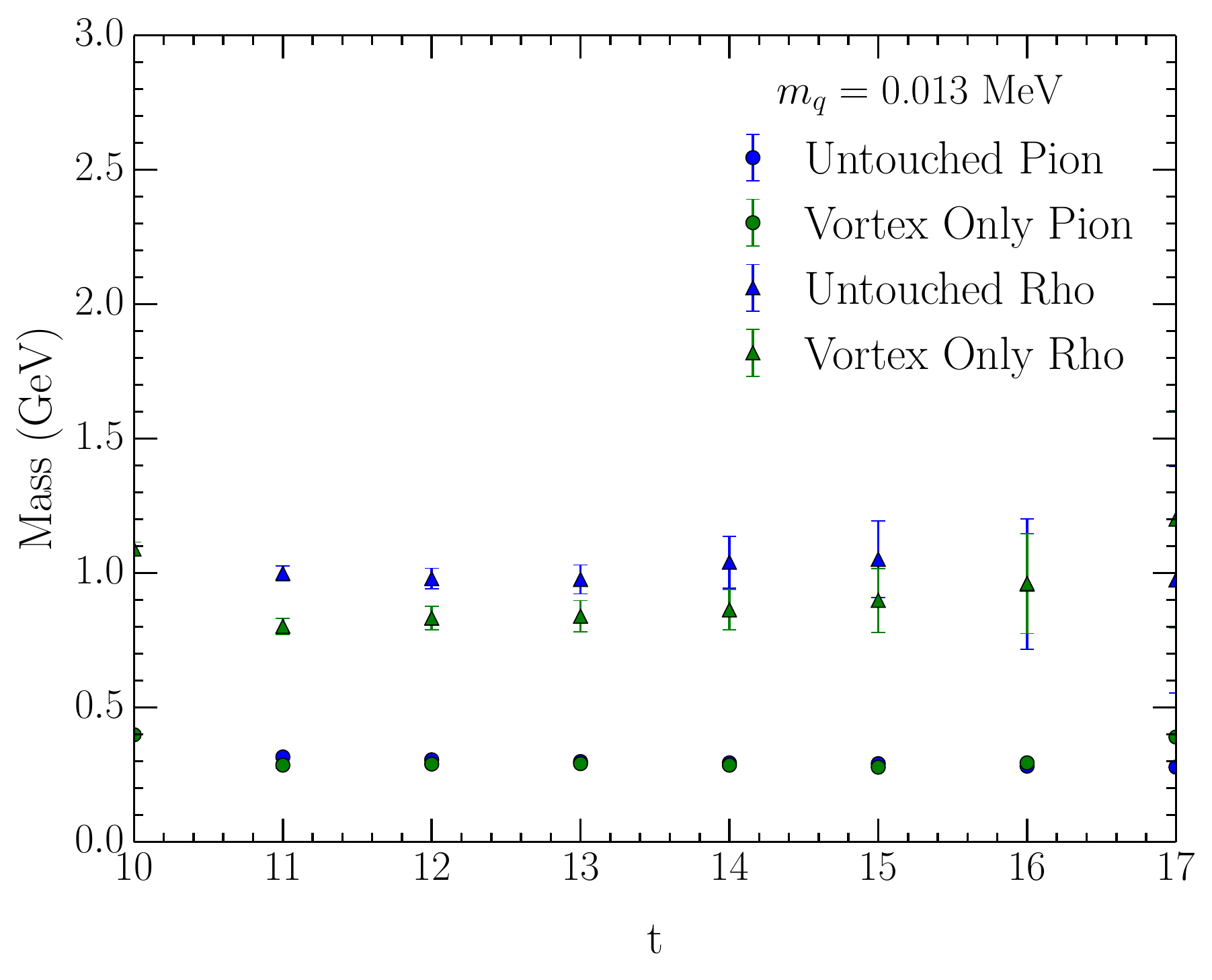}\includegraphics[width=0.4\columnwidth]{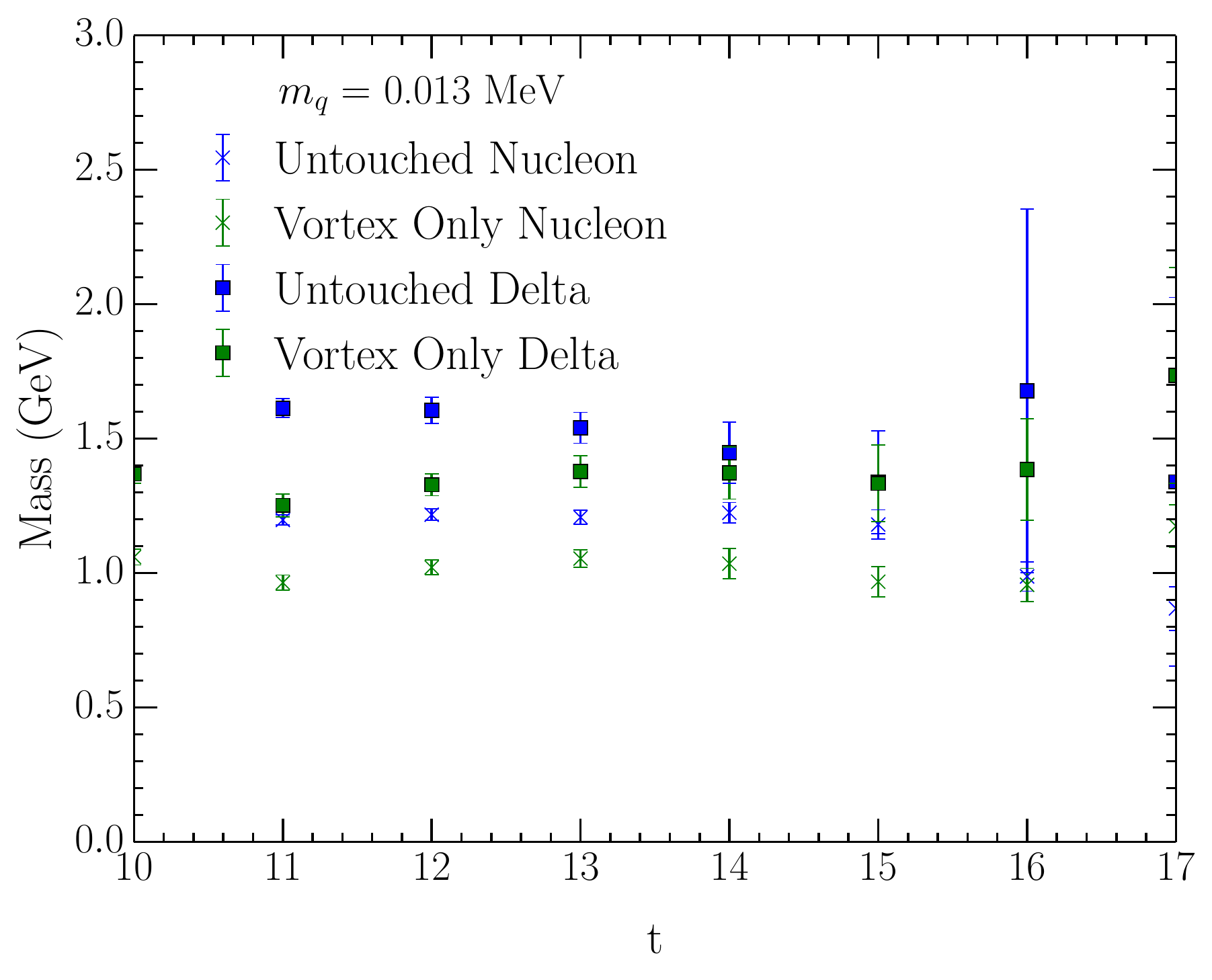}\\
  \includegraphics[width=0.4\columnwidth]{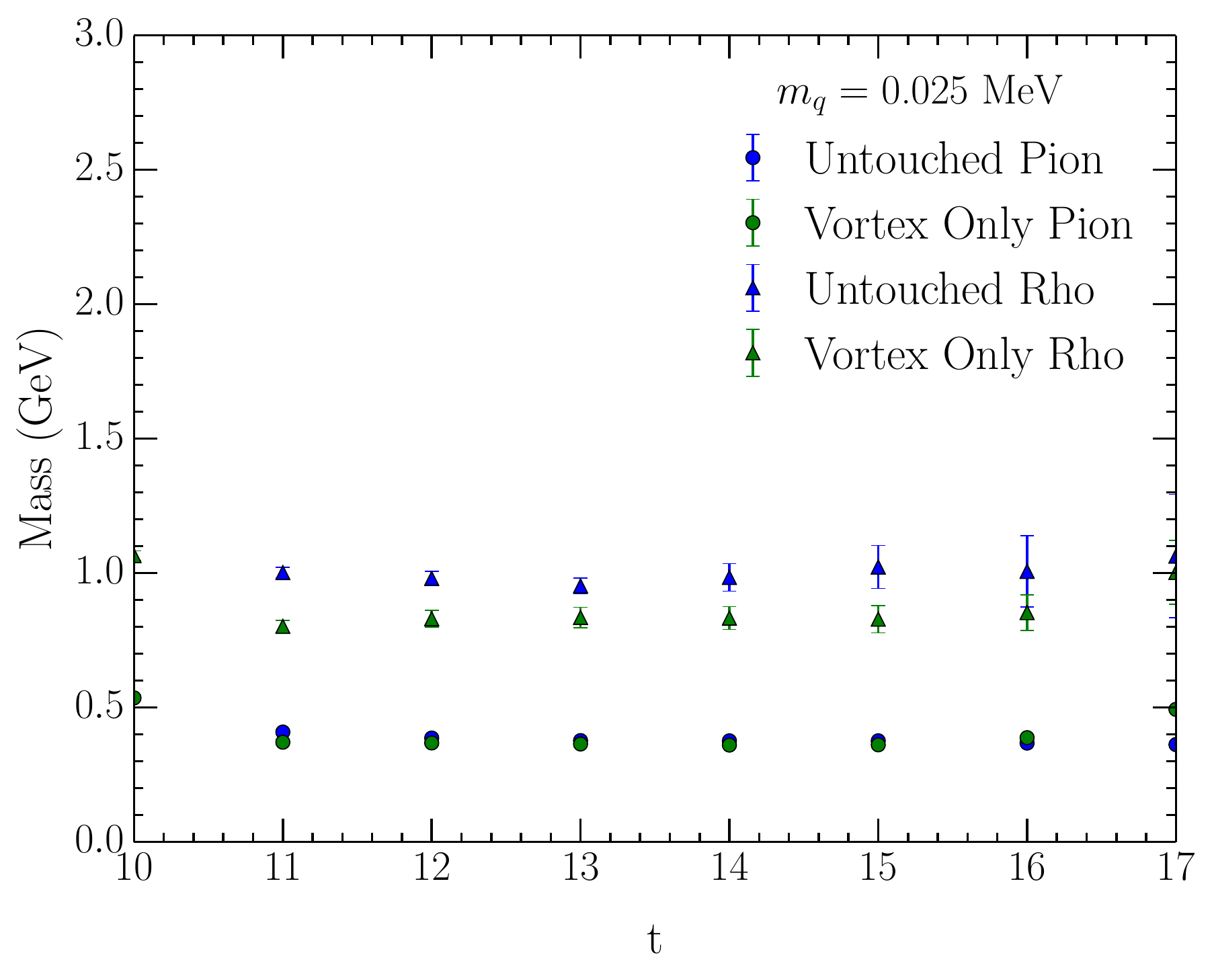}\includegraphics[width=0.4\columnwidth]{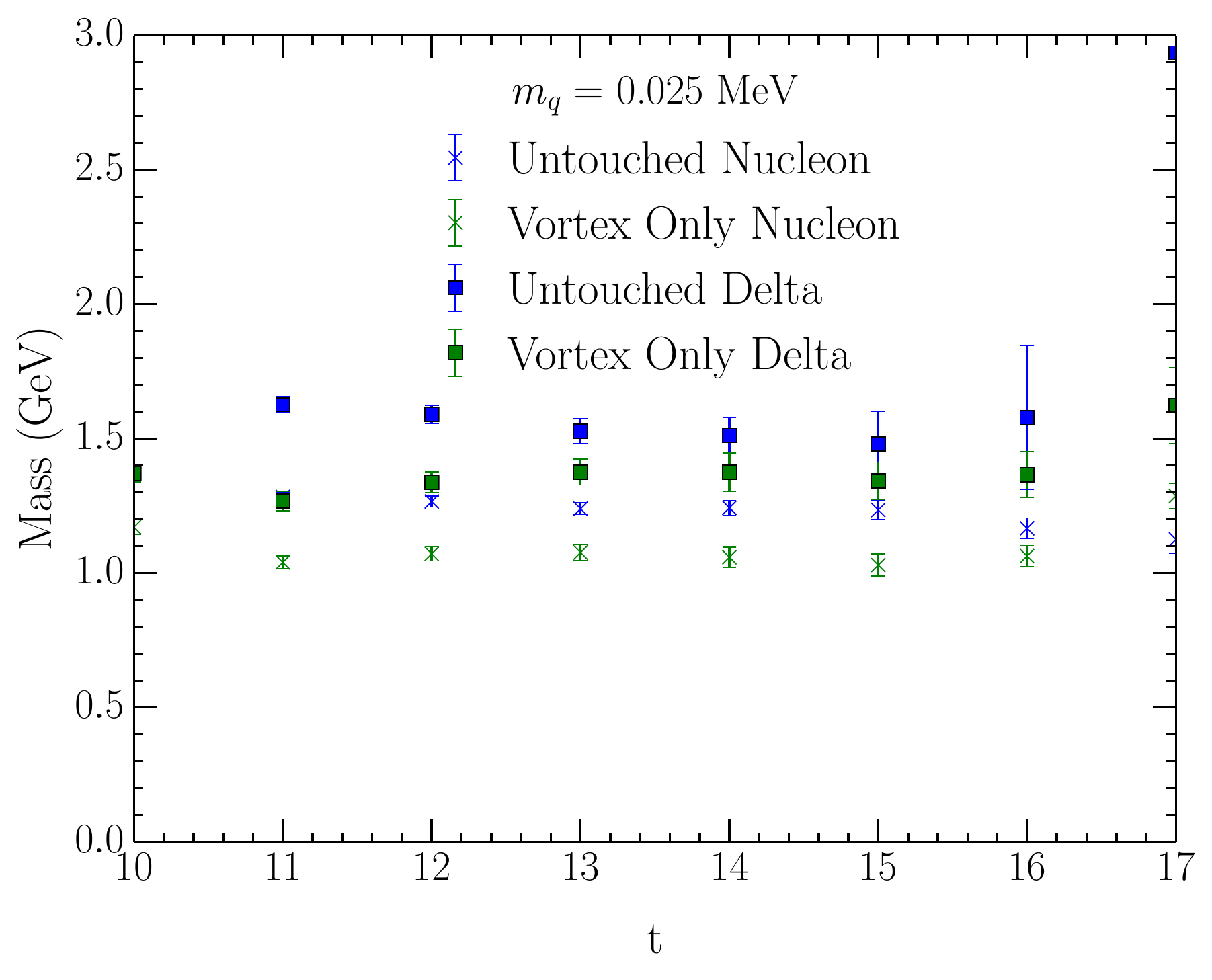}\\
  \includegraphics[width=0.4\columnwidth]{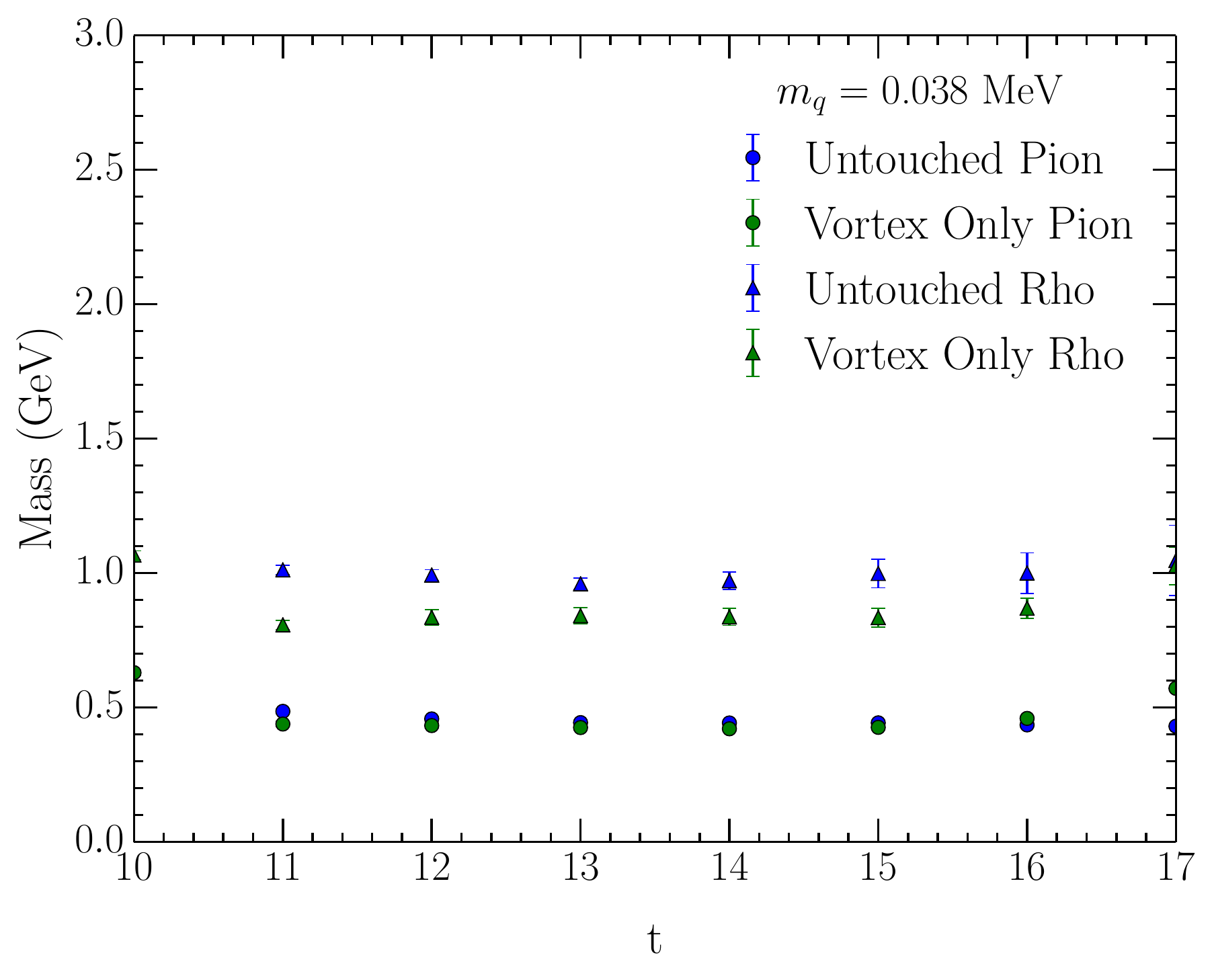}\includegraphics[width=0.4\columnwidth]{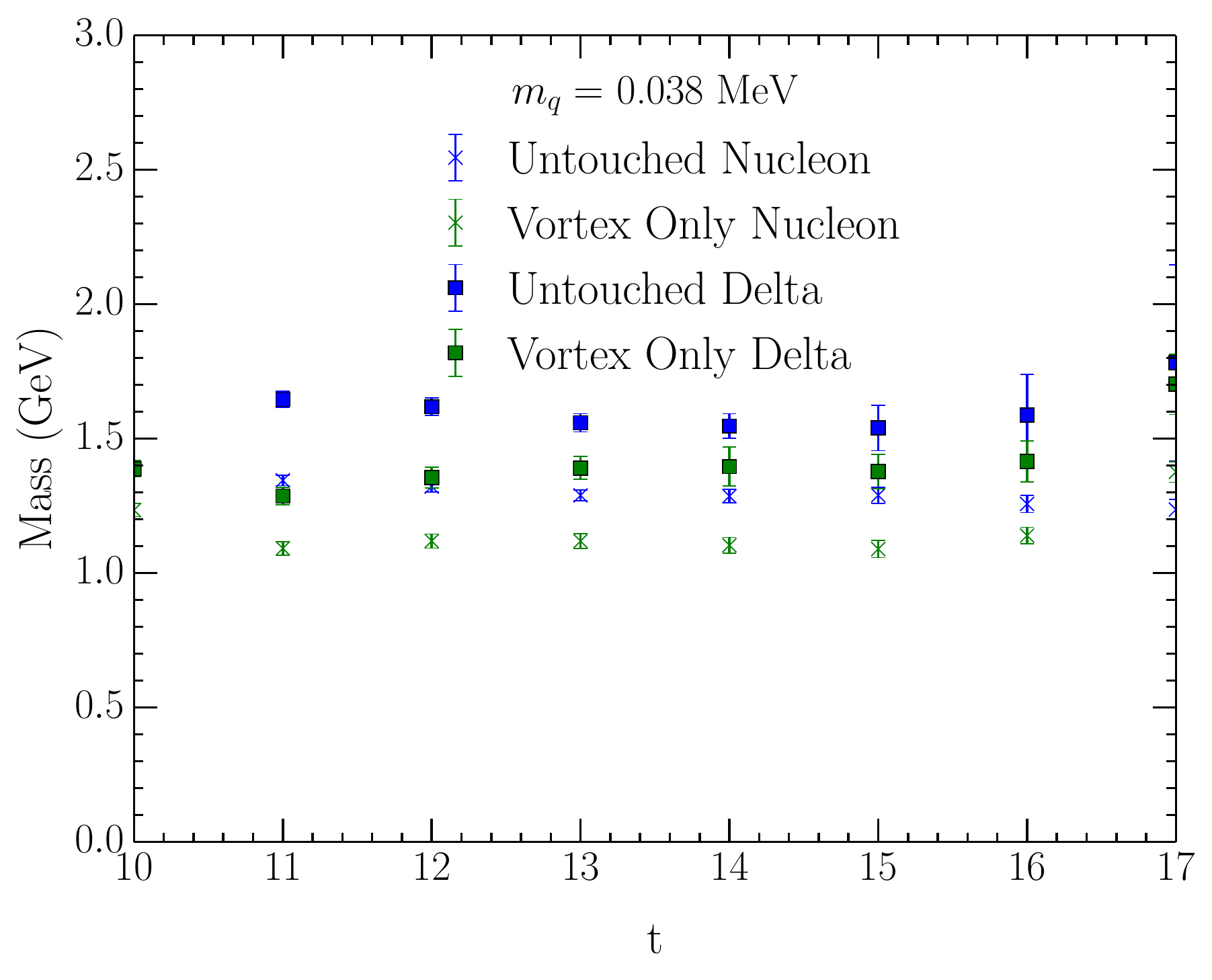}\\
  \includegraphics[width=0.4\columnwidth]{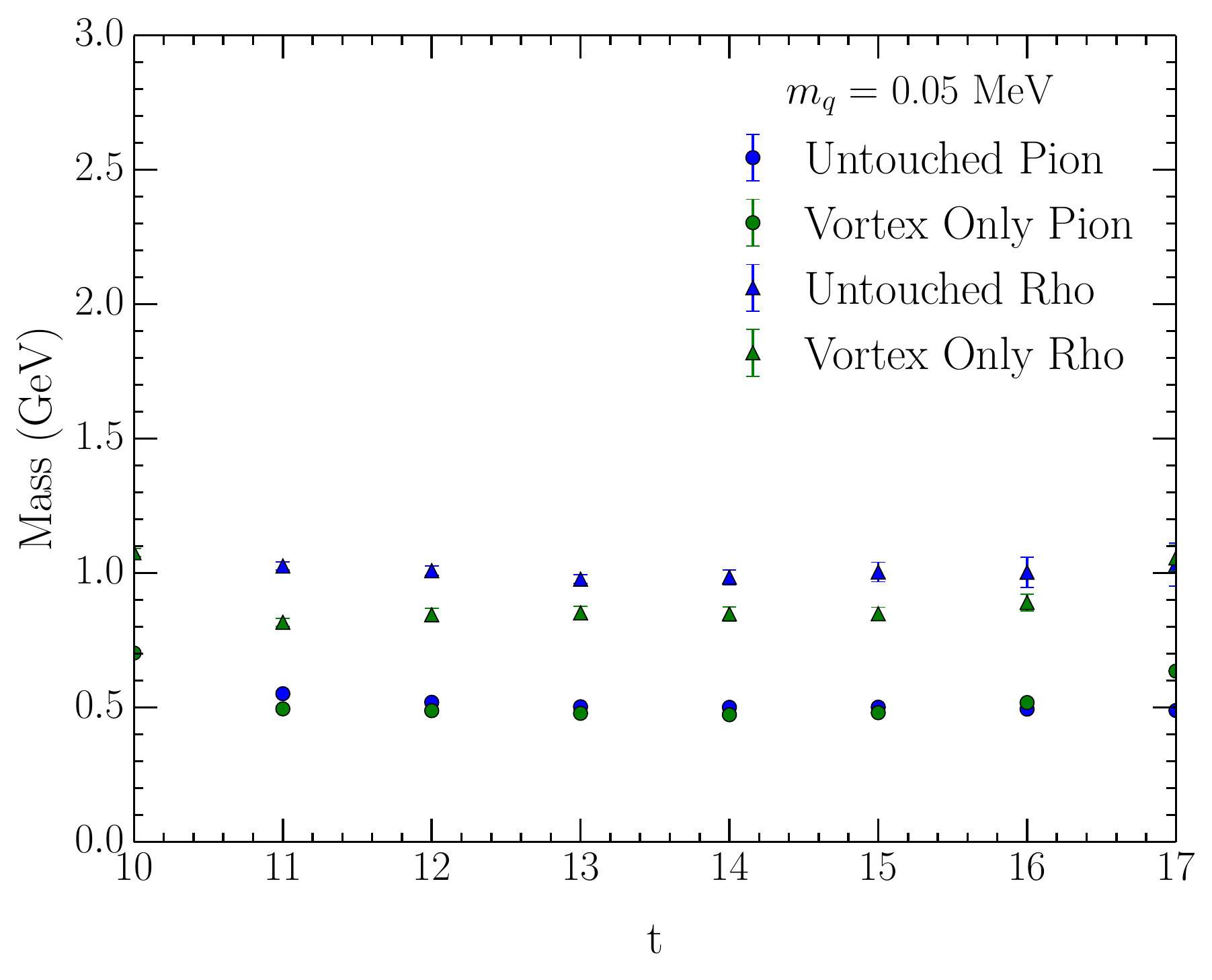}\includegraphics[width=0.4\columnwidth]{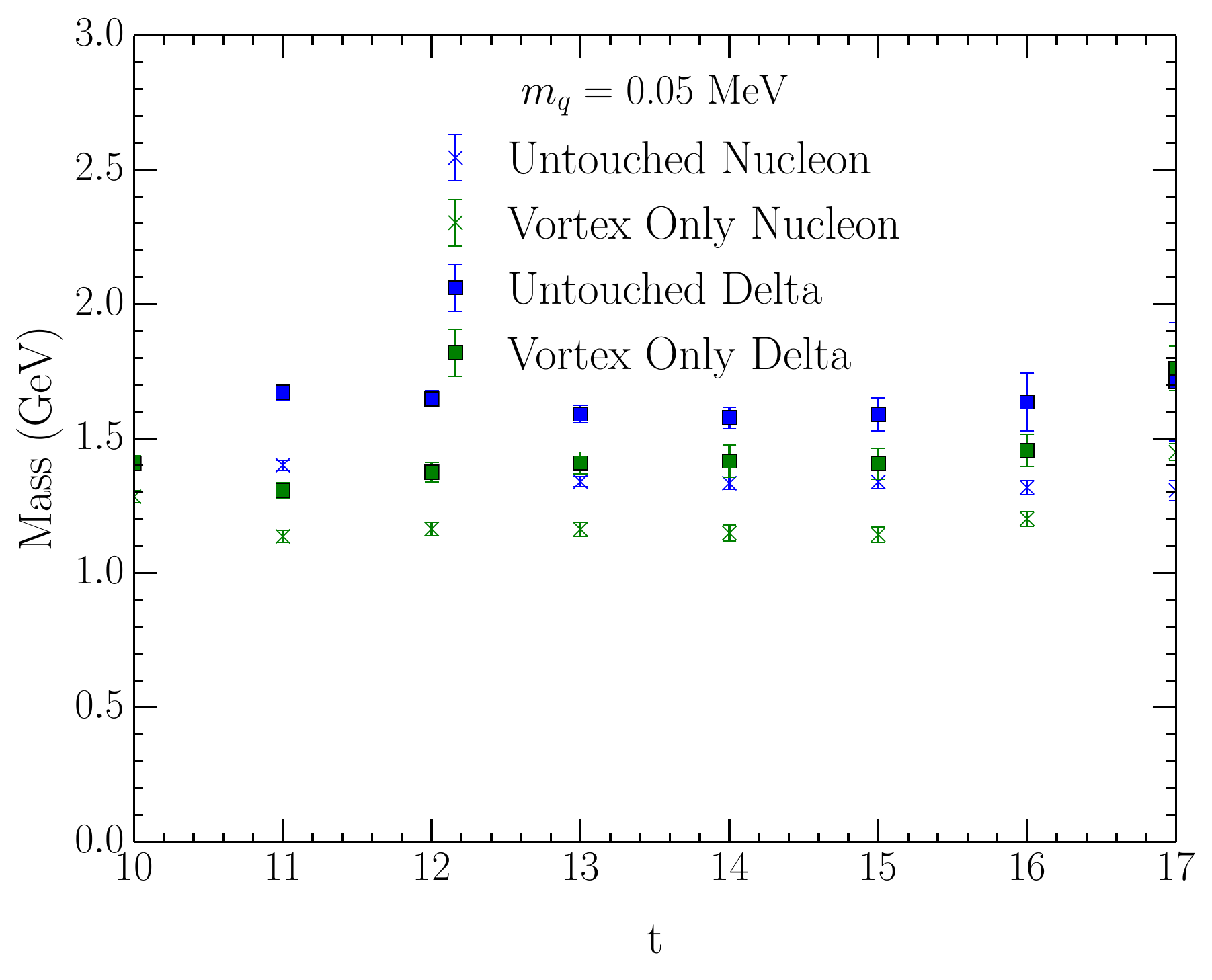}
  \caption{The effective masses for the low-lying mesons (left) and baryons (right) on the untouched (blue) and vortex only (green) ensembles. Results are shown for light bare quark masses with values of $m_q=13,\ 25,\ 38,\ 50$ MeV from top to bottom respectively.}
  \label{Fig:utVOlight}
\end{figure*}

The light hadron spectrum on the untouched and vortex-only ensembles is presented for the four light quark masses in Fig.~\ref{Fig:utVOlight}.
%% , with results for the four heavy quark masses
%% shown in Fig.~\ref{Fig:utVOheavy}. 
Turning first to the lightest quark mass at $m_q=13$ MeV, results for the untouched spectrum are as expected. The pion, rho, nucleon, and Delta all have clear signals, and sit slightly heavier than their physical values. The vortex-only ensemble is able to reproduce all qualitative features of the spectrum, although masses are slightly lower. This is most likely an artifact of cooling \cite{Thomas:2014tda,Trewartha:2015nna,Trewartha:2015ida}. Notably, in both cases the pion is much lighter than the rho, a clear signal that it retains its nature as a pseudo-Goldstone boson, and thus that dynamical chiral symmetry breaking is present on the vortex-only ensemble. A clear separation between the nucleon and Delta baryons is also maintained.

%% \begin{figure*}[p]
%%   \centering
%%   \includegraphics[width=0.4\columnwidth]{Plots/03200utVOPGpirho.pdf}\includegraphics[width=0.4\columnwidth]{Plots/03200utVOPGnucdel.pdf}\\
%%   \includegraphics[width=0.4\columnwidth]{Plots/04000utVOPGpirho.pdf}\includegraphics[width=0.4\columnwidth]{Plots/04000utVOPGnucdel.pdf}\\
%%   \includegraphics[width=0.4\columnwidth]{Plots/04800utVOPGpirho.pdf}\includegraphics[width=0.4\columnwidth]{Plots/04800utVOPGnucdel.pdf}\\
%%   \includegraphics[width=0.4\columnwidth]{Plots/05600utVOPGpirho.pdf}\includegraphics[width=0.4\columnwidth]{Plots/05600utVOPGnucdel.pdf}
%%   \caption{The effective masses for the low-lying mesons (left) and
%%     baryons (right) on the untouched (blue) and vortex only (green)
%%     ensembles. Results are shown for heavy bare quark masses with
%%     values of $m_q=101,126,151,177$ MeV from top to bottom
%%     respectively.}
%%   \label{Fig:utVOheavy}
%% \end{figure*}

The behaviour seen at the lightest quark mass is replicated across the remaining seven quark masses considered. There are clear signals for all four hadrons in both the untouched and vortex-only cases. Again, the masses on the vortex-only ensemble are slightly smaller than those in the untouched ensemble, but the qualitative features with regard to the ordering of the different hadrons are reproduced.

%% Good signal is obtained at the four light quark masses up to $t=17.$
%% The four heavy quark masses enable an exploration of the vortex-only
%% correlators at very large Euclidean times, extending to $t=24.$ An
%% interesting artefact in the vortex-only effective mass plots appears
%% in the range $t=17-20$, where the effective masses rise temporarily
%% before returning to their previous values as the signal
%% degrades. Understanding the nature of this asymptotic behaviour will
%% require further study. If the effect is robust, then possible
%% underlying candidates for the temporary rise include artefacts

%% selection of boundary conditions.  \textcolor{red}{In the following we
%%   focus on results for $t<17$.}

\begin{figure*}[tb]
  \centering
  \includegraphics[width=0.32\columnwidth]{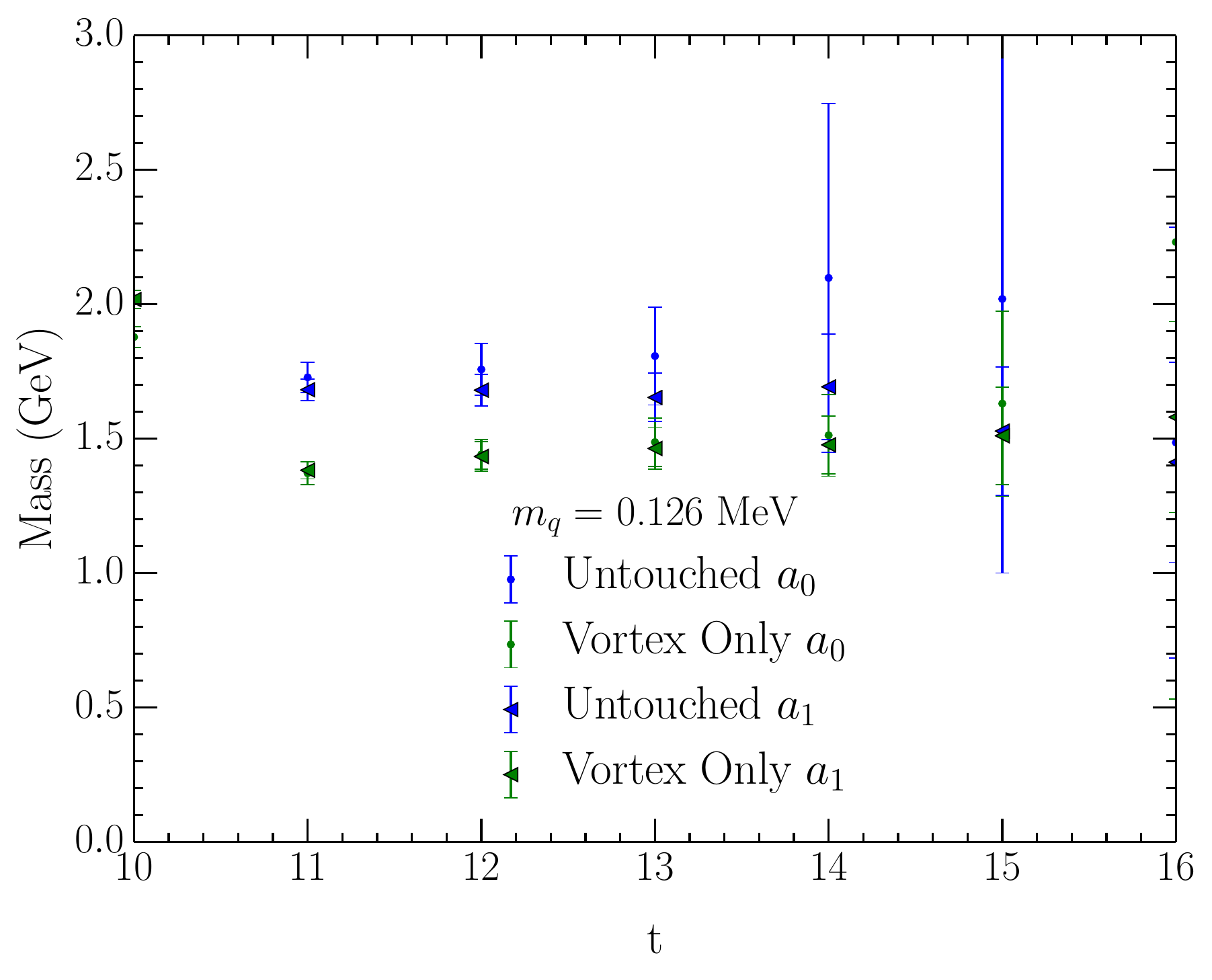}
  \includegraphics[width=0.32\columnwidth]{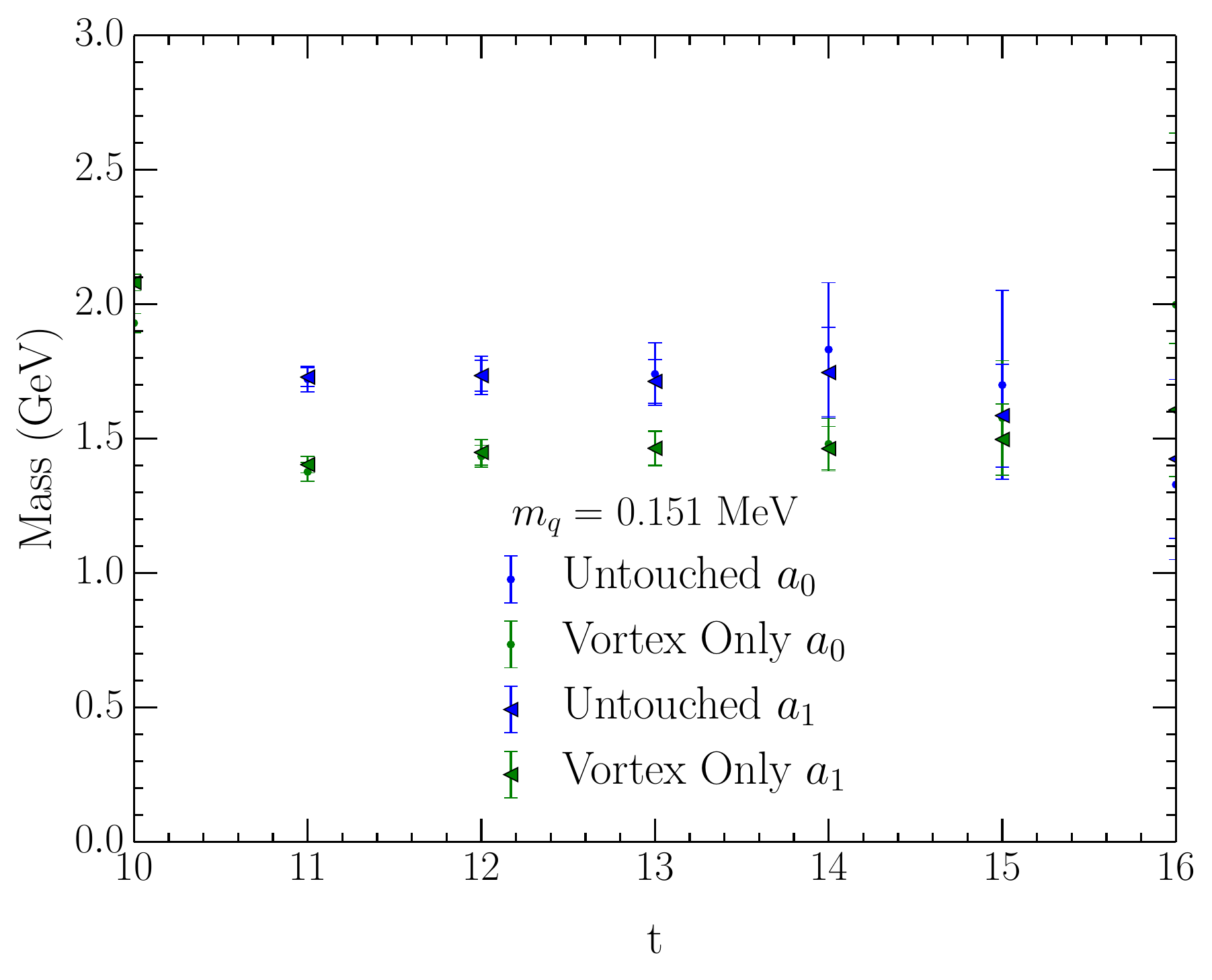}
  \includegraphics[width=0.32\columnwidth]{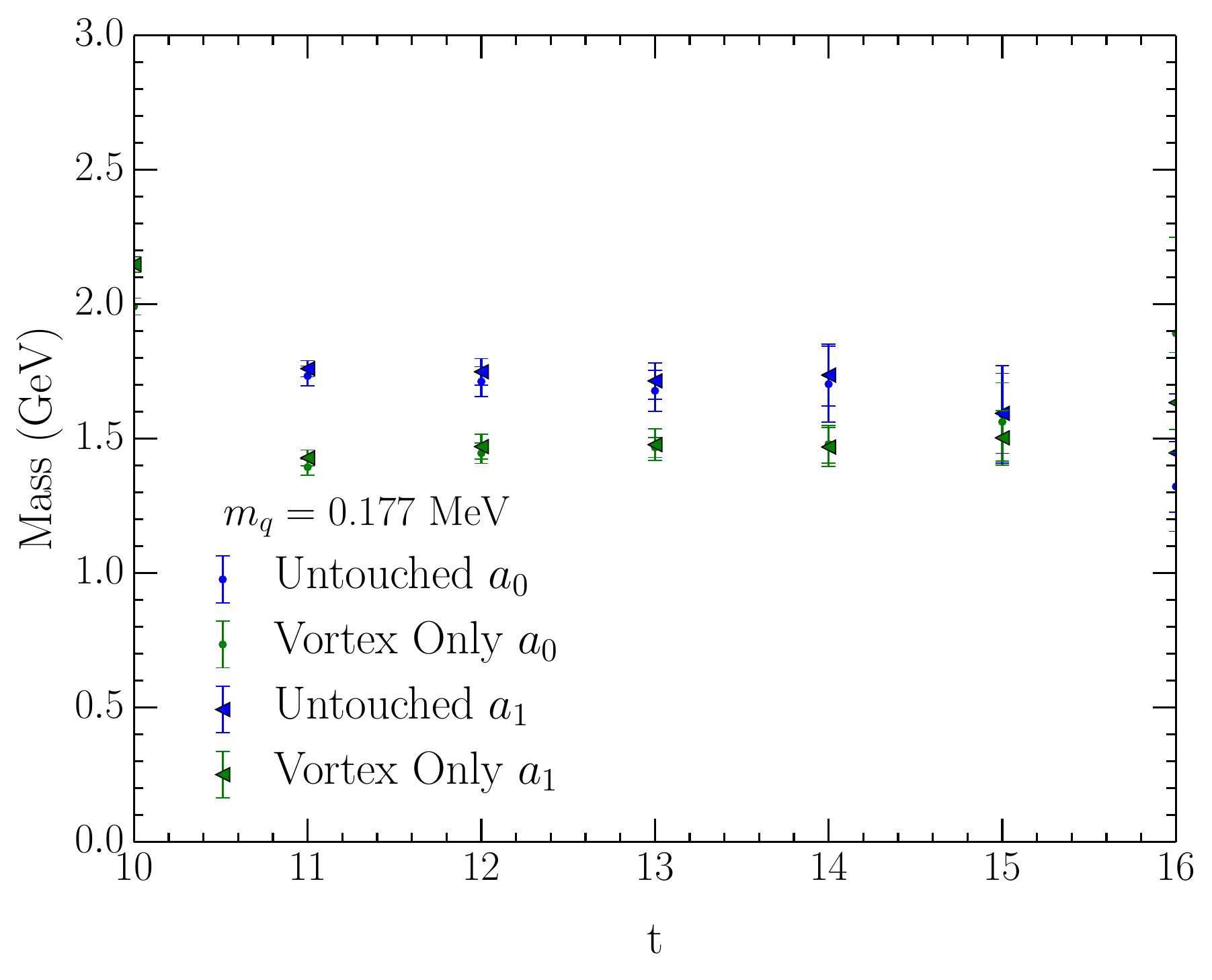}
  \caption{The effective masses for the $a_{0}$ and $a_{1}$ mesons on the untouched (blue) and vortex-only (green) ensembles, at bare quark masses of $126$ (left), $151$ (middle), and $177$ (right) MeV.}
  \label{Fig:utVOa0a1}
\end{figure*}

At low quark masses, the signal for the $a_{0}$ and $a_{1}$ mesons is too poor to allow study.  We show these results exclusively at the three heaviest quark masses, $126$, $151$, and $177$ MeV, in Fig.~\ref{Fig:utVOa0a1}. At these masses, the $a_{0}$ and $a_{1}$ are approximately degenerate on the untouched ensemble, and this is reproduced on the vortex-only ensemble. Again, while the masses are slightly lower, the qualitative features of the hadron spectrum are reproduced.

In pure gauge theory on the lattice, the $\eta'$ meson cannot gain mass from repeated $q\,\bar{q}$ annihilation due to the lack of disconnected fermion loops.  Therefore it does not have the relatively large mass given to it by the axial anomaly in QCD. This leads to an $\eta' - \pi$ `ghost' state in the scalar meson channel, which gives a negative-metric contribution to the two-point correlator~\cite{Bardeen:2001jm}. This effect is most pronounced at low quark masses; at higher quark masses this two-particle ghost state is much higher in energy than the single-particle state, and thus does not contribute at large Euclidean times.  This phenomenon is responsible for the difficulties in measuring the $a_{0}$ and $a_{1}$ at low quark masses.

In summary, gauge field configurations created from the centre vortices identified in the original gauge field configurations in Maximal Centre Gauge (MCG) are able to capture the essence of the QCD vacuum structure. We observe the spin splittings between the nucleon and Delta baryons to be preserved and the pion maintains its pseudo-Goldstone boson nature in the light quark-mass regime. While some reduction in the mass of the low-lying hadron spectrum is seen, it can be attributed to the small amount of cooling applied that is necessary to evolve the thin P-vortices (identified from the plaquette values in MCG) towards the physical \emph{thick} vortices that describe the topologically nontrivial QCD vacuum. Simultaneously, the cooling ensures that the smoothness condition required for the overlap operator is satisfied.

  %This suggests that the \emph{thin} centre vortices identified on the lattice are the seeds of instantons, evolved under cooling to realise the physical \emph{thick} vortices , and

\section{Chiral Symmetry Restoration}

Before presenting results for the vortex-removed ensemble, it is worth carefully considering our expectations for the ground-state hadron spectrum upon removal of dynamical chiral symmetry breaking. 

Under the complete restoration of chiral symmetry, we expect baryon currents related by chiral transformations to become degenerate. The massless QCD Lagrangian has an $\SU{2}_{L} \times \SU{2}_{R} \times \mathrm{U}(1)_{A}$ symmetry. The $\mathrm{U}(1)_{A}$ symmetry is, however, explicitly broken by the axial anomaly. We must therefore admit the possibility that the $\mathrm{U}(1)_{A}$ and $\SU{2}_{L} \times \SU{2}_{R}$ symmetries are restored separately. The complete restoration of chiral symmetry would imply the following degeneracies~\cite{Cohen:1996sb},
\begin{eqnarray}
\pi & \leftrightarrow & a_{0} \quad[\UA] \nonumber \\
\rho & \leftrightarrow & a_{1} \quad[\CS] \nonumber \\
\mathrm{N} & \leftrightarrow &\, \Delta \quad [\CS]\,. \nonumber \\
\end{eqnarray}
In lattice simulations a non-zero bare quark mass is used. Chiral symmetry is thus explicitly broken even in the absence of dynamical chiral symmetry breaking. At small bare quark masses, the explicit breaking of chiral symmetry is negligible and so we expect these degeneracies to be manifest.

At larger masses, chiral symmetry no longer holds even approximately. We thus expect to see something close to a non-interacting constituent-quark like model, where the mass of each state is simply the sum of the dressed quark masses composing it, possibly with some momentum. This is the result seen in Ref.~\cite{OMalley:2011aa}; degenerate $\pi$- and $\rho$-meson masses were observed, even though the two are not related by a chiral transformation. The mesons had a mass of approximately $2/3$ of the mass of the baryons.

\begin{figure*}[!b]
\centerline{\includegraphics[width=0.3\columnwidth]{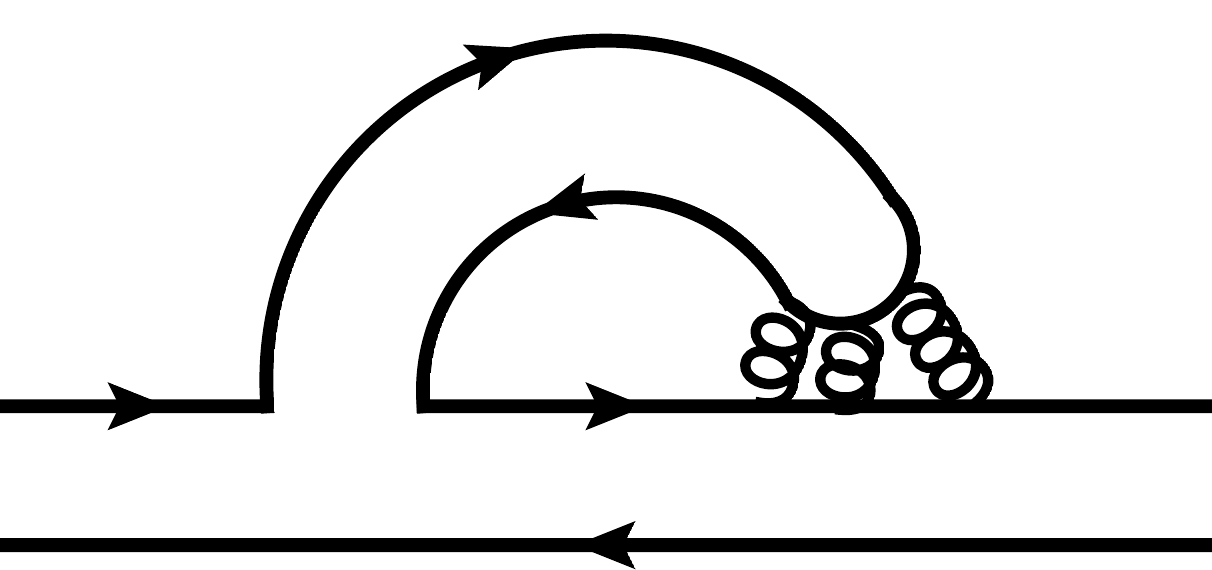}}
\caption{The ``hairpin'' diagram, showing a $\pi$-$\eta'$ or $\rho$-$\eta'$ intermediate state with the quantum numbers of the $a_{0}$ or $a_{1}$ mesons respectively.}
\label{Fig:hairpin}
\end{figure*}

As we are considering pure gauge theory, one must also consider multi-particle states contributing to the $a_{0}$ and $a_{1}$ correlators. In the pure gauge sector, quark flows such as the ``hairpin'', illustrated in Fig.~\ref{Fig:hairpin}, can carry the quantum numbers of the $a_{0}$ or $a_{1}$ through $\pi$-$\eta'$ or $\rho$-$\eta'$ intermediate states respectively. Because the sea-quark loops vital to generating the mass of the singlet $\eta'$ meson are absent, $m_{\eta'} = m_{\pi}$. The associated mass thresholds of these multi-particle states carrying the quantum numbers of the $a_{0}$ and $a_{1}$ are thus $2\,m_{\pi}$ and $m_{\pi} + m_{\rho}$ respectively. We will refer to these multi-particle states as $\pi$-$\eta'$ and $\rho$-$\eta'$ states.

% Double hairpin
%
%
\begin{figure*}[!t]
\centerline{\includegraphics[width=0.3\columnwidth]{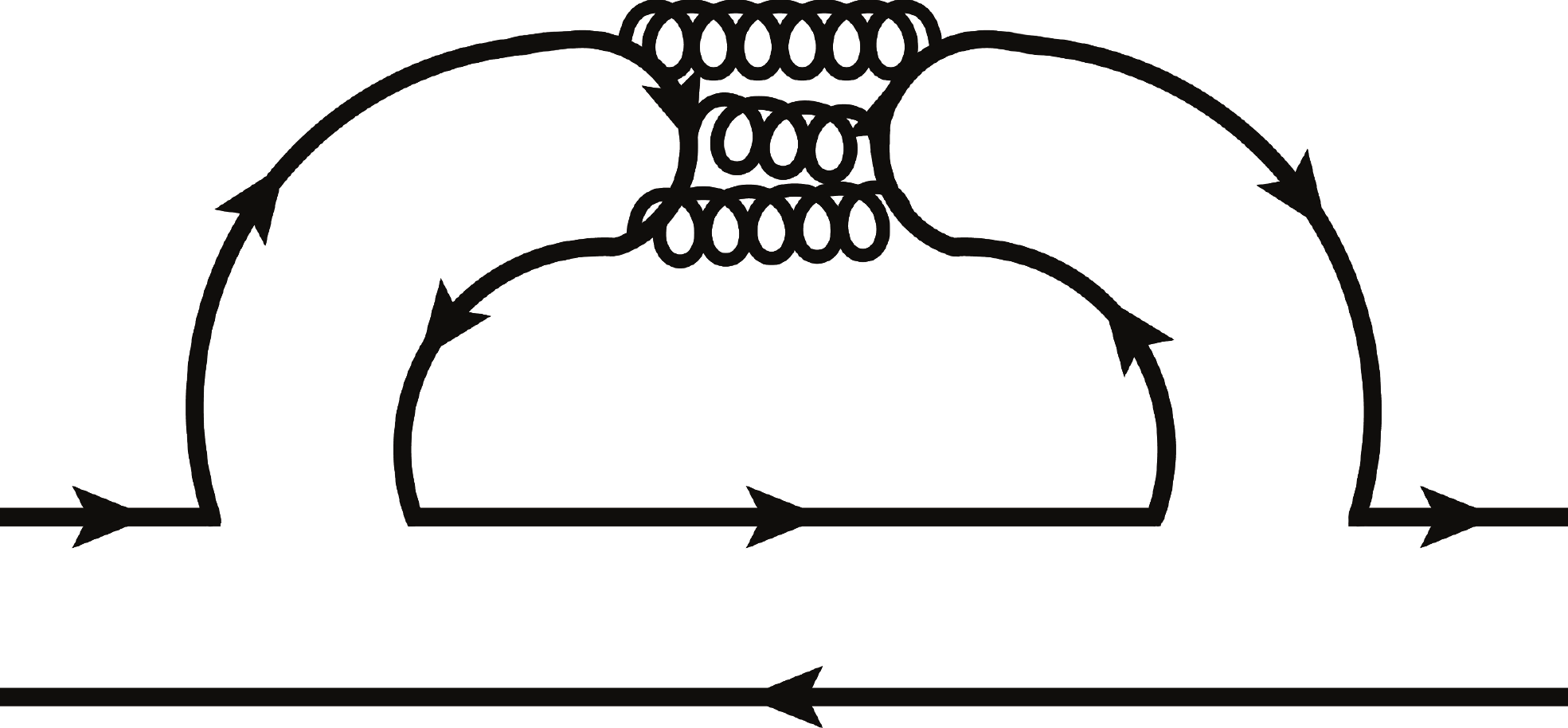}}
\caption{The ``double-hairpin'' diagram, associated with a negative-metric contribution to the $a_{0}$ and $a_{1}$ correlators.}
\label{Fig:doublehairpin}
\end{figure*}

One might also be concerned about the ``double-hairpin'' graph illustrated in Fig.~\ref{Fig:doublehairpin} that can provide a negative-metric contribution to the correlators. However, all of our correlators on the vortex-removed configurations remain positive. There is some evidence of a nontrivial contribution in the $a_{0}$ correlator at the lightest quark mass considered, as its effective mass function rises sharply from below at the earliest Euclidean times.  This is highlighted in the discussion below.

In summary, if vortex removal does indeed result in the loss of dynamical chiral symmetry breaking, we can make a number of predictions. At light quark masses, we expect a chiral regime to hold where the hadron spectrum has the following qualities:
\begin{itemize}
\item
In the absence of dynamical chiral symmetry breaking, the pion is no longer a pseudo-Goldstone boson, and so there is no \textit{a priori} reason for it to have a much lower mass than the other mesons.
\item
The restoration of the $\mathrm{U}(1)_{A}$ symmetry will be shown by the degeneracy of the $\pi$ and ground state $a_{0}$ at low quark masses.
\item
The restoration of the $\SU{2}_{L} \times \SU{2}_{R}$ symmetry will be shown by the degeneracy of the $\rho$ and ground state $a_{1}$ at low quark masses.
\item
The $N$ and $\Delta$ should also be degenerate via $\SU{2}_{L} \times \SU{2}_{R}$ symmetry.
\item
There is no chiral transformation relating the $\pi$ and $\rho$ mesons, and so at light quark masses we expect the two to differ in mass.
\end{itemize}
At heavy quark masses, we expect a constituent regime to hold where the light hadron masses should simply be estimated by counting quarks. However, it should be noted that, due to their positive parity, there is no way to make the quantum numbers of the $a_{0}$ or $a_1$ with two constituent quarks at rest. To create overlap with an $l=1$ orbital angular momentum state needed for positive parity, we must excite at least one of the constituent quarks with the lowest non-trivial momentum available on the lattice. Hence, for the hadron spectrum in the constituent regime we predict the following:
\begin{itemize}
\item
  The $\pi$ should be degenerate with the $\rho$ at high quark masses. Likewise, the $N$ and $\Delta$ should be degenerate, each with a mass $3/2$ times that of the mesons.
\item
  The $a_0$ mass will be the lower of two possibilities: a $\pi$-$\eta'$ state with mass $2\,m_{\pi}$, or a two quark state excited with the lowest non-trivial momentum  
\item
  Similarly, the $a_{1}$ mass should be the lower of two possibilities: a $\rho$-$\eta'$ state, or a two quark state excited with the lowest non-trivial momentum.
\end{itemize}
Note that the most interesting predictions are within the meson spectrum. The baryon spectrum is simple, as the nucleon and Delta are expected be degenerate at all quark masses.  In the constituent regime they are both composed of three dressed quarks, while in the chiral regime they are related through symmetry restoration.

\section{Vortex-Removed Hadron Spectrum}

%% \begin{figure*}[thpb]
%% \subfigure[]{
%% \label{00400VRmesons}
%% \includegraphics[width=0.5\columnwidth]{Plots/00400VRmesons.pdf}}
%% \subfigure[]{
%% \label{00400VRbaryons}
%% \includegraphics[width=0.5\columnwidth]{Plots/00400VRbaryons.pdf}}
%% \caption{The low-lying mesons (left) and baryons (right) considered on the vortex-removed ensemble, at a bare quark mass of $13$ MeV.}
%% \label{Fig:00400VRhadrons}
%% \end{figure*}

\begin{figure*}[p]
  \centering
  \includegraphics[width=0.4\columnwidth]{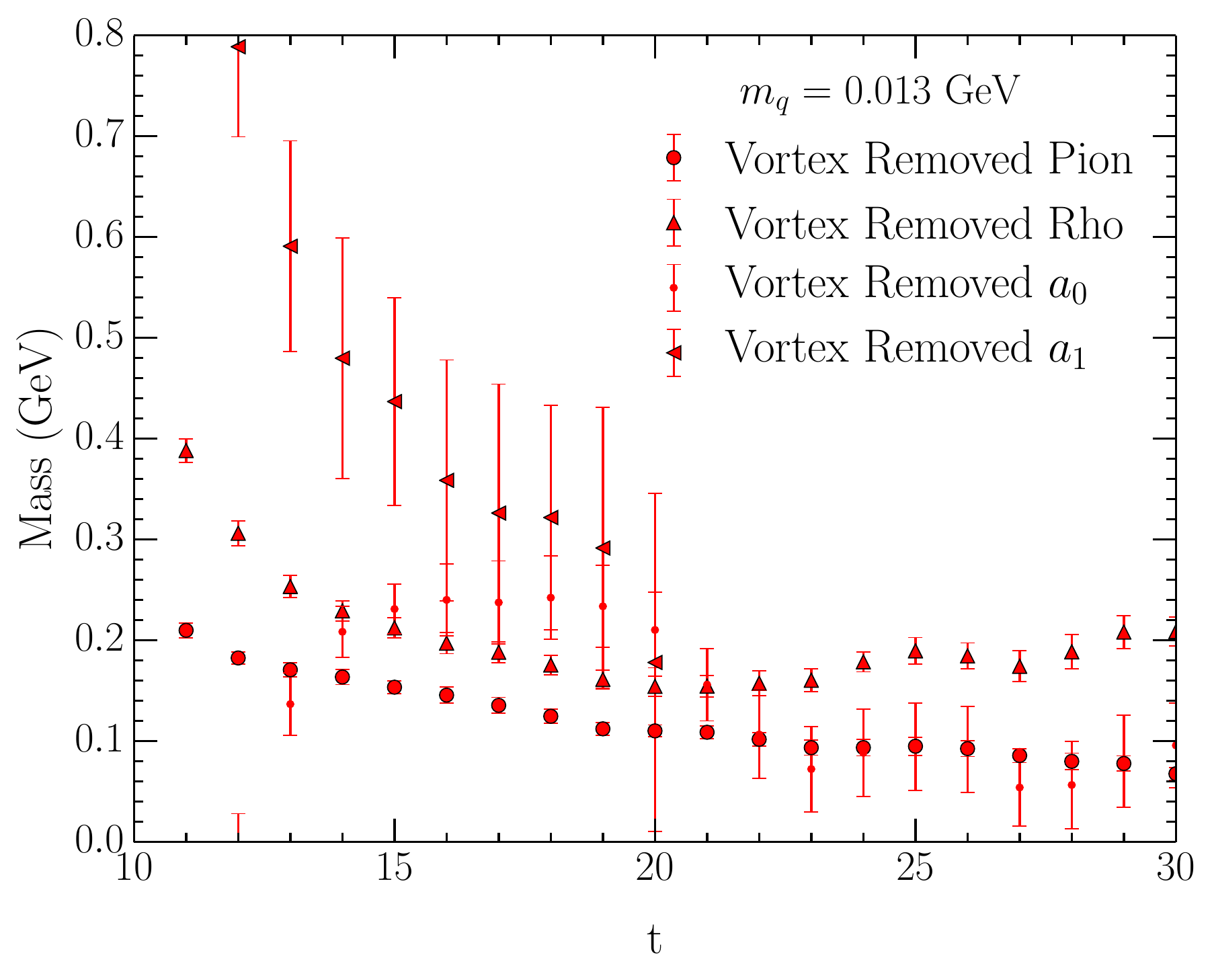}\includegraphics[width=0.4\columnwidth]{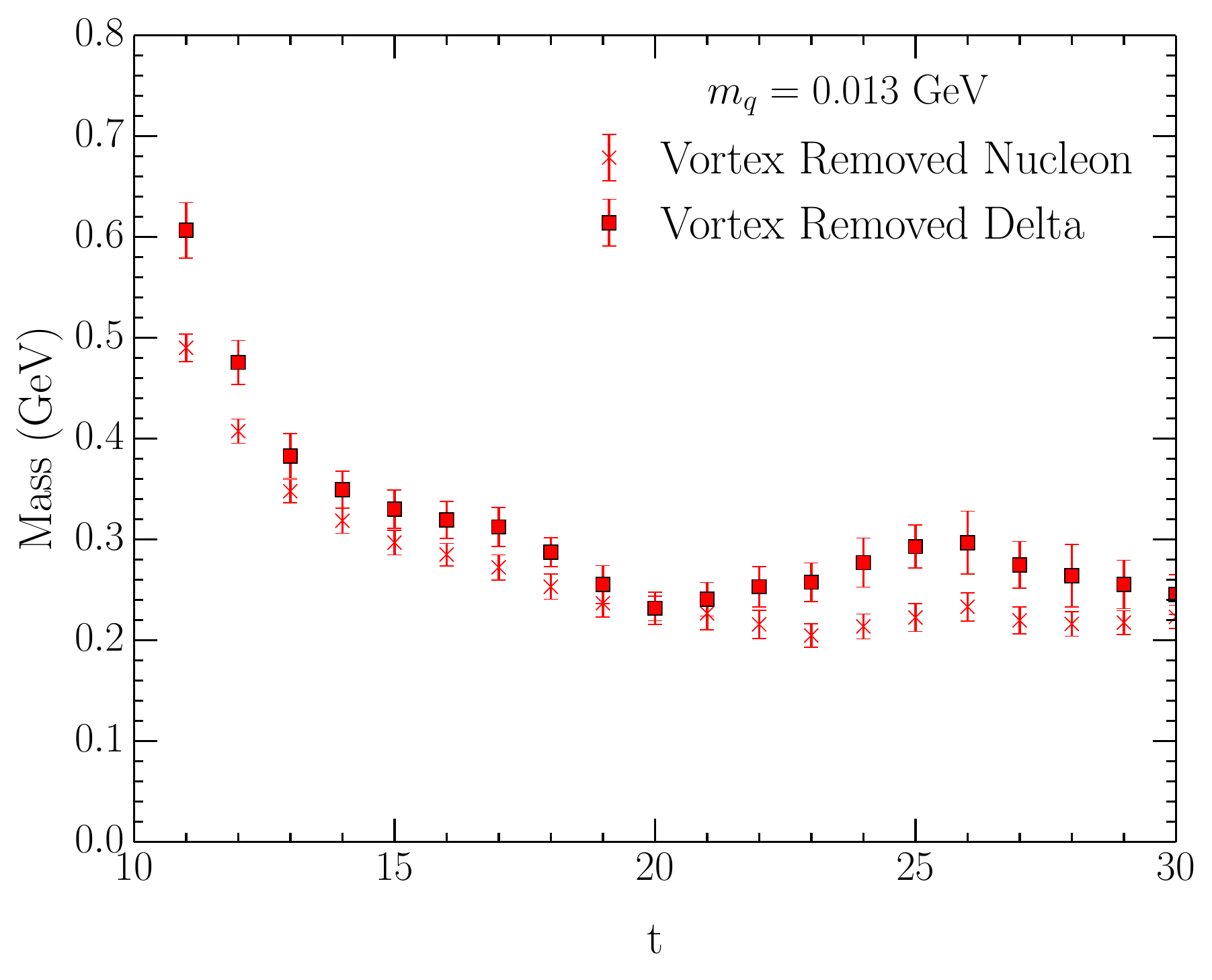}\\
  \includegraphics[width=0.4\columnwidth]{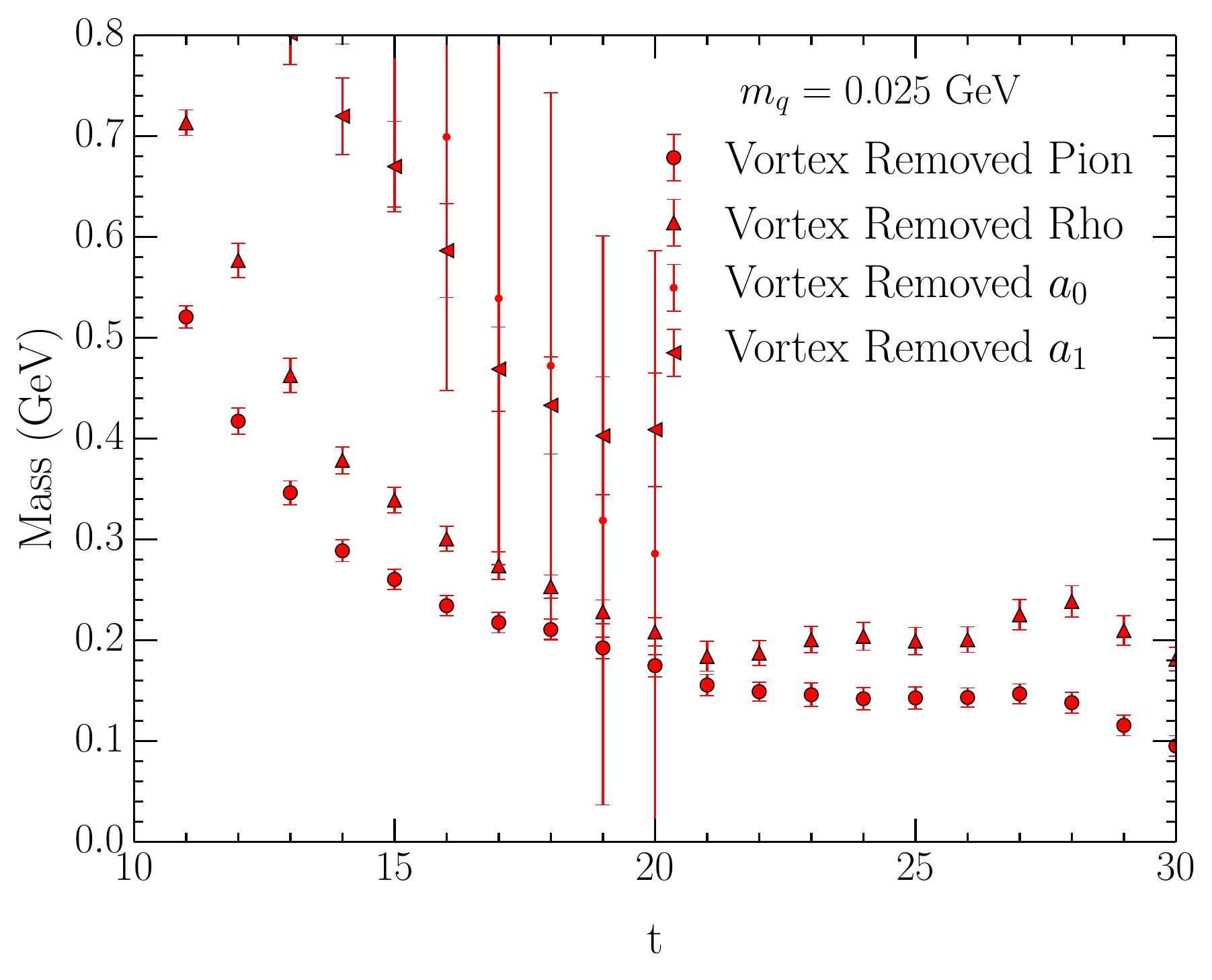}\includegraphics[width=0.4\columnwidth]{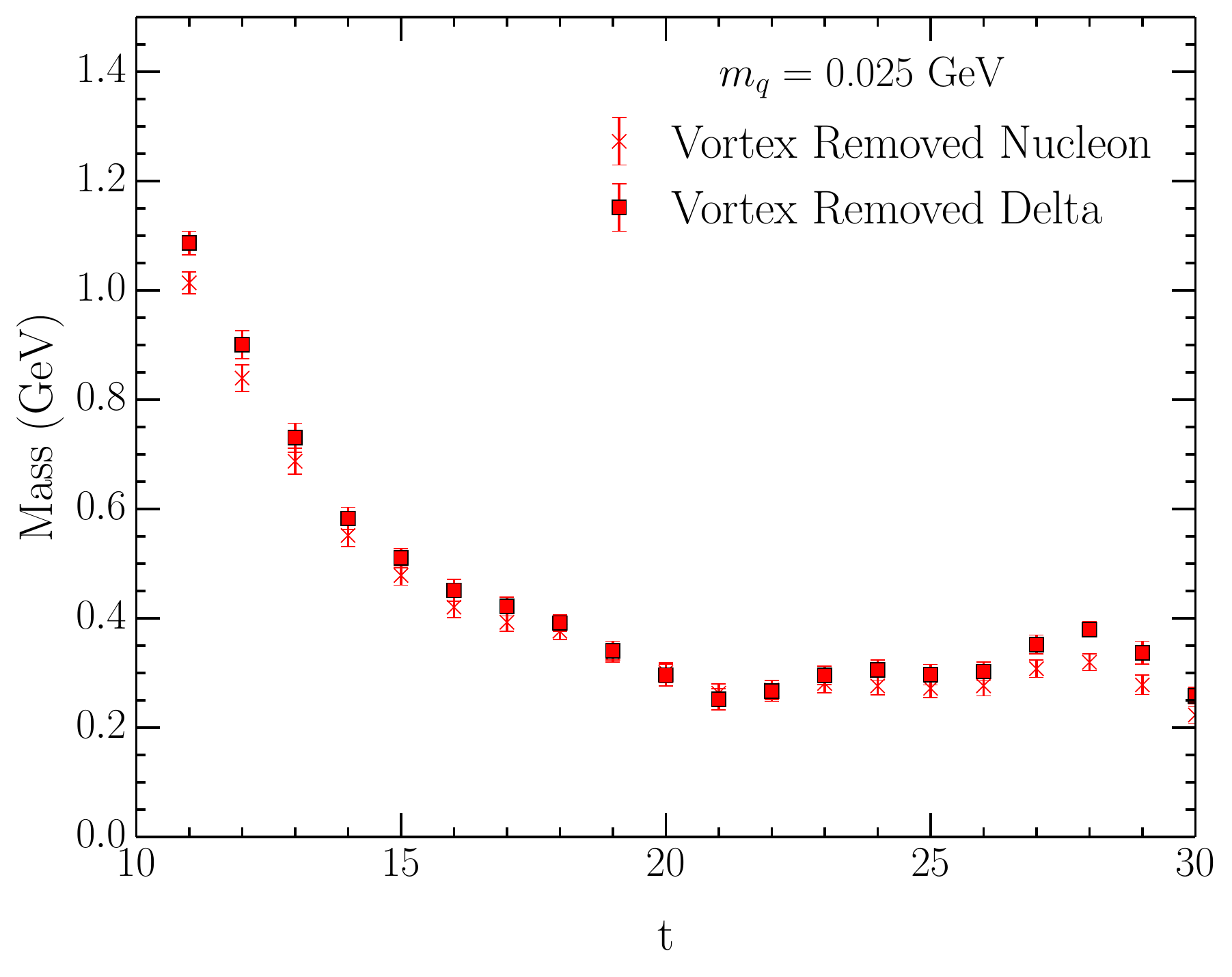}\\
  \includegraphics[width=0.4\columnwidth]{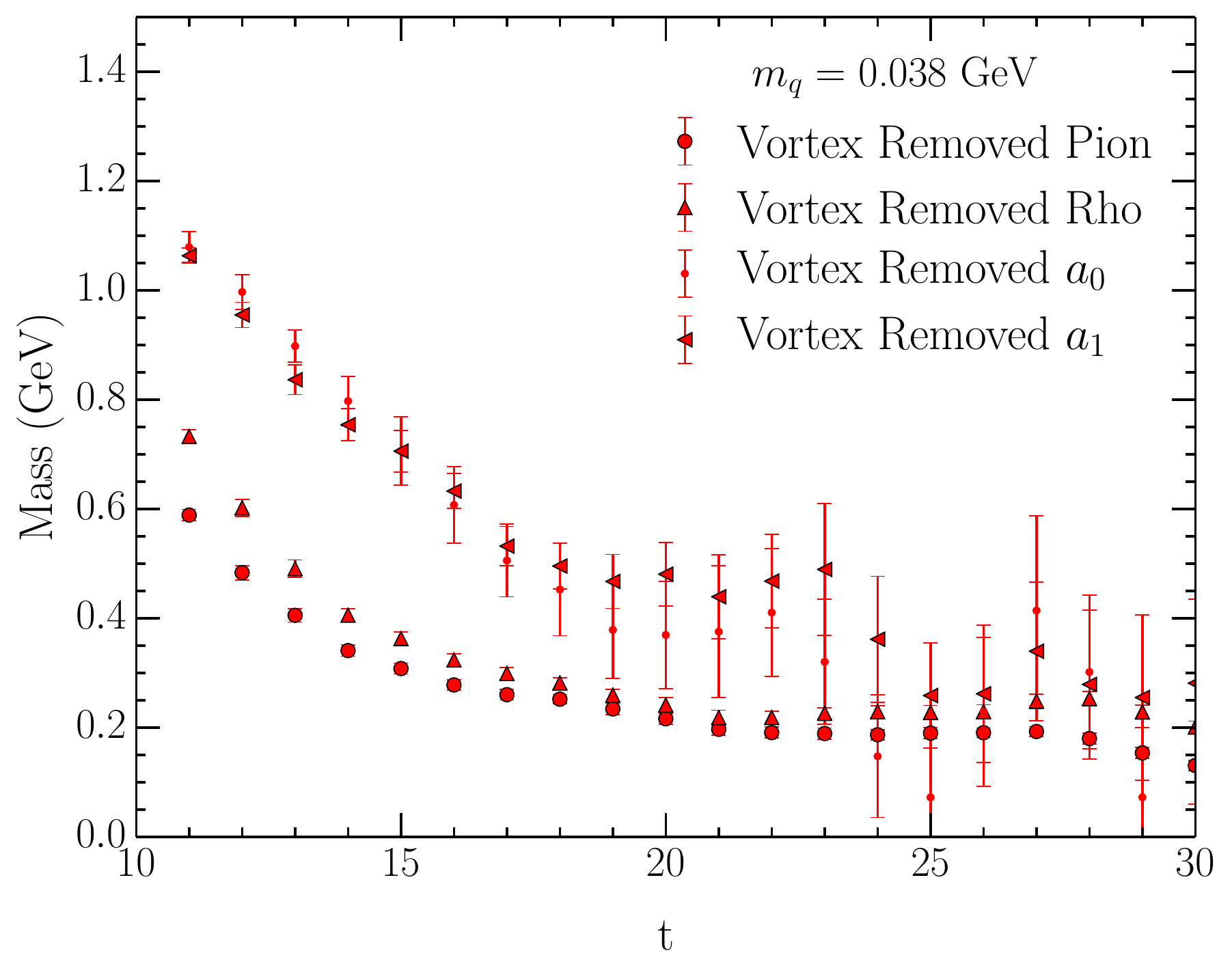}\includegraphics[width=0.4\columnwidth]{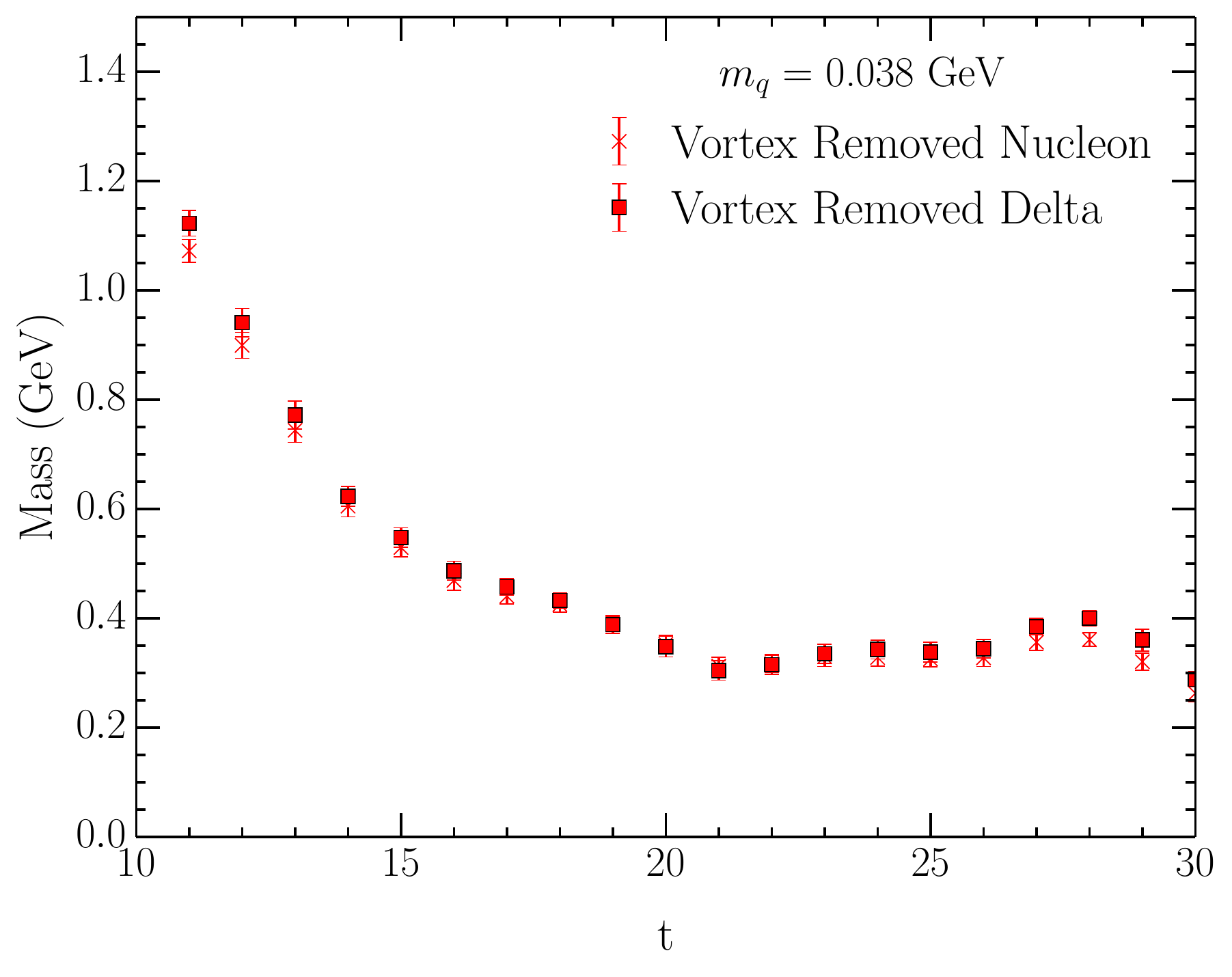}\\
  \includegraphics[width=0.4\columnwidth]{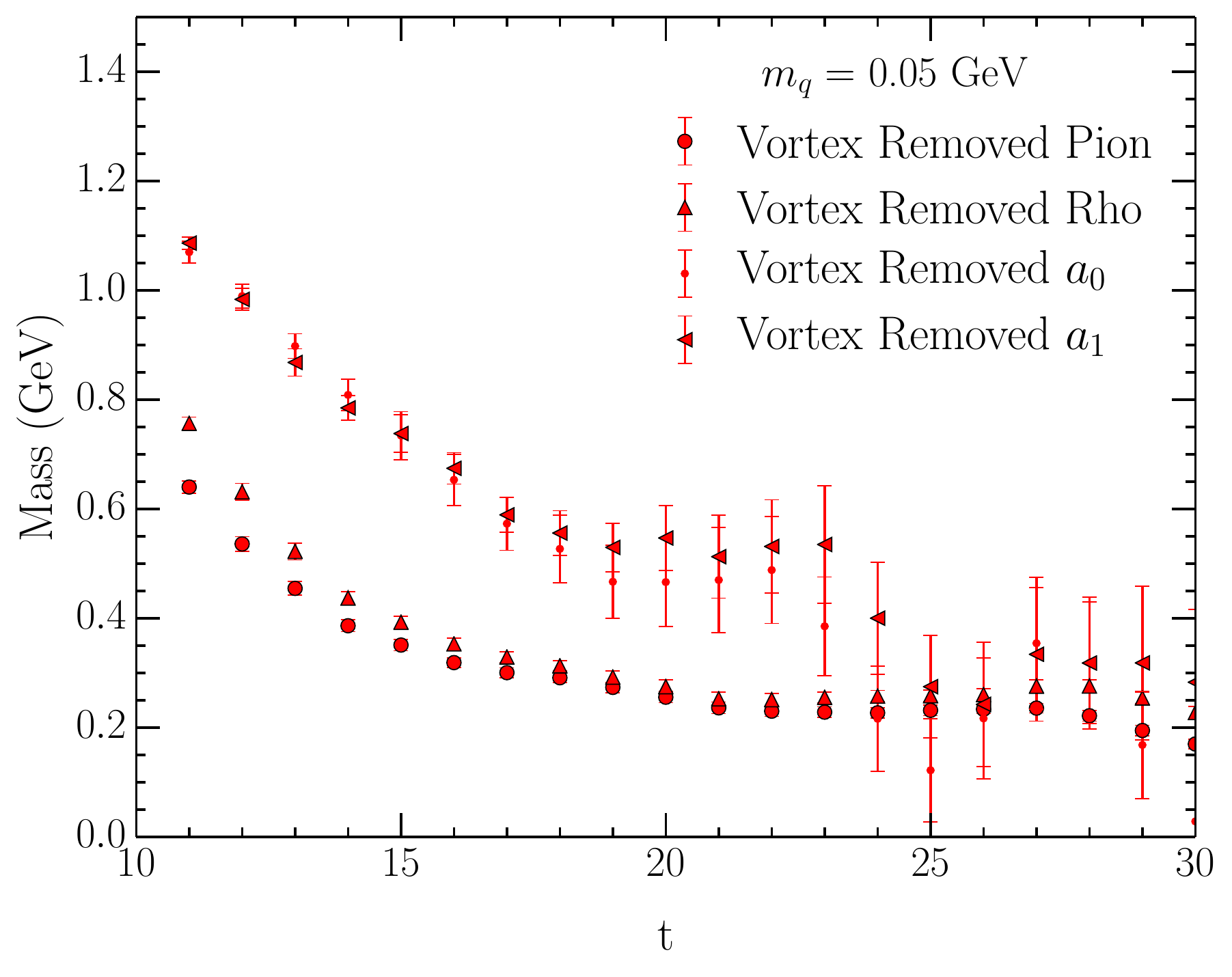}\includegraphics[width=0.4\columnwidth]{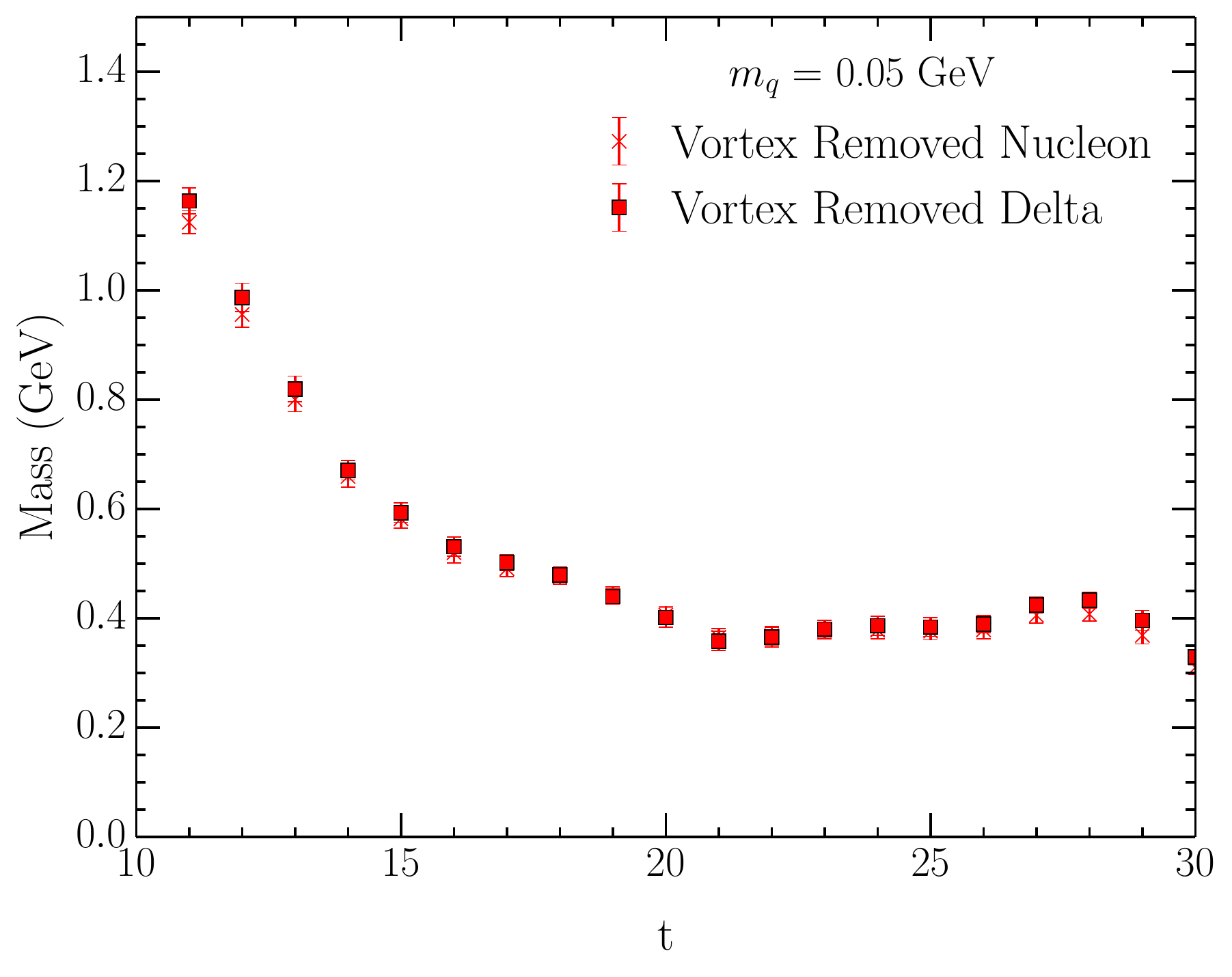}
  \caption{The effective masses for the low-lying mesons (left) and baryons (right) on the vortex-removed ensemble. Results are shown for light bare quark masses, with values of $m_q=13,\ 25,\ 38,\ 50$ MeV from top to bottom respectively. Note the smaller scale on the vertical axis in plots on the top row, and for the meson plot in the second row.}
  \label{Fig:VRlight}
\end{figure*}

Results for the four light quark masses on the vortex removed ensemble are plotted in Fig.~\ref{Fig:VRlight}. We first turn our attention to the meson spectrum, starting with the lightest mass ($m_q = 13$ MeV). The pion on the vortex-removed ensemble is below 100 MeV, compared to a ground state mass of over 200 MeV for the untouched case. The change for the rho is much more drastic, having a mass of around 170 MeV as compared to around 1000 MeV in the untouched ensemble. Both the pion and the rho now have masses smaller than their physical values. On the vortex-removed ensemble, the majority of dynamical mass generation is gone, with only a small remnant reflected by both the pion and the rho having masses larger than twice the bare quark mass. This is consistent with our results for the vortex-removed quark propagator~\cite{Trewartha:2015nna, Trewartha:2015ida}. We note that while the rho is greatly reduced in mass, it is not degenerate with the pion, providing the first indication that we are within the chiral regime where physics beyond simple quark counting can contribute.

\label{sec:VRhadspec}

In the vortex-removed case, the $a_{0}$ effective mass is observed to behave differently from the untouched case.  This time the correlator remains positive.  However, evidence of a negative-metric contribution to the source-sink-symmetric correlator is manifest as the effective mass rises from below at the earliest Euclidean times. Thus, short-distance quenched artifacts survive the process of vortex removal as anticipated, noting that the static quark potential indicates that Coulombic interactions associated with one-gluon exchange also remain present~\cite{Bowman:2010zr}. Thereafter, the effective mass stabilises to an approximate plateau for time slices 15 to 20.  Here the mass in the $a_{0}$ channel is higher than the $\rho$ meson.  However, this meta-stable plateau eventually gives way to a low-lying effective-mass plateau associated with a relatively small coupling to an eigenstate degenerate with the pion.  This ultimate degeneracy provides evidence of the effective restoration of the $\mathrm{U}(1)_{A}$ symmetry. The excited state seen at earlier Euclidean times has a mass consistent with the two-particle $\pi$-$\eta'$ state which can contribute in the $a_{0}$ channel.

The $a_{1}$ behaves similarly to the $a_{0}$, showing an excited state consistent with a $\rho$-$\eta'$ state, then a ground state consistent with the $\rho$ meson. The $a_{1}$, however, has much larger error bars and is not shown for $t > 22$.

Turning now to the remaining three light quark masses ($m_q=25,\ 38,\ 50$ MeV) we see similar trends continue for the $\pi$ and $\rho$ mesons; both have lower masses than in the untouched case, increasing with increasing bare quark mass. The pion continues to be lighter than the rho, although the gap is reduced at higher values of $m_{q}$. As the bare quark mass is increased, so is the explicit chiral symmetry breaking, and so the results move towards to the predictions for the constituent quark regime. While the non-degeneracy of the $\pi$ and $\rho$ mesons at these masses reveals that chiral physics remains manifest, by $m_{q} = 50$ MeV, the $\pi$ and $\rho$ mesons have become almost degenerate, signaling the start of the transition to the constituent quark regime.

At $m_{q} = 25$ MeV, both the $a_{0}$ and $a_{1}$ are too noisy to extract a clean signal, while at the other two quark masses ($m_q=38,\ 50$ MeV), a similar result is seen to that at the lightest mass. There is an excited state with a mass higher than the other two mesons, followed by a ground state plateau similar to the $\pi$ and $\rho$ mesons. The degeneracy of the $a_{0}$ with the pion, and the $a_{1}$ with the $\rho$, is a signal of the restoration of both the $\mathrm{U}(1)_{A}$ and $\SU{2}_{L} \times \SU{2}_{R}$ symmetries. This, combined with the non-degeneracy of the $\pi$ and $\rho$ mesons, suggests that at $m_{q} = 50$ MeV, explicit chiral symmetry breaking is still small enough that the predictions of chiral symmetry restoration hold.

In the baryon spectrum, at the lightest mass the nucleon and Delta both have masses around 220 - 260 MeV, dramatically lower than in the untouched cases. Notably, they are also approximately degenerate. At the three remaining light masses the nucleon and $\Delta$ effective masses show remarkably similar behaviour, and are degenerate within error bars. This is consistent with the restoration of the $\SU{2}_{L} \times \SU{2}_{R}$ symmetry and our predictions for the chiral regime. Similar to the case for the $\pi$ and $\rho$ mesons, the baryons show a loss of almost all dynamical mass generation, with a much lower plateau reached than in the untouched case (but larger than three times the bare quark mass). Unlike in the meson channel, as the nucleon and $\Delta$ baryons are predicted to be degenerate in both the chiral and constituent regime, there is no signal of a transition as the bare quark mass is increased.

We also note that all four of the light hadrons $(\pi,\ \rho,\ N,\ \Delta)$ show a slow approach to the mass plateau, indicating a dense tower of excited states. This echoes results seen using a Wilson fermion action in Ref.~\cite{OMalley:2011aa}. Also of note is the stability of the ground state seen in both the mesons and baryons, a reflection of the near-empty gauge field background in the vortex removed case.

\begin{figure*}[p]
  \centering
  \includegraphics[width=0.4\columnwidth]{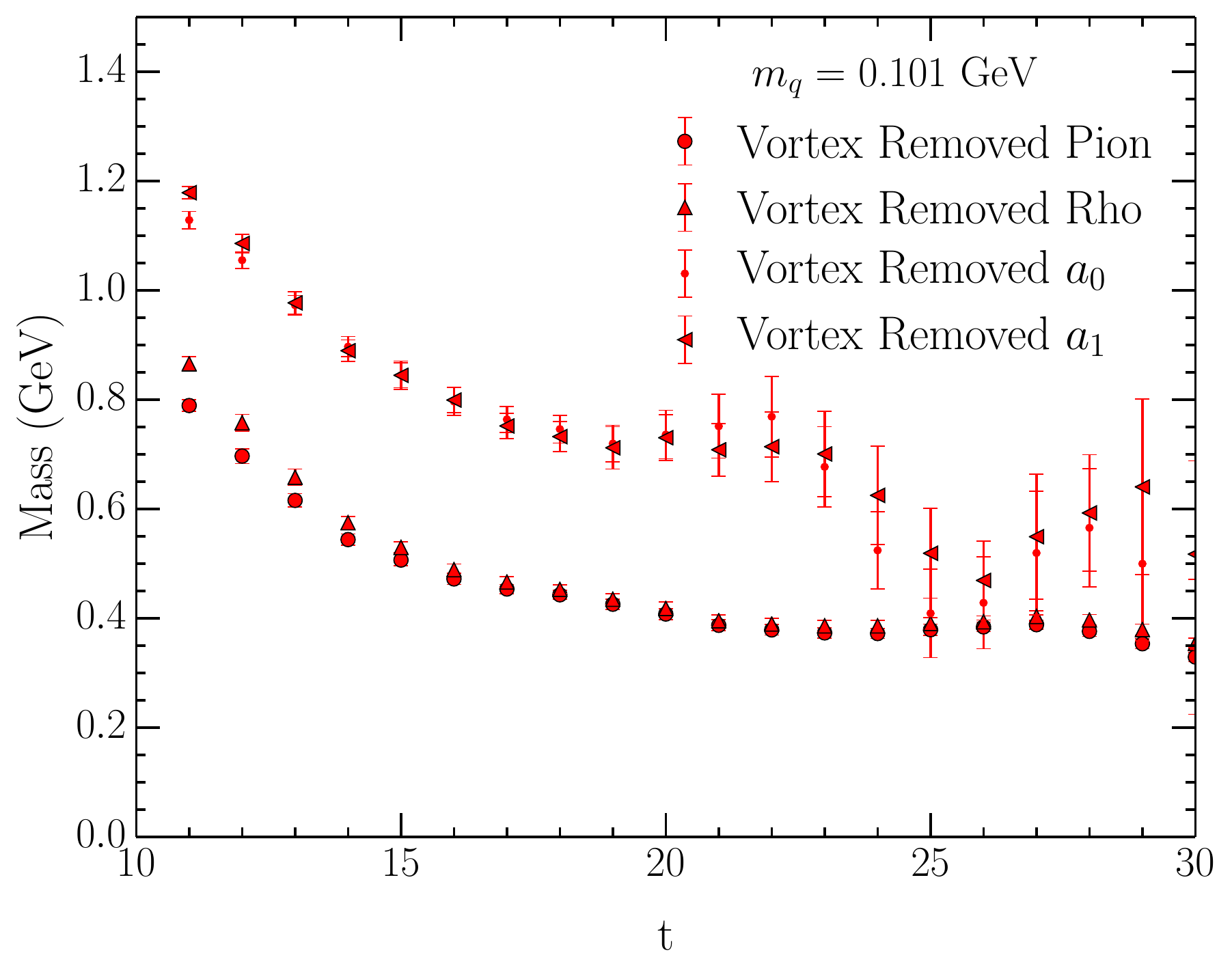}\includegraphics[width=0.4\columnwidth]{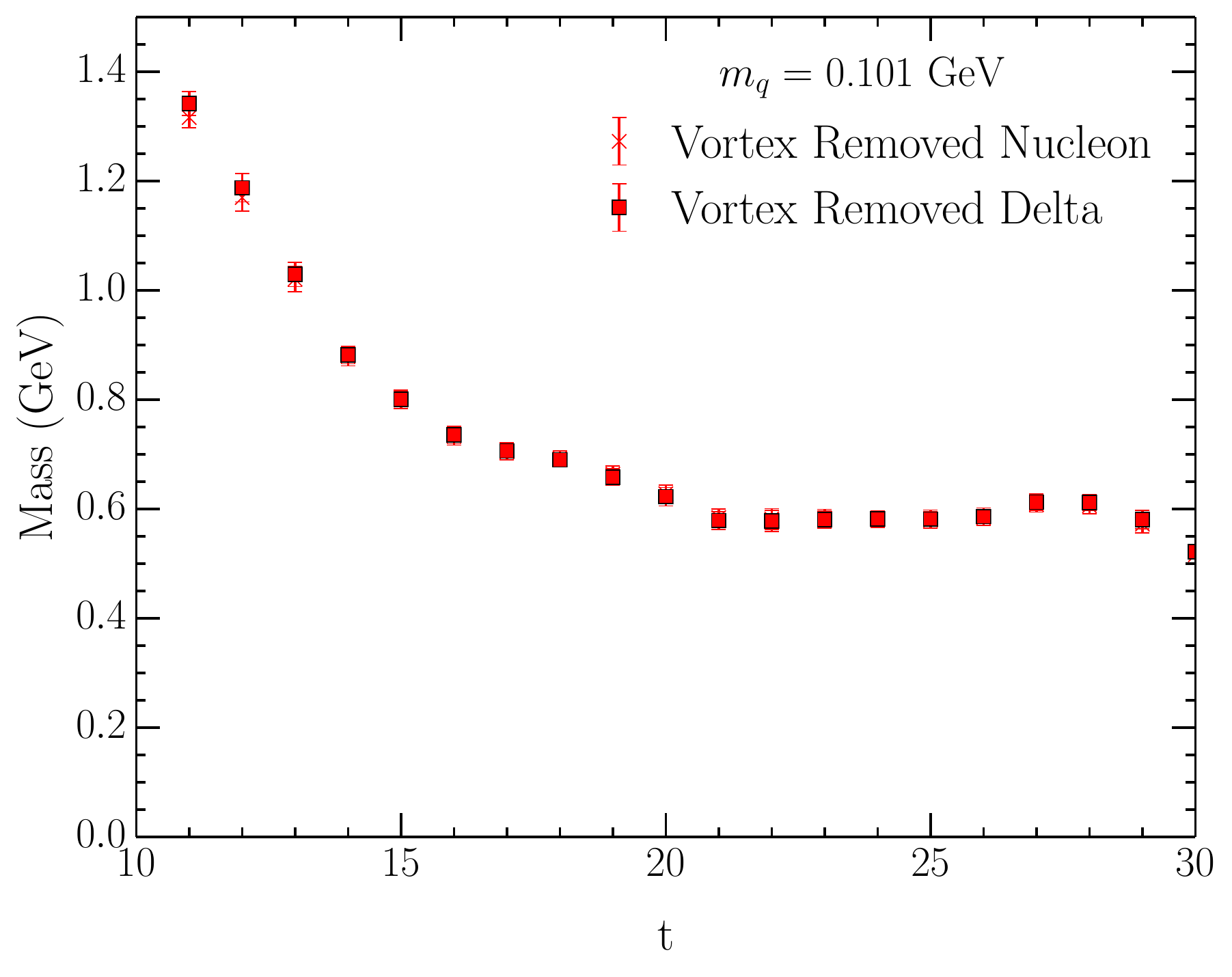}\\
  \includegraphics[width=0.4\columnwidth]{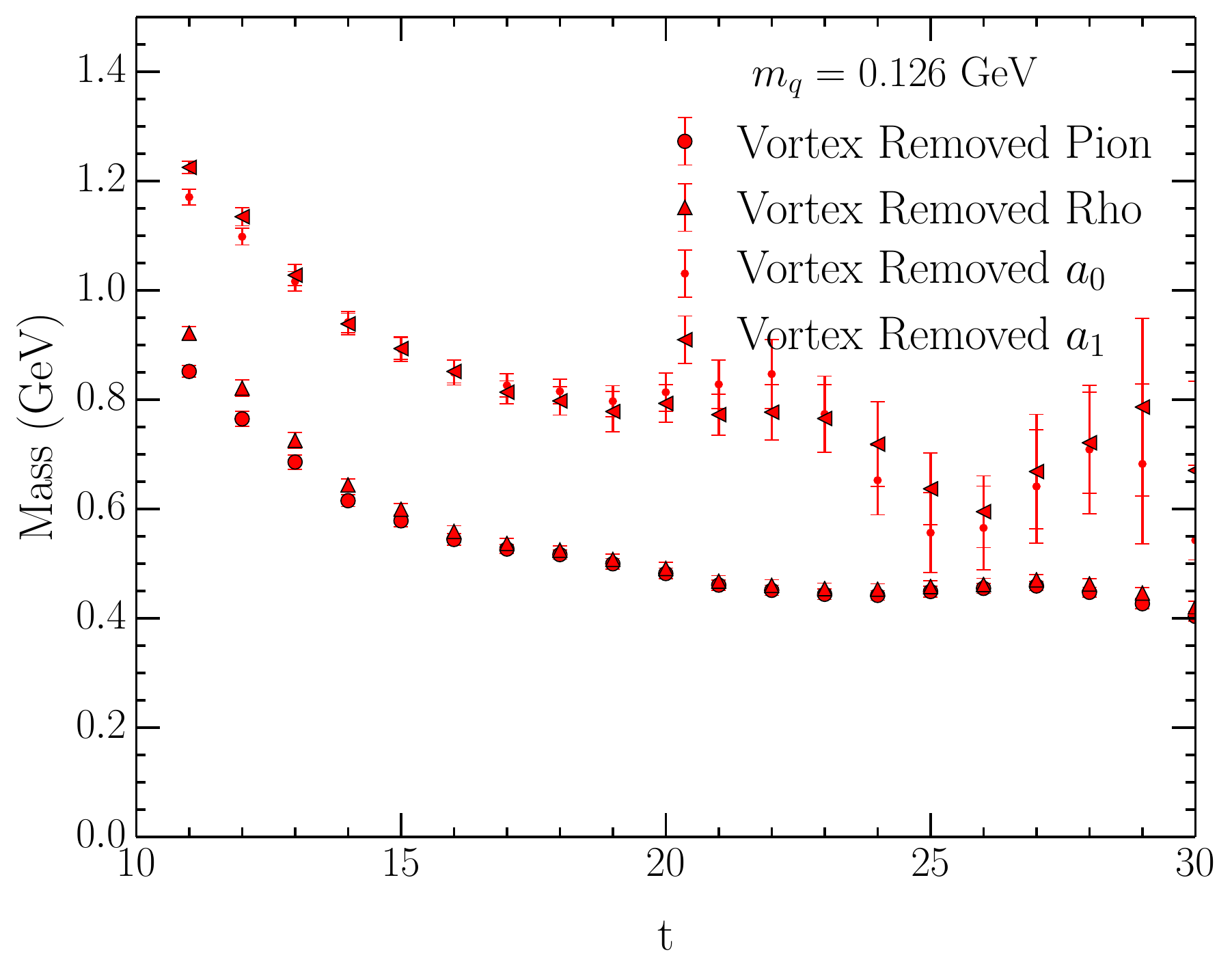}\includegraphics[width=0.4\columnwidth]{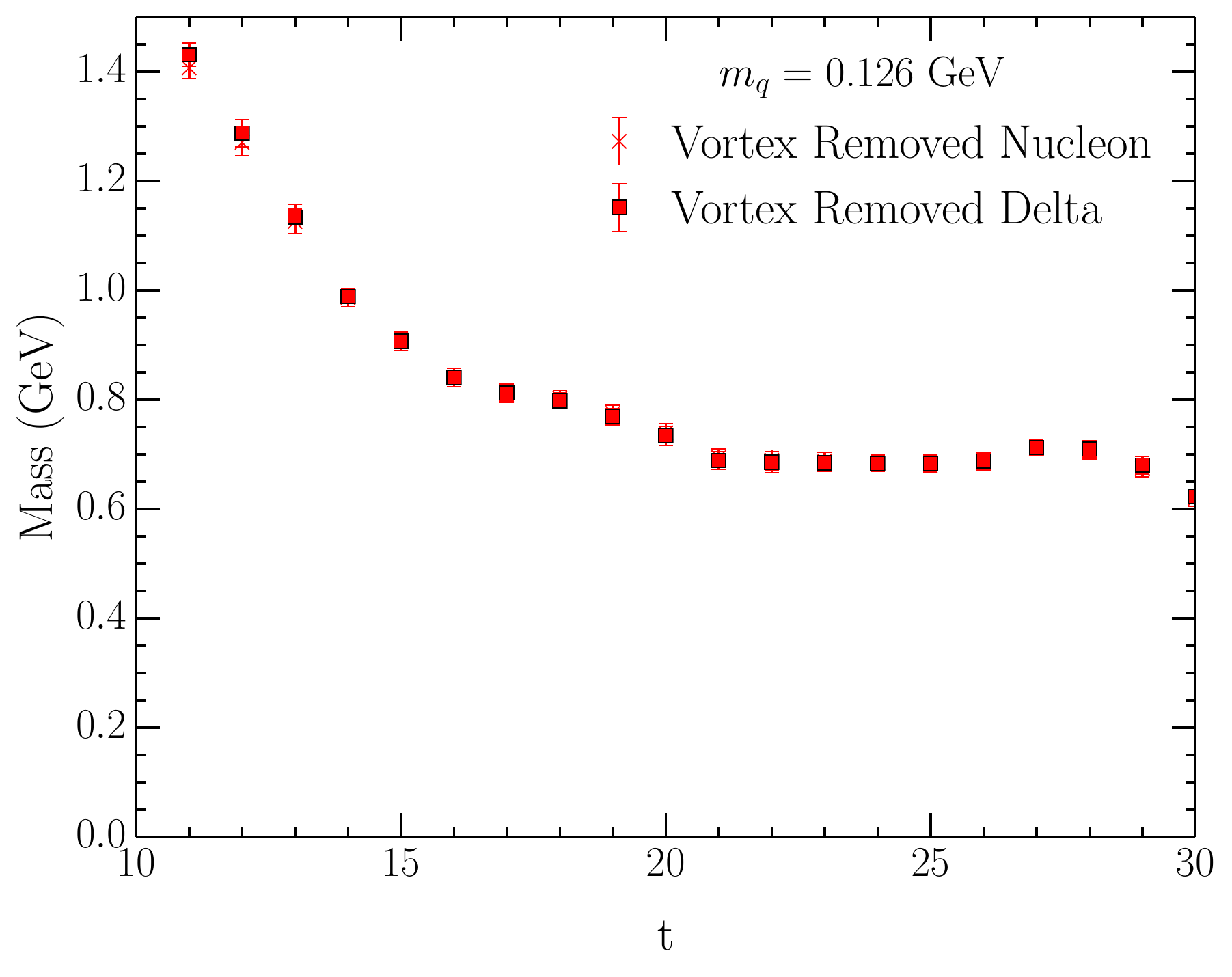}\\
  \includegraphics[width=0.4\columnwidth]{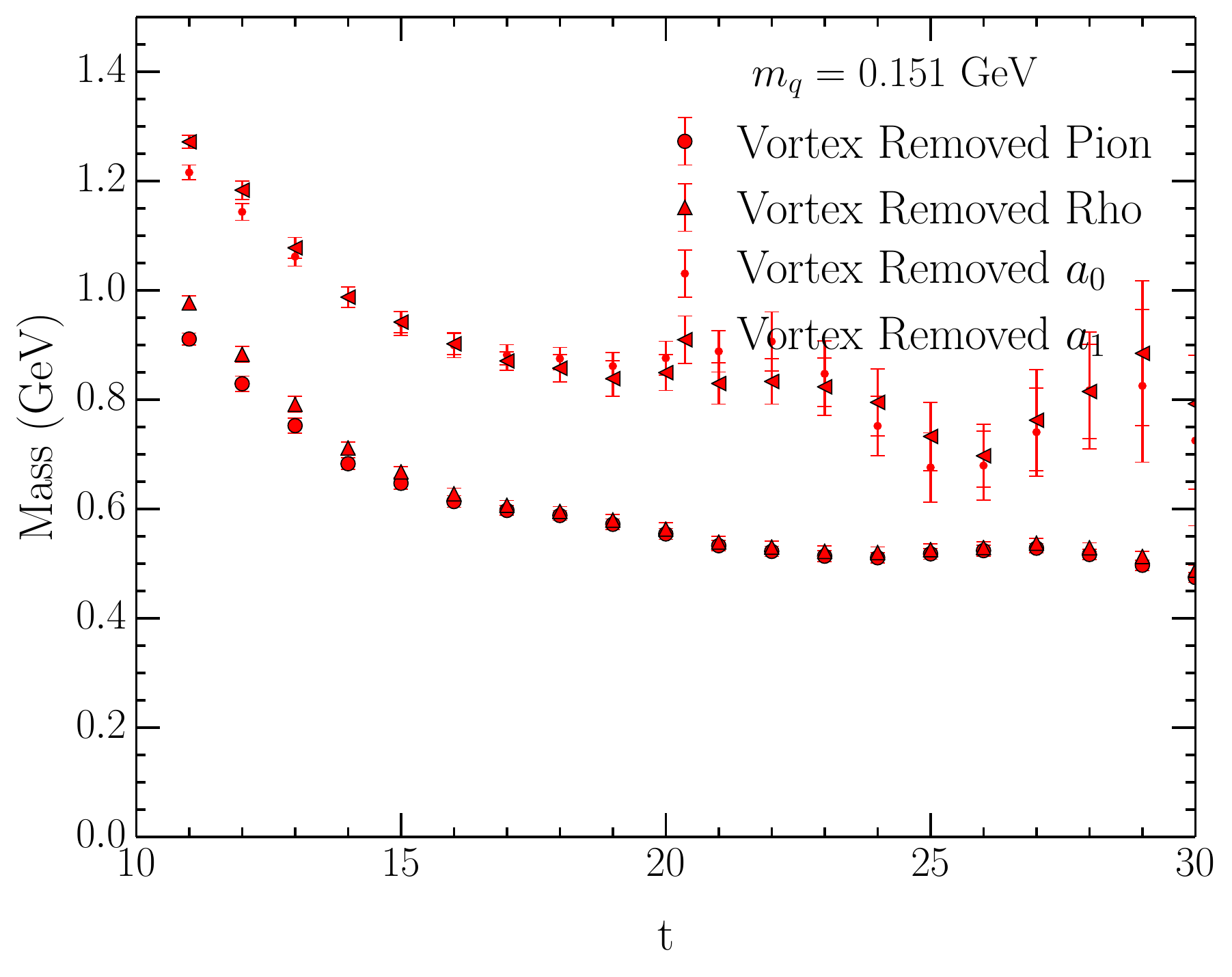}\includegraphics[width=0.4\columnwidth]{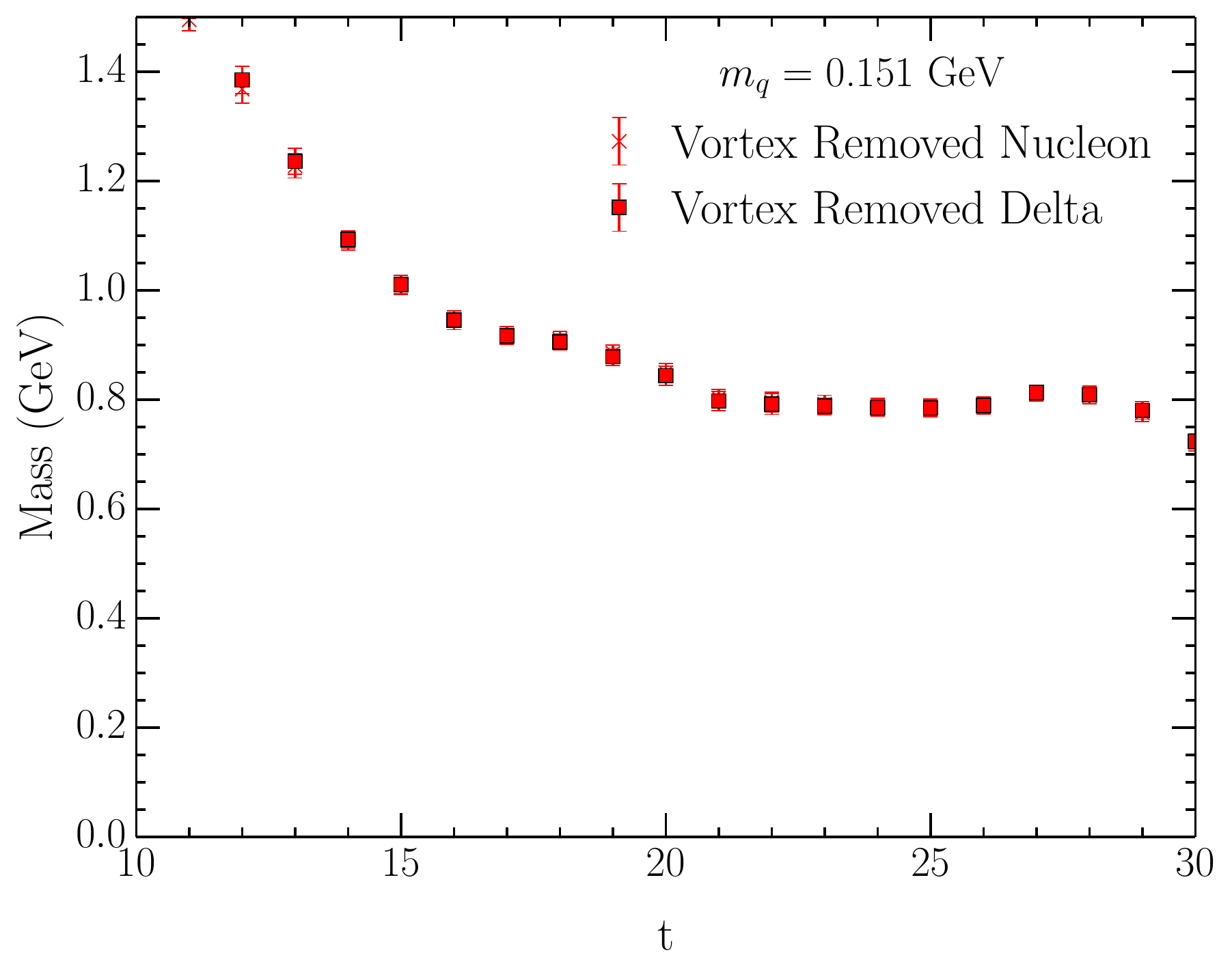}\\
  \includegraphics[width=0.4\columnwidth]{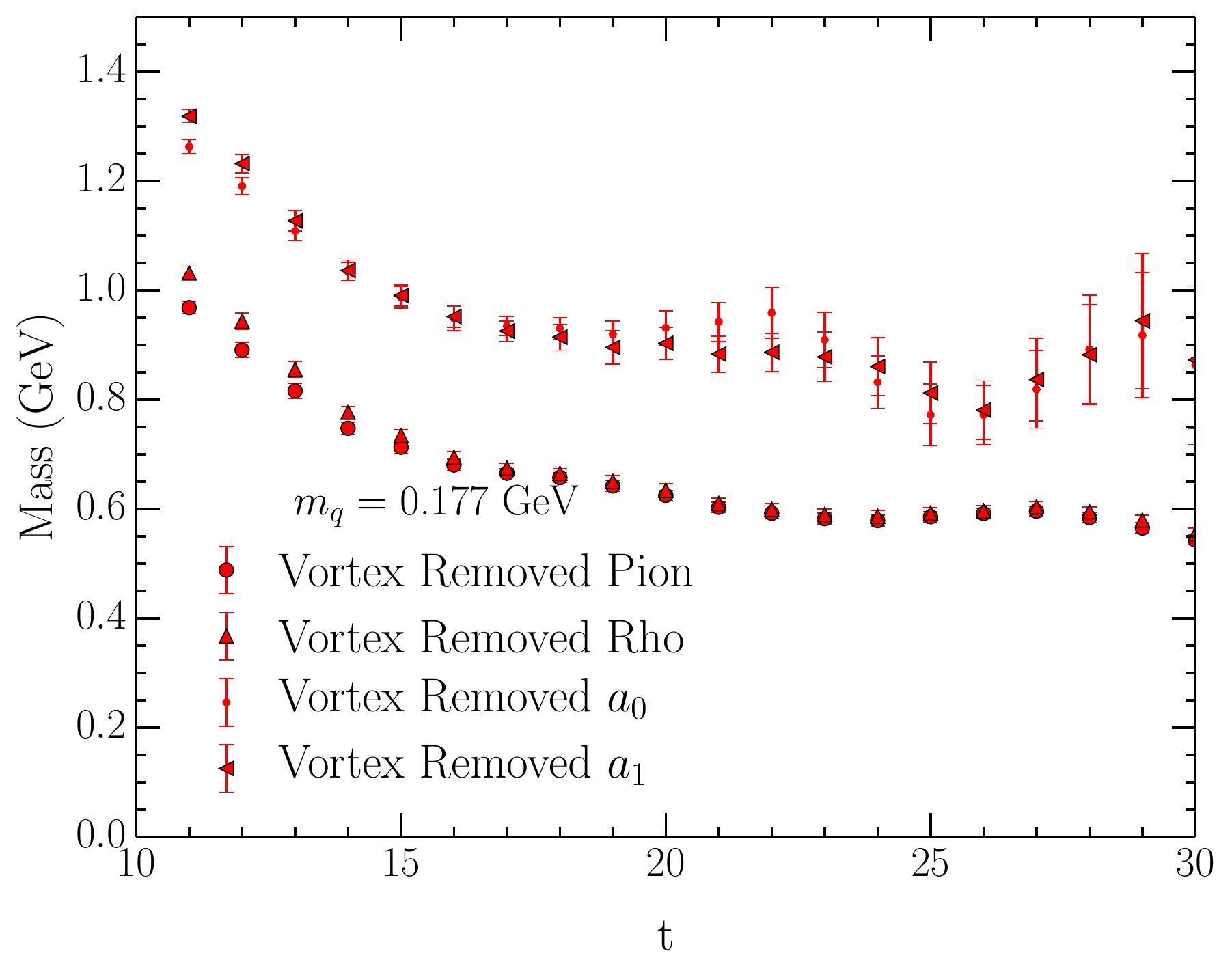}\includegraphics[width=0.4\columnwidth]{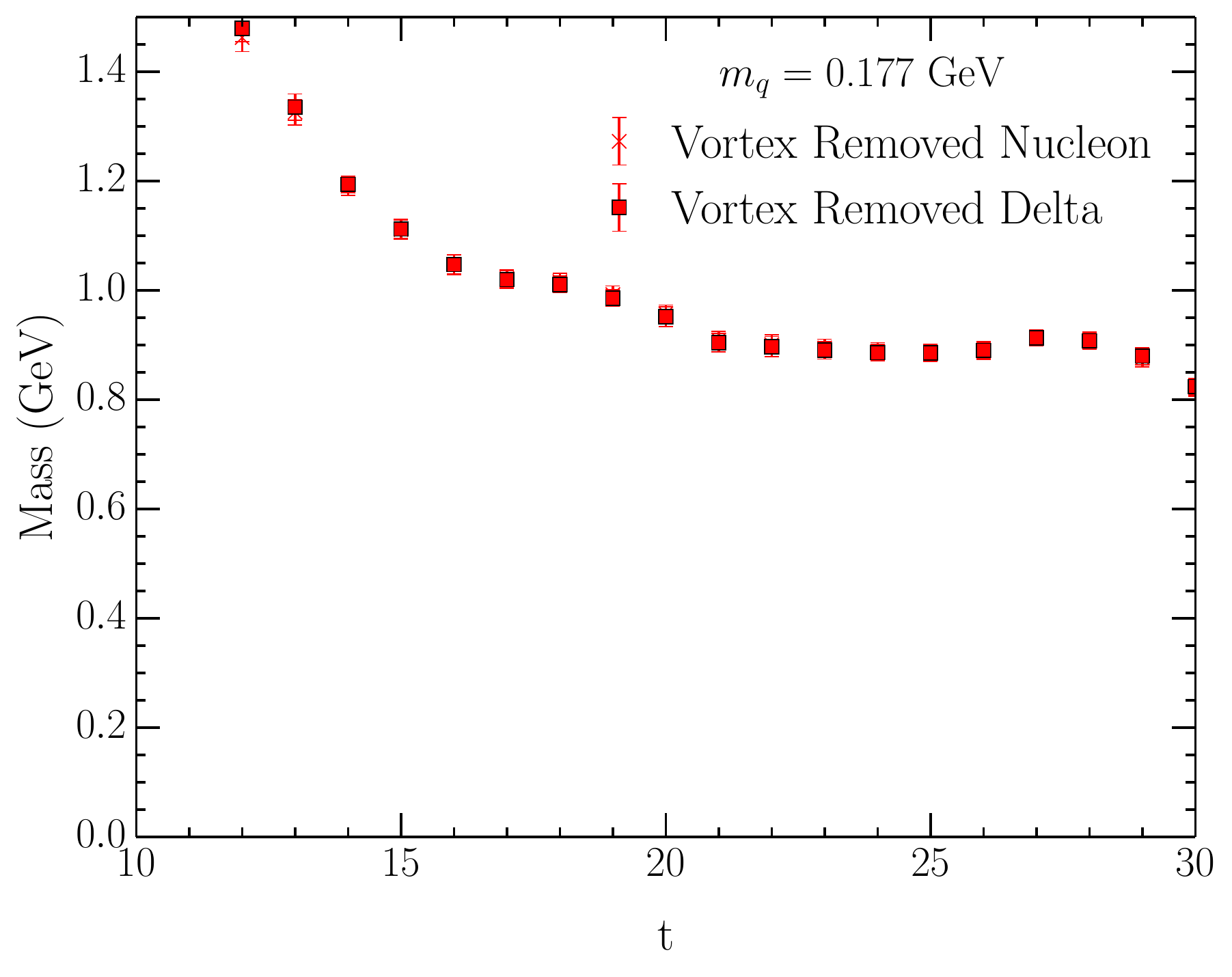}
  \caption{The effective masses for the low-lying mesons (left) and
    baryons (right) on the vortex-removed ensemble. Results are shown
    for heavy bare quark masses, with values of
    $m_q=101,\ 126,\ 151,\ 177$ MeV from top to bottom respectively.}
  \label{Fig:VRheavy}
\end{figure*}

Results at the four heavy quark masses ($m_q = 101,\ 126,\ 151,\ 177$ MeV) are presented in Fig.~\ref{Fig:VRheavy}. At these masses, the $\pi$ and $\rho$ mesons have become approximately degenerate, indicating that explicit chiral symmetry breaking is now large enough that both hadrons behave as though composed of two weakly-interacting constituent quarks. The nucleon and Delta baryons remain degenerate in the constituent regime.

% as at the lighter masses (though the underlying reason for the degeneracy differs in the constituent regime).

At these higher quark masses, the $a_{0}$ and $a_{1}$ no longer reach a plateau degenerate with the $\pi$ and $\rho$ mesons respectively, as chiral symmetry is no longer approximately restored. Instead, in the constituent regime the $a_0$ and $a_1$ are degenerate with each other, and the lightest state in these channels is now a two quark state excited with the lowest non-trivial momentum.

\begin{table}[tb]
\caption{Fitted masses of the pion, rho, nucleon, and Delta on the vortex-removed ensemble as a function of the bare quark mass, $m_{q}.$}
\centerline{\begin{tabular}{ ccccc }
\hline 
$m_{q}$ (MeV) & $m_{\pi}$ (MeV) & $m_{\rho}$ (MeV) & $m_{N}$ (MeV) & $m_{\Delta}$ (MeV) \\
\hline
\hline
13 & $85(3)$ & $171(7)$ & $219(6)$ & $260(10)$ \\
25 & $132(4)$ & $203(5)$ & $272(7)$ & $295(7)$ \\
38 & $173(4)$ & $228(5)$ & $316(7)$ & $334(6)$ \\
50 & $213(4)$ & $257(4)$ & $365(5)$ & $378(5)$ \\
100 & $366(3)$ & $386(3)$ & $572(5)$ & $575(5)$ \\
126 & $439(3)$ & $453(3)$ & $676(4)$ & $676(4)$ \\
151 & $510(3)$ & $521(3)$ & $780(4)$ & $779(4)$ \\
177 & $578(3)$ & $588(3)$ & $881(4)$ & $880(5)$ \\
\hline 
\end{tabular}}
\label{Tab:hadfitmasses}
\end{table}

Qualitatively, the results seen suggest agreement with the predictions of chiral symmetry restoration below a bare quark mass of $50$ MeV, and above that, agreement with the predictions of a constituent-quark like model. We now turn to quantitative measures of these predictions. Fits of the ground state masses of the pion, rho, nucleon, and Delta on the vortex-removed ensemble across all eight quark masses are summarised in Table~\ref{Tab:hadfitmasses}.

\begin{figure*}[tb]
\centerline{\includegraphics[width=0.9\columnwidth]{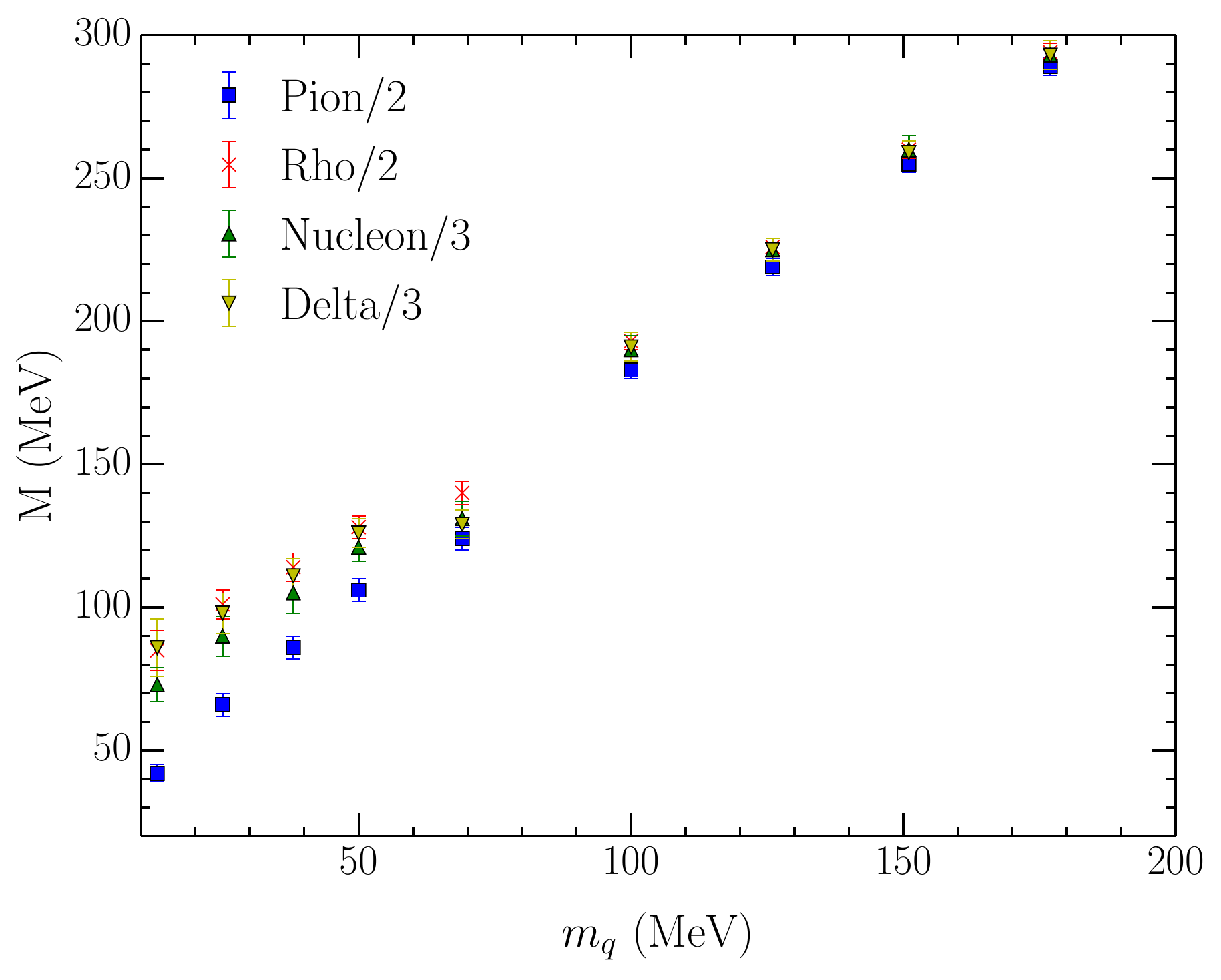}}
\caption{The implied constituent quark mass from each of the hadrons considered as a function of the input bare quark mass.}
\label{Fig:VRConsQ}
\end{figure*}

We first consider the validity of the constituent-quark like model of the hadron spectrum in the heavy quark mass region. In Fig.~\ref{Fig:VRConsQ}, we have plotted the hadron masses divided by the number of valence quarks as a function of the bare quark mass. At masses of $m_{q} = 101\,\mathrm{MeV}$ and beyond, the constituent-quark like model is highly successful in describing the behaviour of the spectrum, with all hadrons approximately degenerate after division by the number of constituent quarks. At these quark masses, all four hadrons can be accurately modelled as weakly interacting dressed quarks. Below this value, while the rho, nucleon and Delta are still in agreement within statistical uncertainties, the pion is lighter. It is in this region, therefore, that we expect the predictions of the chirally restored theory to be valid.

\begin{figure*}[tb]
  \centering
  \includegraphics[width=0.32\textwidth]{Plots/01600VRmesons.pdf}
  \includegraphics[width=0.32\textwidth]{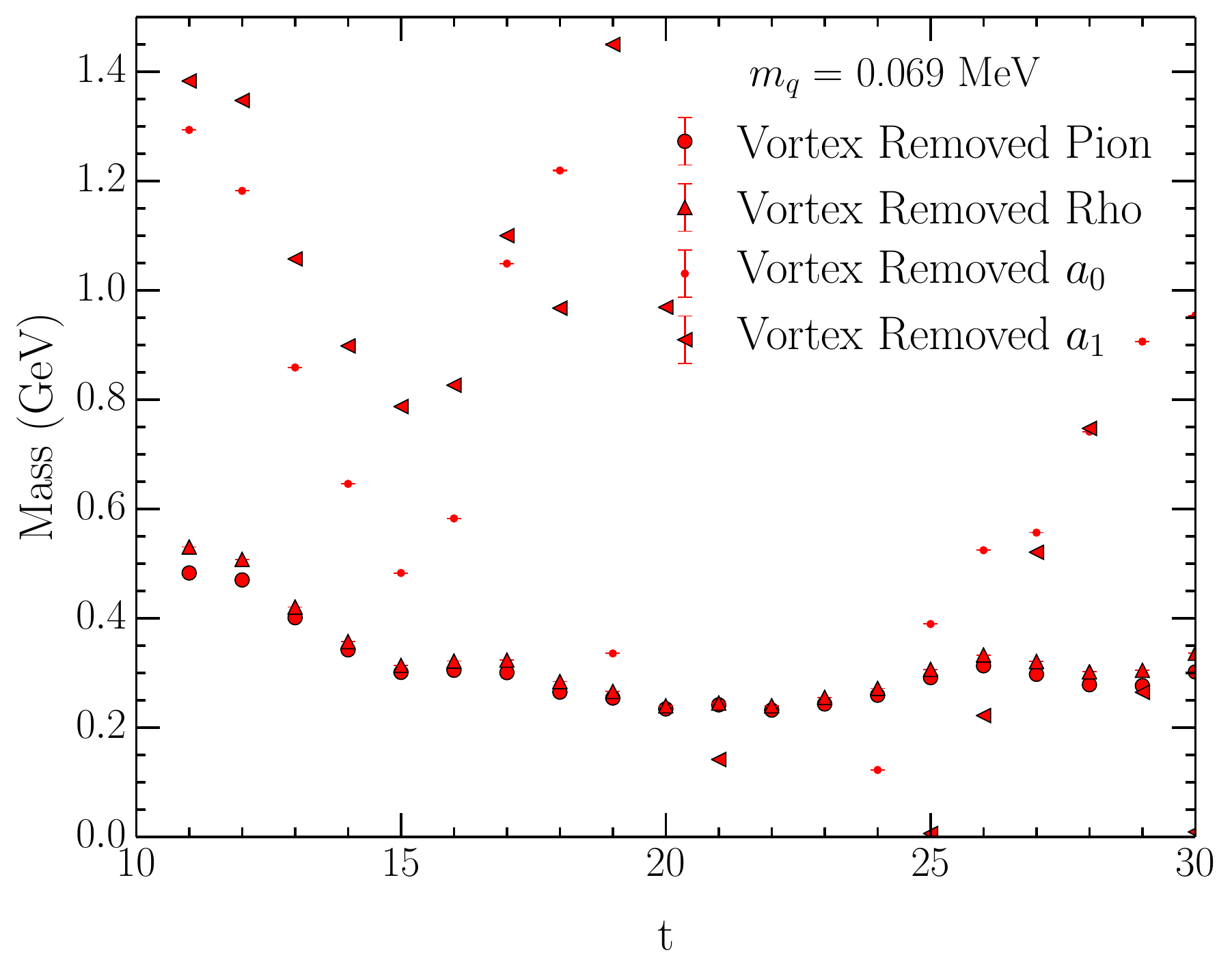}
  \includegraphics[width=0.32\textwidth]{Plots/03200VRmesons.pdf}
  \caption{The effective masses for the $a_{0}$ and $a_{1}$ mesons at bare quark masses of $50$ (left), $69$ (middle), and $100$ (right) MeV. The sequence of plots shows the transition from the chiral regime (left), passing through an intermediate region (middle) before reaching the constituent regime (right).}
  \label{Fig:VRtransition}
\end{figure*}

 In Figure~\ref{Fig:VRConsQ} we also include points at $m_q = 69$ MeV, in the transition region between the chiral and constituent regimes. While we are able to obtain fits for the $\pi,\rho,N,$ and $\Delta$ at this intermediate quark mass in the intermediate region, in Fig.~\ref{Fig:VRtransition} we see the $a_0$ and $a_1$ correlators fluctuate wildly, perhaps indicating that the nature of these two states is ill-defined in the transition region. Indeed, due to their distinct properties in each regime it is on the $a_0$ and $a_1$ mesons that we now focus.

 \section{Mass ratios for the $a_0$ and $a_1$ mesons}
 
 The nature of the $a_0$ and $a_1$ mesons strongly differs between the chiral and constituent regimes. While the large error bars on the $a_{0}$ and $a_{1}$ correlators make fitting a mass value difficult, we can test the predictions for these two mesons in both regimes using ratios of masses.

For the $a_{0}$, the $\mathrm{U}(1)_{A}$ symmetry predicts degeneracy with the pion in the chiral regime. As the two-particle $\pi$-$\eta'$ state has the same quantum numbers as the $a_{0}$, it can appear as an excited state, or possibly as the ground state in the constituent quark regime. In the weakly-interacting constituent-quark regime, we can model the lowest energy two-quark energy eigenstate having overlap with the $l=1$ orbital angular momentum required to form the desired quantum numbers by
\begin{equation}
\ket{\psi_\qq} = \frac{1}{2}\Big(\ket{0,\vec{p}} - \ket{0,-\vec{p}} + \ket{\vec{p},0} - \ket{-\!\vec{p},0}\Big),
\end{equation}
where $|\vec{p}| = 2\pi/L$ is the minimum non-trivial momentum
available to a free constituent quark on a lattice with spatial length
$L.$ The energy $E$ associated with this quark model state is given by
\begin{equation}
\label{eq:EQdef}
E^{2} = \left( M^{2} + \left(\frac{2\pi}{L}\right)^{2} \right),
\end{equation}
where $M = 2\, \cq$ is twice the constituent quark mass.
%%  For the $a_{0}$ and $a_{1}$ mesons in the constituent regime, we can model the energy  as a two-quark state expect their mass to be given by two dressed quarks
%% For the $a_{0}$ mesons in the constituent regime, we can model the energy  as a two-quark state expect their mass to be given by two dressed quarks with one of the quarks excited by the lowest available non-trivial momentum on the lattice.
%%   The energy associated with this is given by
%% \begin{equation}
%% \label{eq:EQdef}
%% E^{2} = \left( M^{2} + \left(\frac{2\pi}{L}\right)^{2} \right),
%% \end{equation}
%% where $\cq$ is the constituent quark mass. We estimate the constituent quark mass as half of the fitted ground state rho mass.
%% to give overlap with the appropriate quantum numbers. lowest non-trivial momentum, $2\,E_{q}$.
%and  We estimate the constituent quark mass as half of the fitted ground state rho mass.

In consideration of all of the above, we define the $a_0$ mass ratio as
\begin{equation}
R_{0} = \frac{m_{a_{0}}}{2\,m_{\pi}}.
\end{equation}
In the chiral regime, we expect this to have a value of $1/2$, as $m_{a_{0}} = m_{\pi}$. If the $a_{0}$ is described by a $\pi$-$\eta'$ state, it will have a value of approximately $1$. For our two quark state in the constituent-quark regime, this ratio will be given by
$\nicefrac{E}{2m_\pi}.$
%\begin{equation}
%  \frac{E}{2 M} = \frac{E_{q}}{2\,\cq},
%\end{equation}
%where we have defined $E_q$ to be the average energy carried by each constituent quark such that $E = 2 E_q.$ 

For the $a_{1}$, in the chiral regime the $\SU{2}_{L} \times \SU{2}_{R}$ symmetry predicts degeneracy with the $\rho$ meson. The quantum numbers of the $a_{1}$ can be produced by a $\rho$-$\eta'$ state, and so we expect this to appear as an excited state in the chiral regime, and as either an excited state or the ground state in the constituent regime. Again, we can construct a model two quark state that describes the $a_1$ in the constituent regime. This model state has the same expected energy $E$ given by Eq.~(\ref{eq:EQdef}), such that the $a_1$ is degenerate with the $a_0$ at heavy quark masses.

%% \begin{table}[t]
%% \caption{The two mass ratios considered, together with their expected values in the chirally restored and constituent quark regimes.}
%% \centerline{\begin{tabular}{ cccc }
%% \hline 
%% Ratio & Definition & Chiral regime value & Constituent quark regime value \\
%% \hline
%% \hline
%% $R_{0}$ & $\frac{m_{a_{0}}}{2\,m_{\pi}}$ & $\frac{1}{2}$ & \begin{tabular}{c} Smaller of $1$ ($\pi$-$\eta'$ state) or \\ $\frac{2\,E_{q}}{4\,\cq}$ (2 quark state) \end{tabular} \\
%% $R_{1}$ & $\frac{m_{a_{1}}}{2\,m_{\rho}}$ & $\frac{1}{2}$ & \begin{tabular}{c} Smaller of $1$ ($\rho$-$\eta'$ state) or \\ $\frac{2\,E_{q}}{4\,\cq}$ (2 quark state) \end{tabular} \\
%% \hline 
%% \end{tabular}}
%% \label{Tab:ratiodefs}
%% \end{table}

We therefore define the $a_1$ mass ratio as
\begin{equation}
R_{1} = \frac{m_{a_{1}}}{2\,m_{\rho}}.
\end{equation}
Again, in the chiral regime we expect this to have a value of $\frac{1}{2}$, as $m_{a_{1}} = m_{\rho}$. In the constituent quark regime, we expect a value of $\nicefrac{E}{2m_\rho}$ for a two-quark state. %$\frac{2\,E_{q}}{4\,\cq}$.
We note that while a $\rho$-$\eta'$ state can create the quantum numbers of the $a_{1}$, in the constituent regime the pion and the rho become approximately degenerate, and so this state will have mass $m_{\rho} + m_{\pi} = 2\,m_{\rho}$ and once again produce a value of $R_{1} \simeq 1$. In the chiral regime, the pion is lighter than the rho, and so $R_{1}$ for this state will be less than $1$, varying from $0.75$ at $m_{q} = 13$ MeV to $0.91$ at $m_{q} = 50$ MeV. This still allows a clean separation from the prediction of restored chiral symmetry, where $R_{1} = \nicefrac{1}{2}$.

\begin{table}[tb]
\caption{The two mass ratios considered, together with their expected values in the chirally-restored and weakly-interacting constituent-quark regimes.}
\smallskip
\centerline{\begin{tabular}{ cccc }
\hline 
Ratio & Definition & Chiral regime value & Constituent quark regime value \\
\hline
\hline
$R_{0}$ & $\frac{m_{a_{0}}}{2\,m_{\pi}}$ & $\frac{1}{2}$ & \begin{tabular}{c} Smaller of $1$ ($\pi$-$\eta'$ state) or \\ $\nicefrac{E}{2m_\pi}$ (2 quark state) \end{tabular} \\
$R_{1}$ & $\frac{m_{a_{1}}}{2\,m_{\rho}}$ & $\frac{1}{2}$ & \begin{tabular}{c} Smaller of $1$ ($\rho$-$\eta'$ state) or \\ $\nicefrac{E}{2m_\rho}$ (2 quark state) \end{tabular} \\
\hline 
\end{tabular}}
\label{Tab:ratiodefs}
\end{table}

\begin{table}[tb]
  \caption{For each bare quark mass $m_q,$ the constituent quark masses $\cq$ inferred from the fitted ground-state rho meson masses and the corresponding energy $E$ of a two quark state with the smallest non-trivial lattice momentum are indicated. Also shown are the values of the ratios $\nicefrac{E}{2m_\pi},\nicefrac{E}{2m_\rho},$ and $\nicefrac{(m_\pi+m_\rho)}{2m_\rho}.$}
  \smallskip
\centerline{\begin{tabular}{ cccccc }
\hline 
$m_{q}$ (MeV) & $\cq$ (MeV) & $E$ (MeV) & $\nicefrac{E}{2m_\pi}$ & $\nicefrac{E}{2m_\rho}$ & $\frac{m_\pi+m_\rho}{2m_\rho}$ \\
\hline
\hline
13 & $85$ & $525$ & 3.09 & 1.53 & 0.75 \\
25 & $101$ & $536$ & 2.03 & 1.32 & 0.83 \\
38 & $114$ & $546$ & 1.58 & 1.20 & 0.88 \\
50 & $128$ & $559$ & 1.31 & 1.09 & 0.91 \\
100 & $193$ & $628$ & 0.86 & 0.81 & 0.97 \\
126 & $226$ & $672$ & 0.76 & 0.74 & 0.98 \\
151 & $260$ & $719$ & 0.71 & 0.69 & 0.99 \\
177 & $294$ & $769$ & 0.67 & 0.65 & 0.99 \\
\hline 
\end{tabular}}
\label{Tab:consqmasses}
\end{table}

We have summarised these ratios and their expected values in Table~\ref{Tab:ratiodefs}. Based on the results in Fig.~\ref{Fig:VRConsQ}, we have defined the constituent quark mass $\cq$ to be half the fitted mass of the rho meson. These values are listed in Table~\ref{Tab:consqmasses}, together with the corresponding energy $E$ of a two quark state and the values of the mass ratios  $\nicefrac{E}{2m_\pi},\nicefrac{E}{2m_\rho},$ and $\nicefrac{(m_\pi+m_\rho)}{2m_\rho}.$. Interestingly, a comparison of the different mass ratios reveals that at all four heavy quark masses, the two quark state is lighter than the corresponding $\pi$-$\eta'$ or $\rho$-$\eta'$ multi-particle state (while at the four light quark masses the reverse is true). Hence, we predict that in the constituent regime the value of $R_0$ and $R_1$ should approach $\nicefrac{E}{2m_\pi}$ or $\nicefrac{E}{2m_\rho}$ respectively.

\begin{figure*}[t]
  \centering
  \includegraphics[width=0.5\columnwidth]{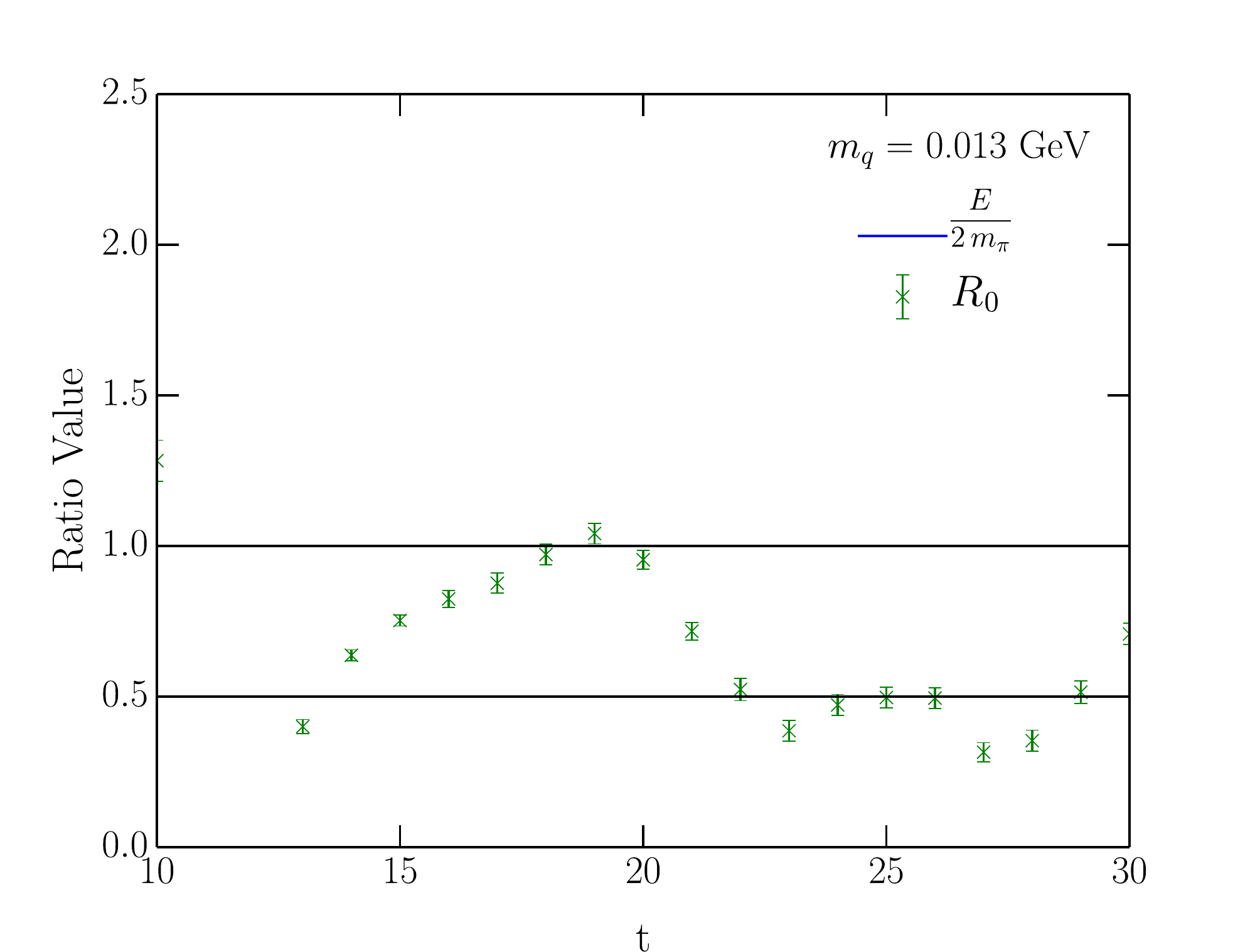}\includegraphics[width=0.5\columnwidth]{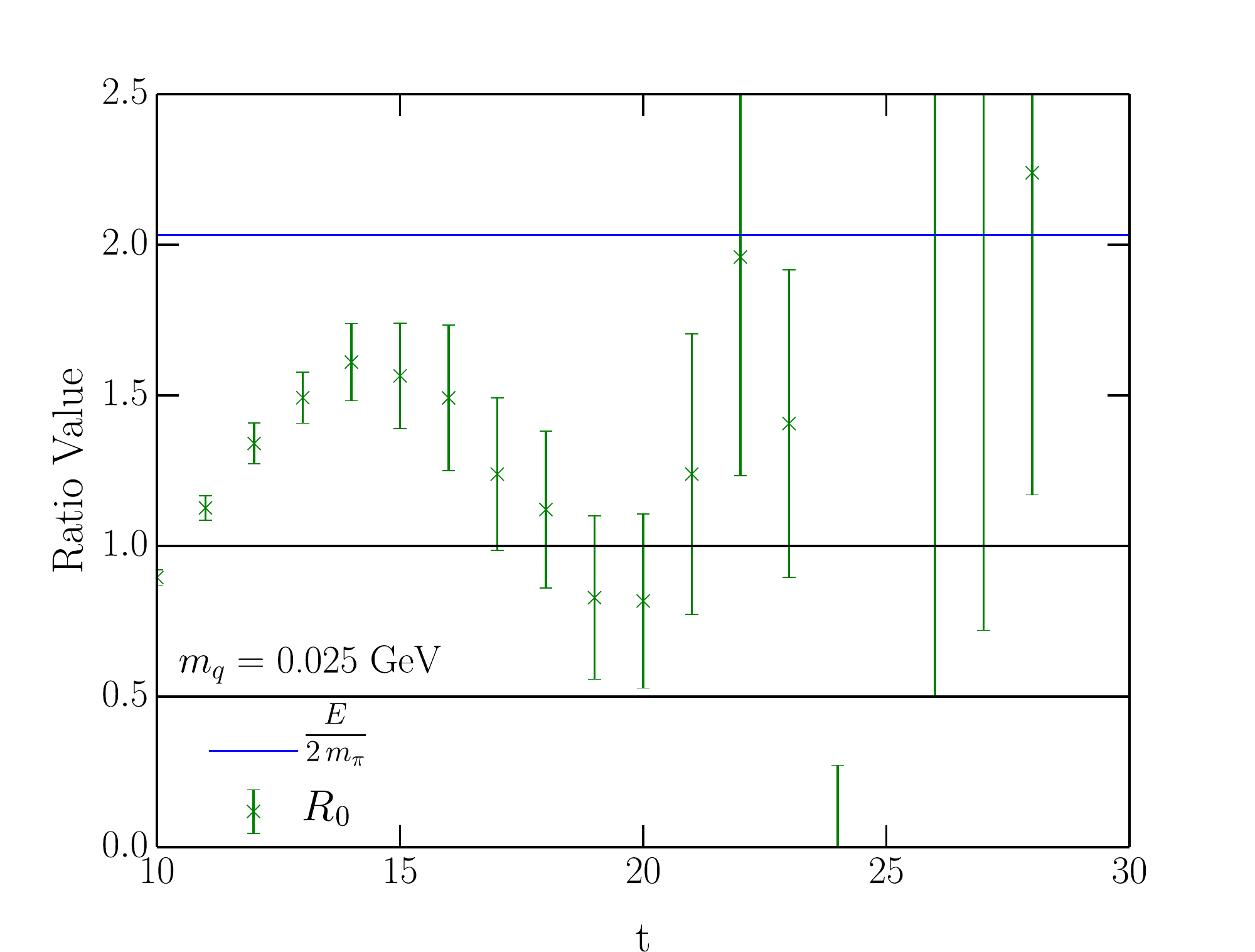}\\
  \includegraphics[width=0.5\columnwidth]{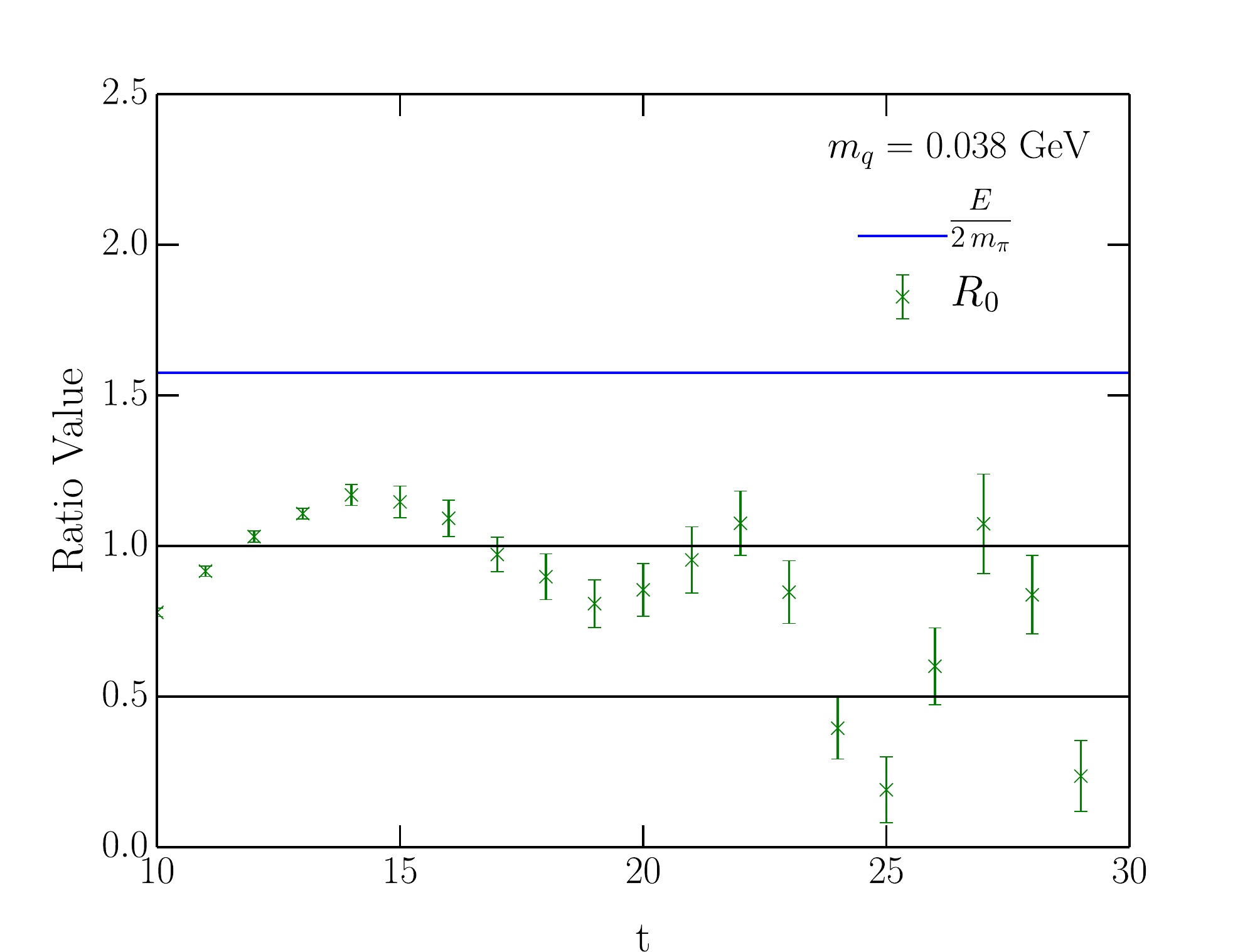}\includegraphics[width=0.5\columnwidth]{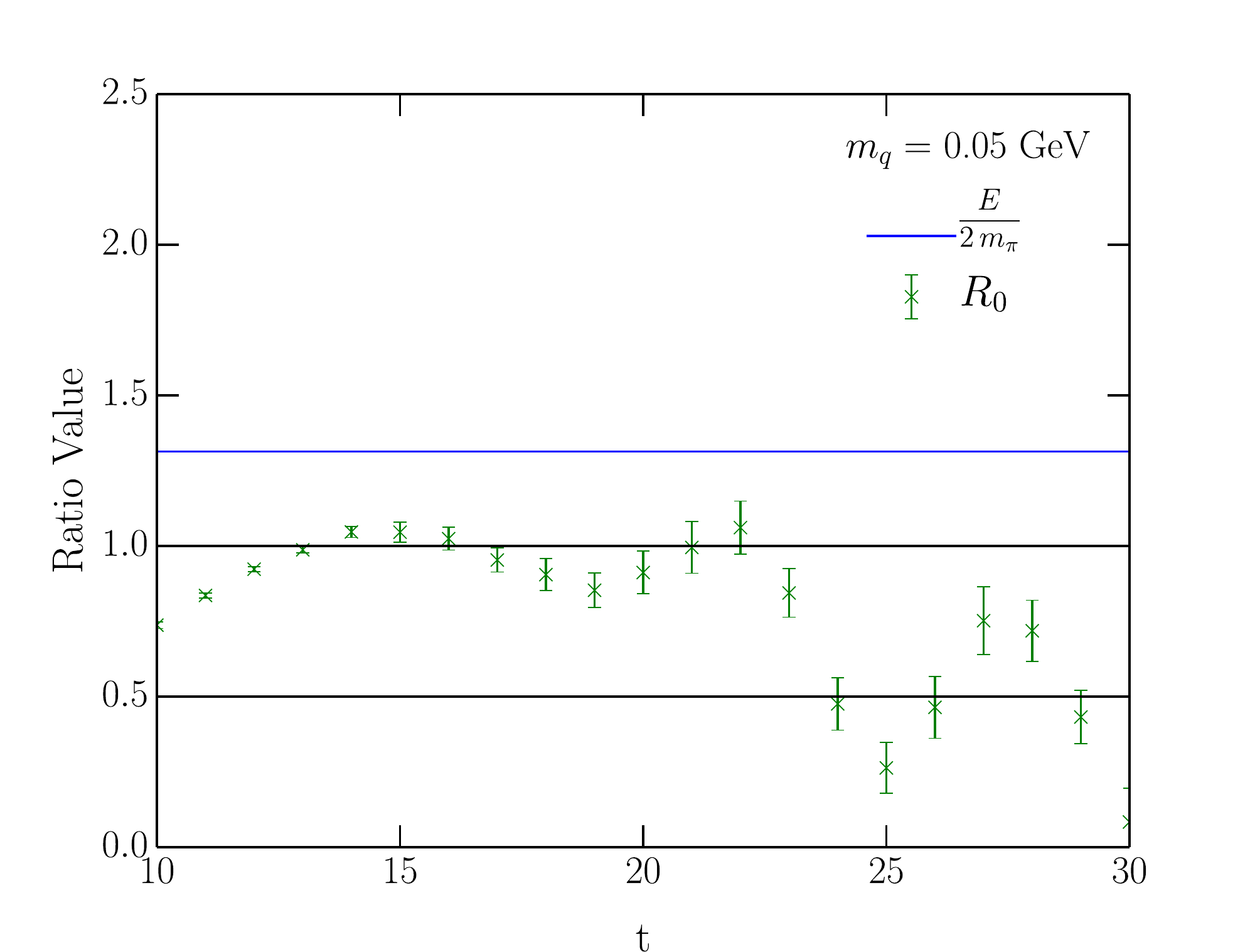}\\
  \caption{The ratio $R_{0}$ for the $a_{0}$ meson on the vortex-removed ensemble in the chiral regime, at light bare quark masses with values of $m_q=13,\ 25,\ 38,\ 50$ MeV increasing from left to right then top to bottom. Horizontal lines are drawn at $\frac{1}{2}$ ($\mathrm{U}(1)_{A}$ symmetry), $1$ ($\pi$-$\eta'$ state), and $\frac{E}{2m_\pi}$ (two-quark state) to guide the eye. Note that at the lightest mass the value of $\nicefrac{E}{2m_\pi}$ is above the range of the vertical axis.}
  \label{Fig:VRR0low}
\end{figure*}

%In the left column of
In Fig.~\ref{Fig:VRR0low} we present the ratio $R_{0}$ for the $a_{0}$ meson at the four light bare quark masses. At the lightest mass ($m_q = 13$ MeV), $R_{0}$ touches $1$, before dropping down to a stable value at $\frac{1}{2}$. The plateau at $\frac{1}{2}$ shows a restoration of the $\mathrm{U}(1)_{A}$ symmetry; degeneracy of the $a_{0}$ and pion. There is also evidence of a $\pi$-$\eta'$ state in the same channel, reflected by the value around $1$ at earlier time slices.

At $m_{q} = 25$ MeV the signal for $R_0$ is poor, with large fluctuations in the central value. By contrast, at the next two masses ($m_{q} = 38,50$ MeV), the ratio hovers around $1$ at early time slices, providing evidence of the formation of a multi-particle $\pi$-$\eta'$ state.  At later time slices, however, while there is some evidence of the value decreasing, the signal becomes too noisy to see a clear plateau at $\frac{1}{2}$.  It may be that due to the additional symmetry breaking from the axial anomaly, the restoration of the $\mathrm{U}(1)_{A}$ symmetry is particularly sensitive to explicit symmetry breaking from the bare quark mass. 

%  Seems too speculative...
%
%% In this case, the $a_{0}$ would be in an interim region, where the $\mathrm{U}(1)_{A}$ symmetry is broken, but the hadron spectrum does not yet behave like a constituent quark model. In this case, the lowest-lying state in this channel is a $\pi$-$\eta'$ threshold state.

%% \begin{figure*}[thpb]
%% \subfigure[]{
%% \label{03200VRR0}
%% \includegraphics[width=0.5\columnwidth]{Plots/03200VRR0.pdf}}
%% \subfigure[]{
%% \label{04000VRR0}
%% \includegraphics[width=0.5\columnwidth]{Plots/04000VRR0.pdf}}
%% \subfigure[]{
%% \label{04800VRR0}
%% \includegraphics[width=0.5\columnwidth]{Plots/04800VRR0.pdf}}
%% \subfigure[]{
%% \label{05600VRR0}
%% \includegraphics[width=0.5\columnwidth]{Plots/05600VRR0.pdf}}
%% \caption{The ratio $R_{0}$ for the $a_{0}$ meson on the vortex-removed ensemble, at bare quark masses of $100$ \subref{03200VRR0}, $126$ \subref{04000VRR0}, $151$ \subref{04800VRR0}, and $177$ MeV \subref{05600VRR0}. Horizontal lines are drawn at $\frac{1}{2}$ ($\mathrm{U}(1)_{A}$ chiral regime), $1$ ($\pi$-$\eta'$ state), and $\frac{2\,E_{q}}{4\,\cq}$ (two-quark state) to guide the eye.}
%% \label{Fig:VRR0high}
%% \end{figure*}

\begin{figure*}[t]
  \centering
  \includegraphics[width=0.5\columnwidth]{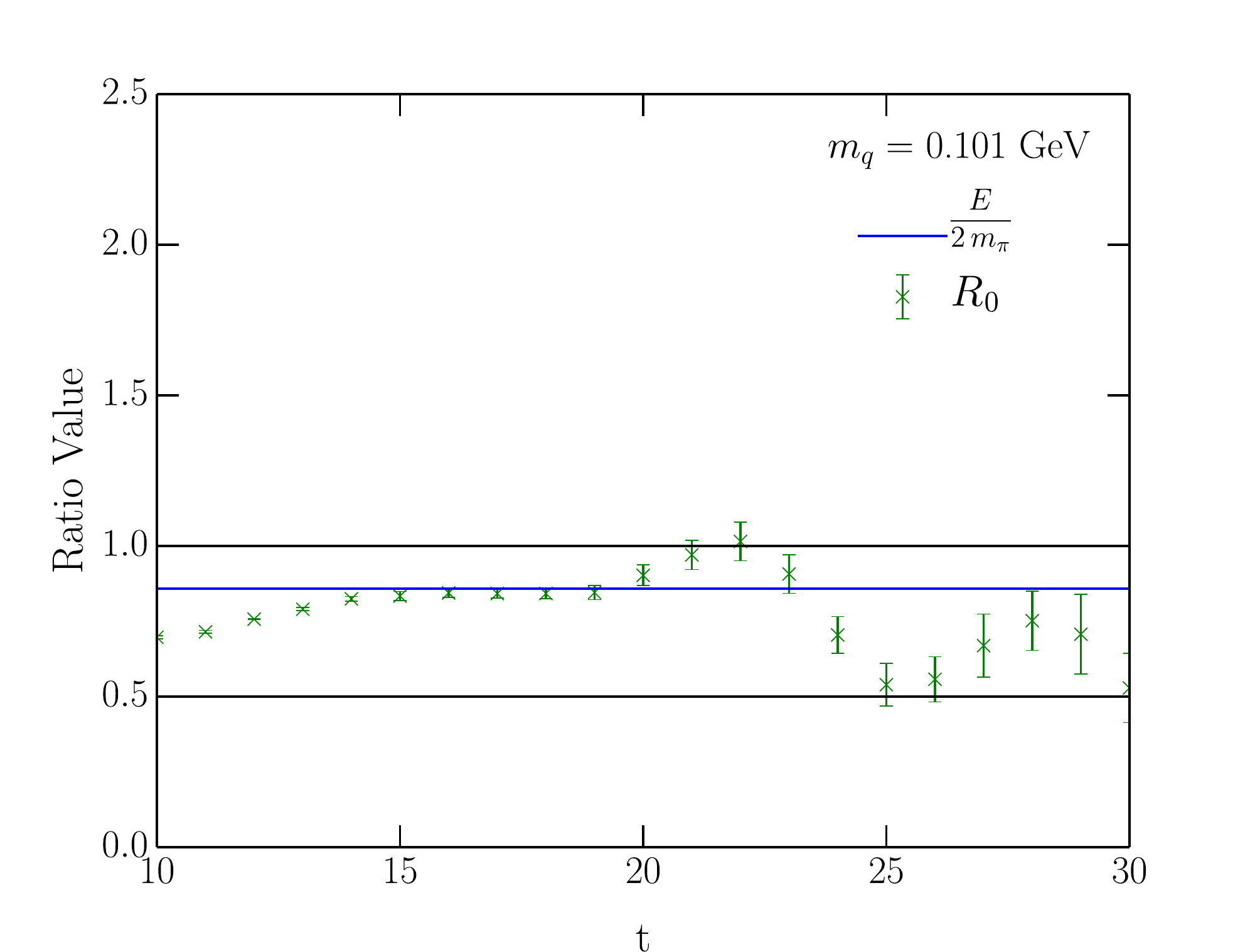}\includegraphics[width=0.5\columnwidth]{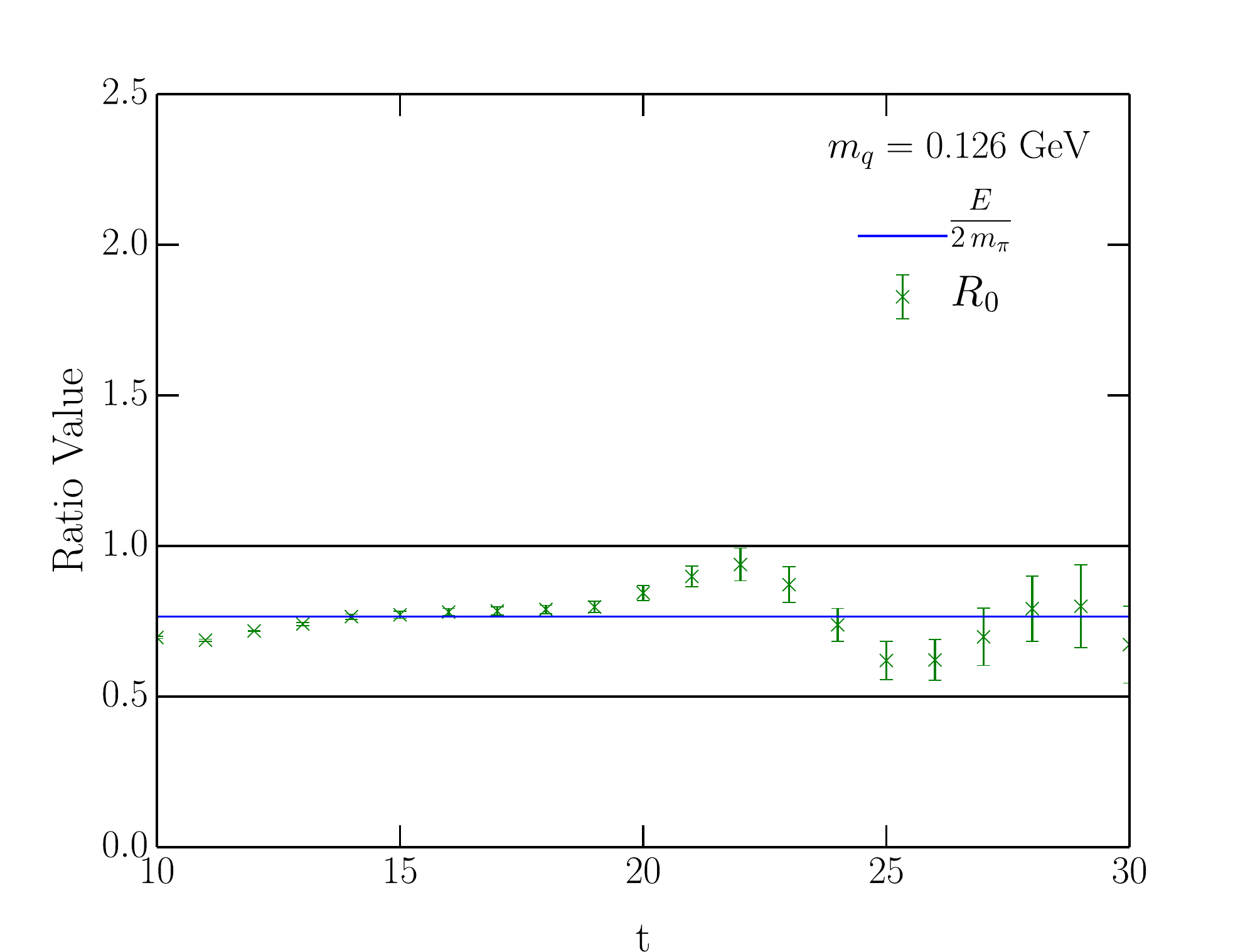}\\
  \includegraphics[width=0.5\columnwidth]{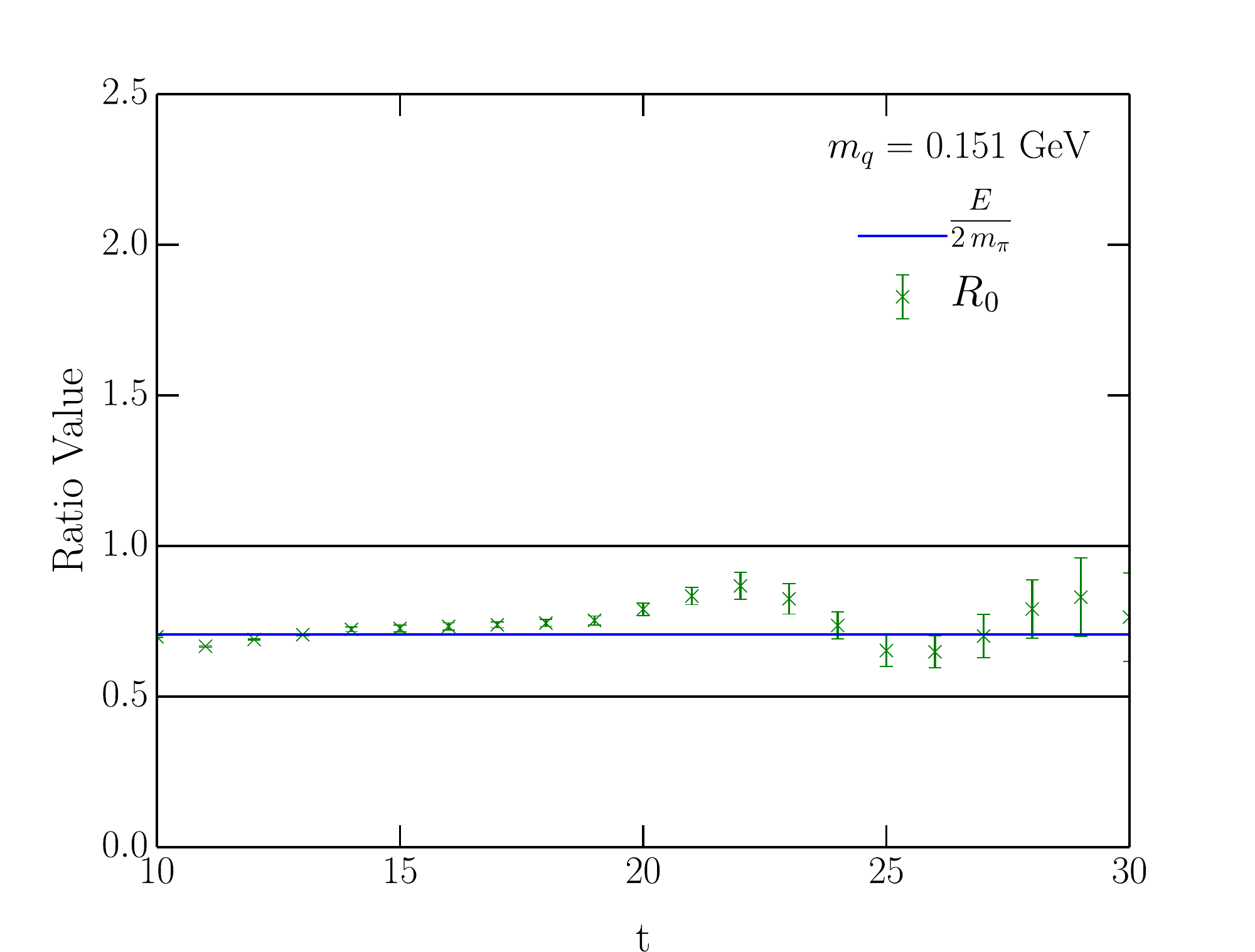}\includegraphics[width=0.5\columnwidth]{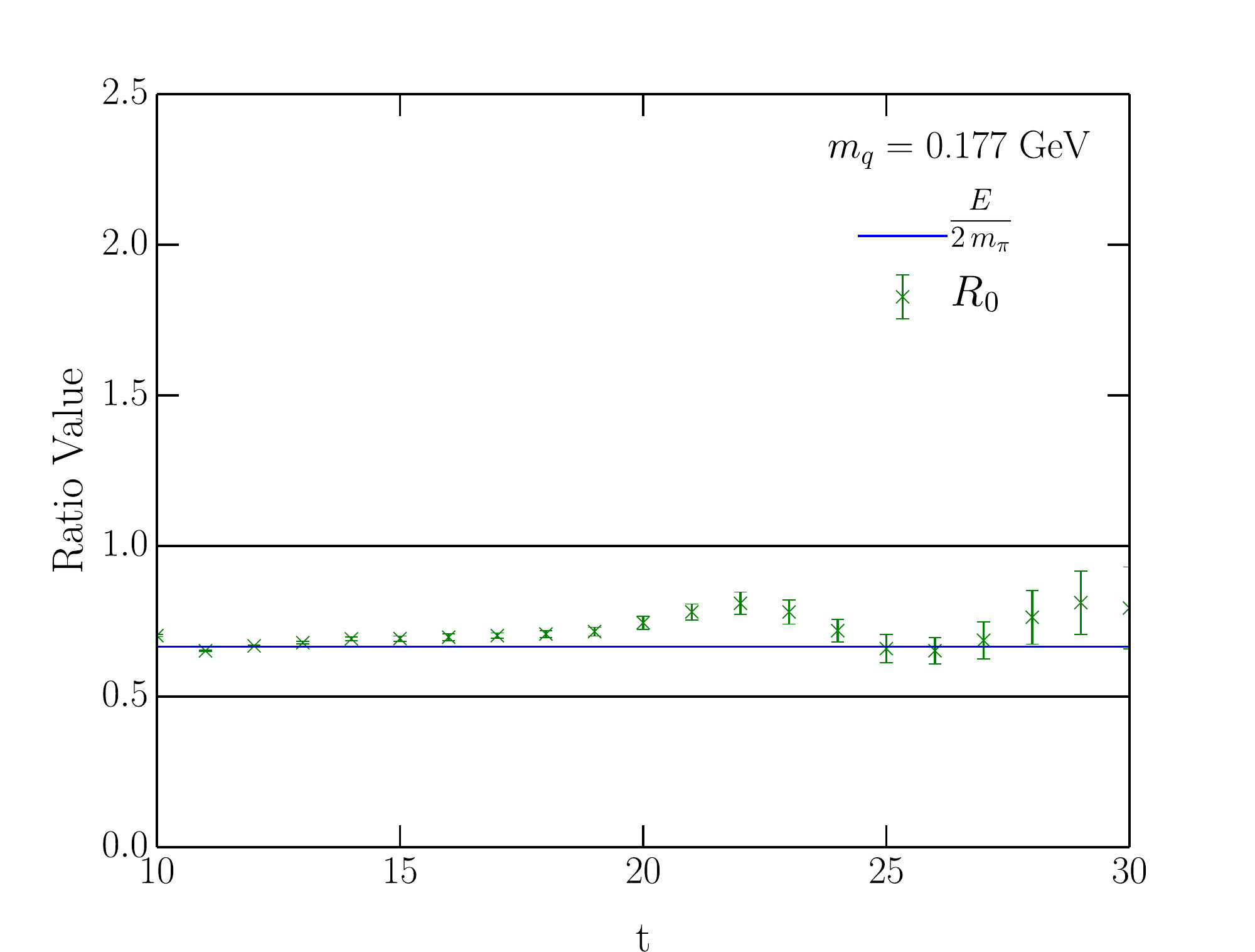}\\
  \caption{The ratio $R_{0}$ for the $a_{0}$ meson on the vortex-removed ensemble in the the constituent regime, at heavy bare quark masses with $m_q=101,\ 126,\ 151,\ 177$ MeV increasing from left to right then top to bottom. Horizontal lines are drawn at $\frac{1}{2}$ ($\mathrm{U}(1)_{A}$ symmetry), $1$ ($\pi$-$\eta'$ state), and $\frac{E}{2m_\pi}$ (two-quark state) to guide the eye.}
  \label{Fig:VRR0high}
\end{figure*}

%the right column of
The plots of $R_{0}$ for the $a_{0}$ meson at the four heavy quark masses are shown in Fig.~\ref{Fig:VRR0high}. We have seen previously that at these masses the hadrons behave like weakly-interacting constituent quarks, and this is quantified here. Up to Euclidean times of $t \simeq 20,$ $R_{0}$ lies almost exactly on the line drawn at $\nicefrac{E}{2m_\pi}$, indicating the $a_{0}$ is best described by a two quark state excited with the minimum lattice quantum of momenta. The ratio $\nicefrac{E}{2m_\pi}$ is less than one for all four heavy masses, implying that the two quark state is lighter than the $\pi$-$\eta'$ state.  Hence, the observations are consistent with our predictions for $R_0$ in the constituent regime. At later times, the signal for the ratio oscillates, with no clear evidence of any other states in this region.

%% \begin{figure*}[p]
%%   \centering
%%   \includegraphics[width=0.4\columnwidth]{Plots/00400VRR1.pdf}\includegraphics[width=0.4\columnwidth]{Plots/03200VRR1.pdf}\\
%%   \includegraphics[width=0.4\columnwidth]{Plots/00800VRR1.pdf}\includegraphics[width=0.4\columnwidth]{Plots/04000VRR1.pdf}\\
%%   \includegraphics[width=0.4\columnwidth]{Plots/01200VRR1.pdf}\includegraphics[width=0.4\columnwidth]{Plots/04800VRR1.pdf}\\
%%   \includegraphics[width=0.4\columnwidth]{Plots/01600VRR1.pdf}\includegraphics[width=0.4\columnwidth]{Plots/05600VRR1.pdf}
%%   \caption{The ratio $R_{1}$ for the $a_{1}$ meson on the vortex-removed ensemble in the chiral regime (left) and the constituent regime (right). From top to bottom, the left column shows the four light bare quark masses ($m_q=13,25,40,50$ MeV), and the right column shows the four heavy bare quark masses ($m_q=101,126,151,177$ MeV). Horizontal lines are drawn at $\frac{1}{2}$ ($\SU{2}_{L} \times \SU{2}_{R}$ chiral regime), $\frac{(m_{\pi} + m_{\rho})}{2\,m_{\rho}}$, $1$, and $\frac{E}{2m_\rho}$ (two-quark state) to guide the eye.}
%%   \label{Fig:VRR1}
%% \end{figure*}

%% @@@@@@@@@@@@@@@@@@@@@@@@@@@@@@@@@@@@@@@@@@@@@@@@@@@@@@@@@@@@@@@@@@@@@@@@

\begin{figure*}[t]
  \centering
  \includegraphics[width=0.5\columnwidth]{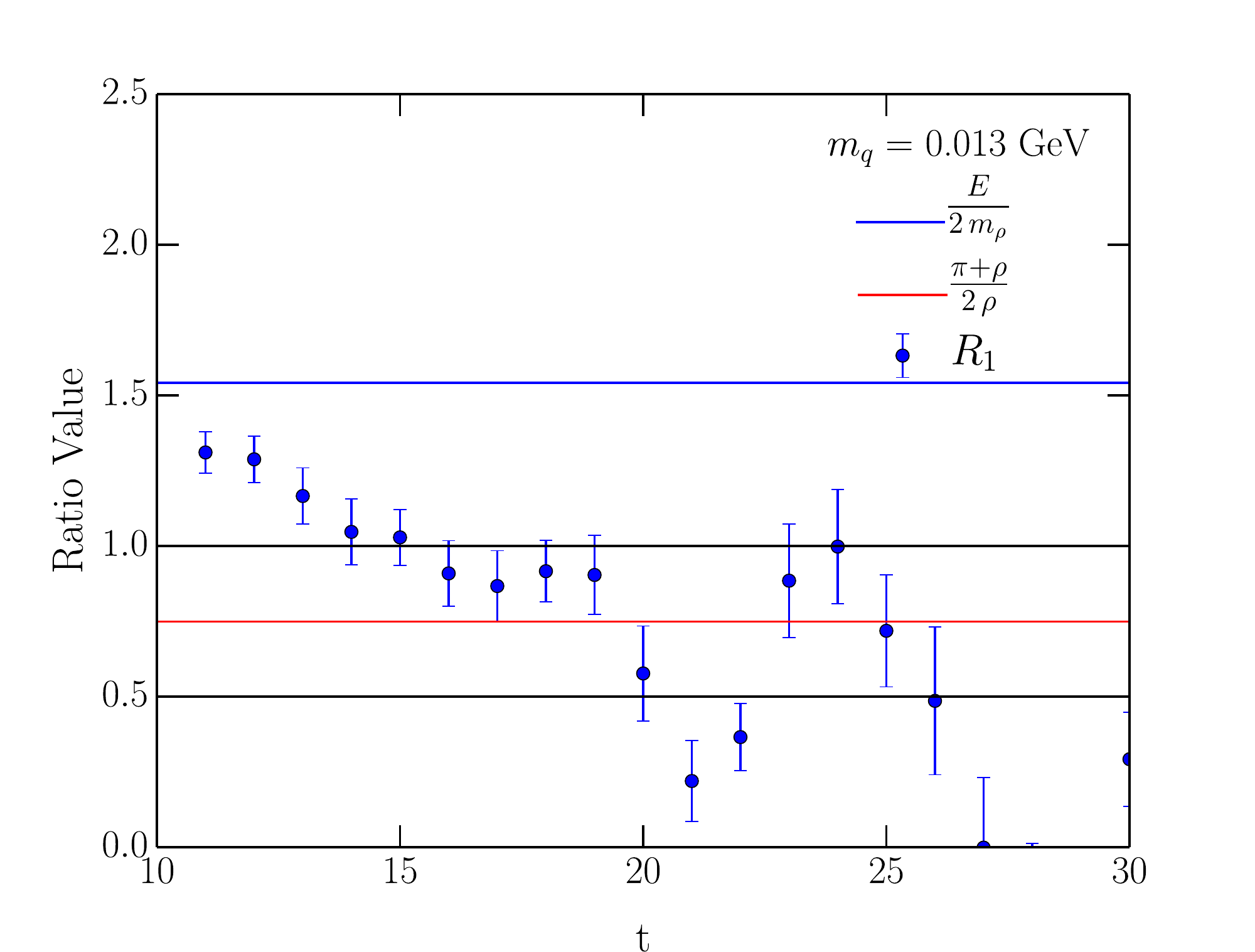}\includegraphics[width=0.5\columnwidth]{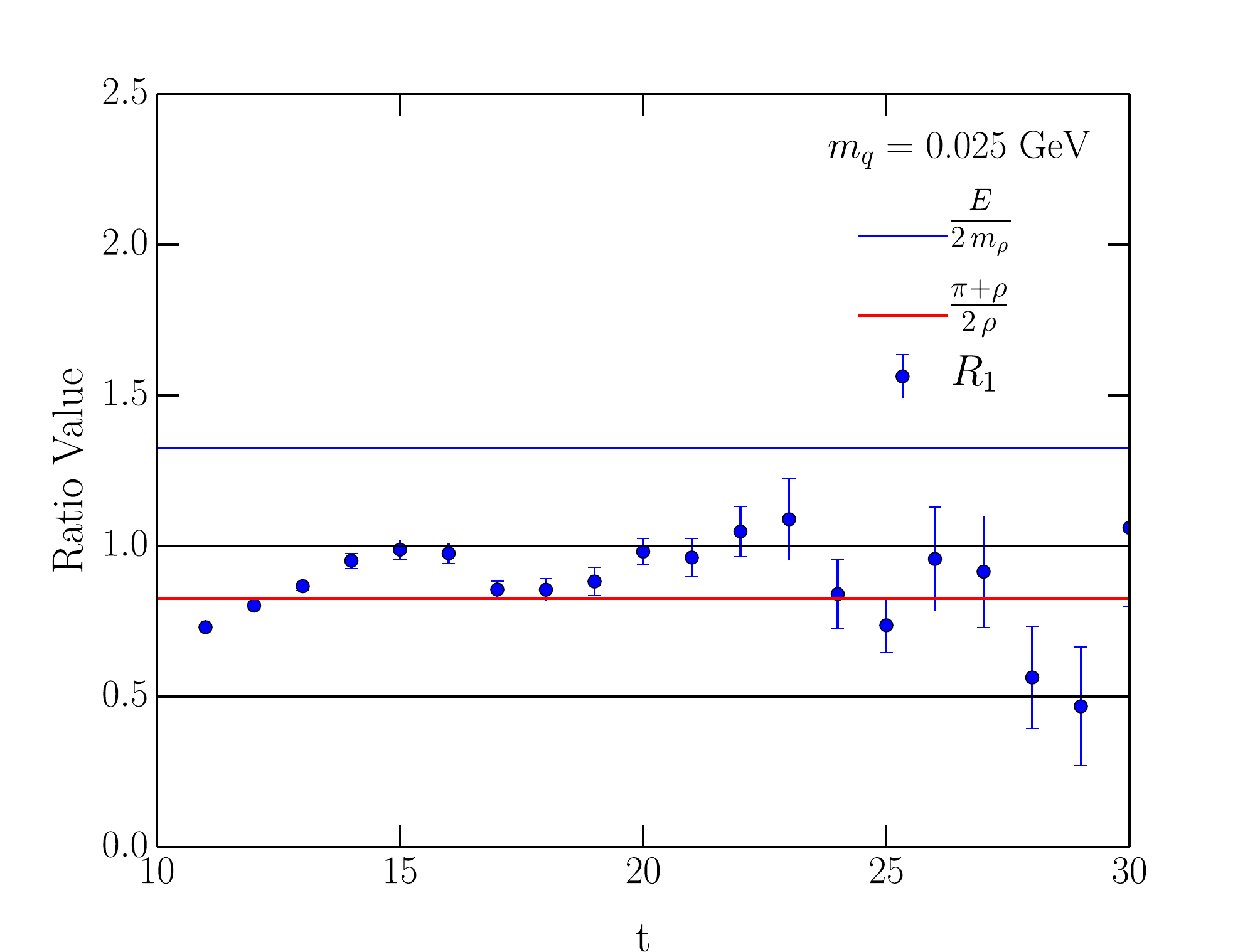}\\
  \includegraphics[width=0.5\columnwidth]{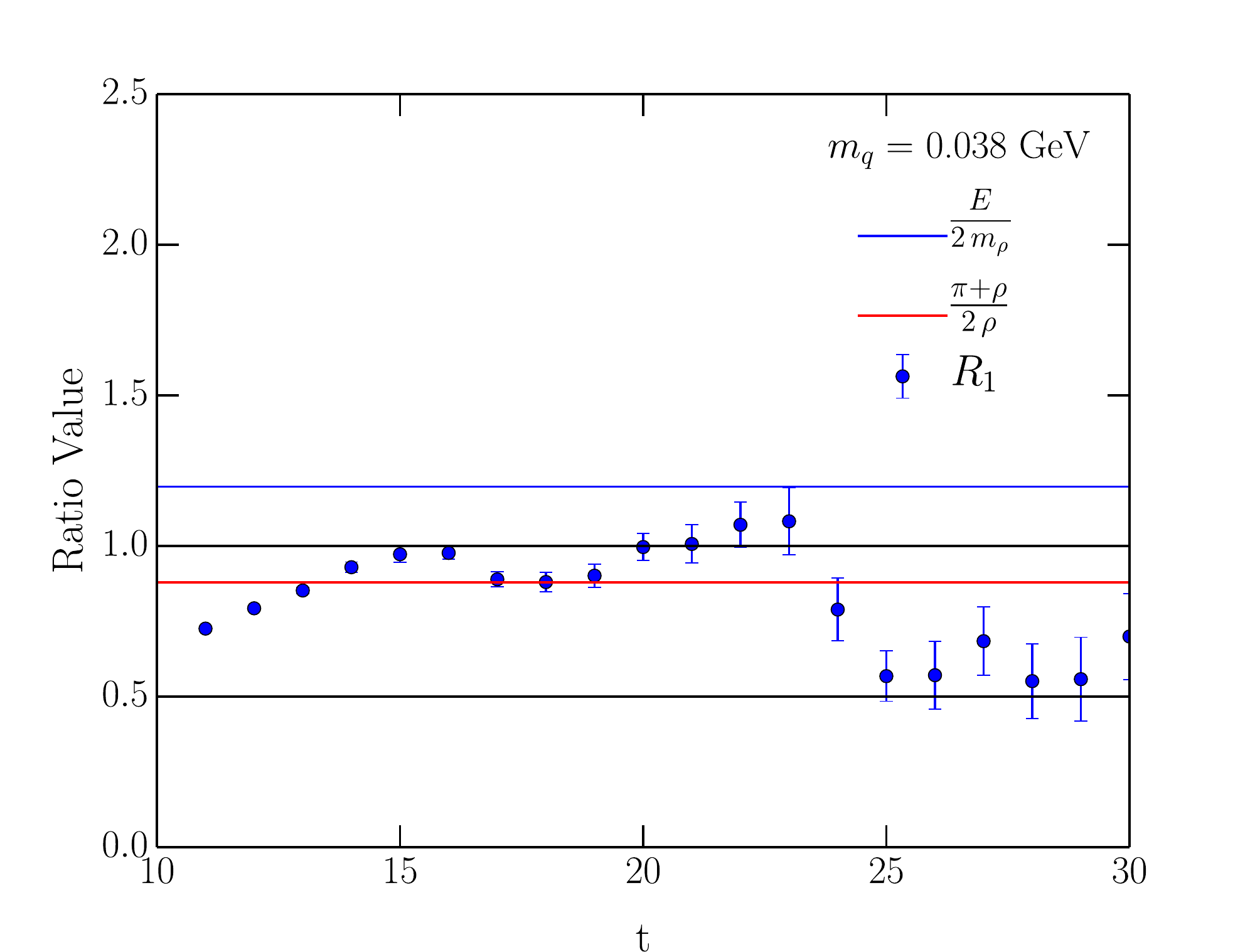}\includegraphics[width=0.5\columnwidth]{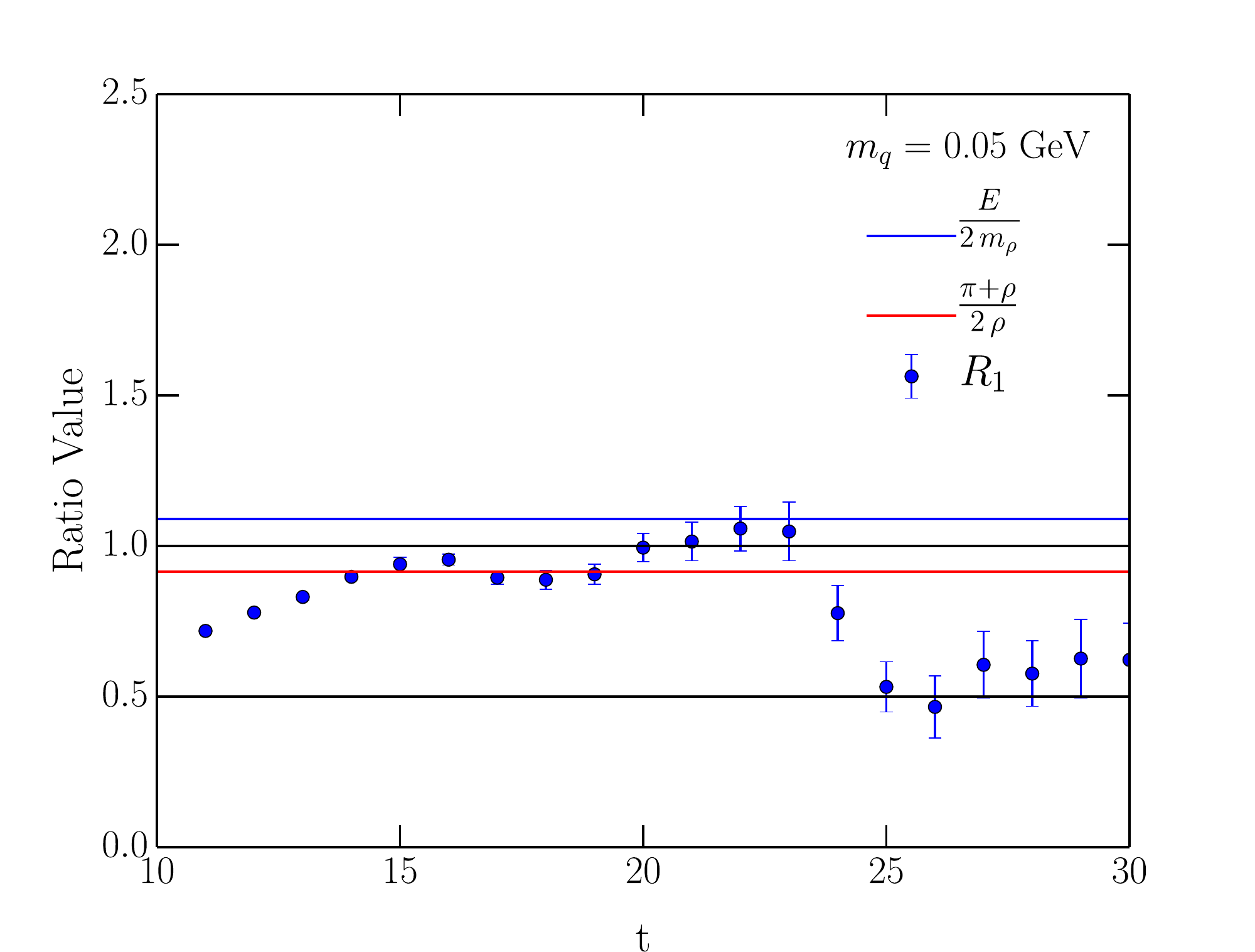}\\
  \caption{The ratio $R_{1}$ for the $a_{1}$ meson on the vortex-removed ensemble in the chiral regime, at light bare quark masses with $m_q=13,\ 25,\ 38,\ 50$ MeV increasing from left to right then top to bottom. Horizontal lines are drawn at $\frac{1}{2}$ ($\SU{2}_{L} \times \SU{2}_{R}$ symmetry), $\frac{(m_{\pi} + m_{\rho})}{2\,m_{\rho}}$, $1$, and $\frac{E}{2m_\rho}$ (two-quark state) to guide the eye.}
  \label{Fig:VRR1low}
\end{figure*}

%% \begin{figure*}[thpb]
%% \subfigure[]{
%% \label{00400VRR1}
%% \includegraphics[width=0.5\columnwidth]{Plots/00400VRR1.pdf}}
%% \subfigure[]{
%% \label{00800VRR1}
%% \includegraphics[width=0.5\columnwidth]{Plots/00800VRR1.pdf}}
%% \subfigure[]{
%% \label{01200VRR1}
%% \includegraphics[width=0.5\columnwidth]{Plots/01200VRR1.pdf}}
%% \subfigure[]{
%% \label{01600VRR1}
%% \includegraphics[width=0.5\columnwidth]{Plots/01600VRR1.pdf}}
%% \caption{The ratio $R_{1}$ for the $a_{1}$ meson on the vortex-removed ensemble, at bare quark masses of $13$ \subref{00400VRR1}, $25$ \subref{00800VRR1}, $38$ \subref{01200VRR1}, and $50$ MeV \subref{01600VRR1}. Horizontal lines are drawn at $\frac{1}{2}$ ($\SU{2}_{L} \times \SU{2}_{R}$ chiral regime), $\frac{m_{\pi} + m_{\rho}}{2\,m_{\rho}}$, $1$, and $\frac{2\,E_{q}}{4\,\cq}$ (two-quark state) to guide the eye.}
%% \label{Fig:VRR1low}
%% \end{figure*}

\begin{figure*}[t]
  \centering
  \includegraphics[width=0.5\columnwidth]{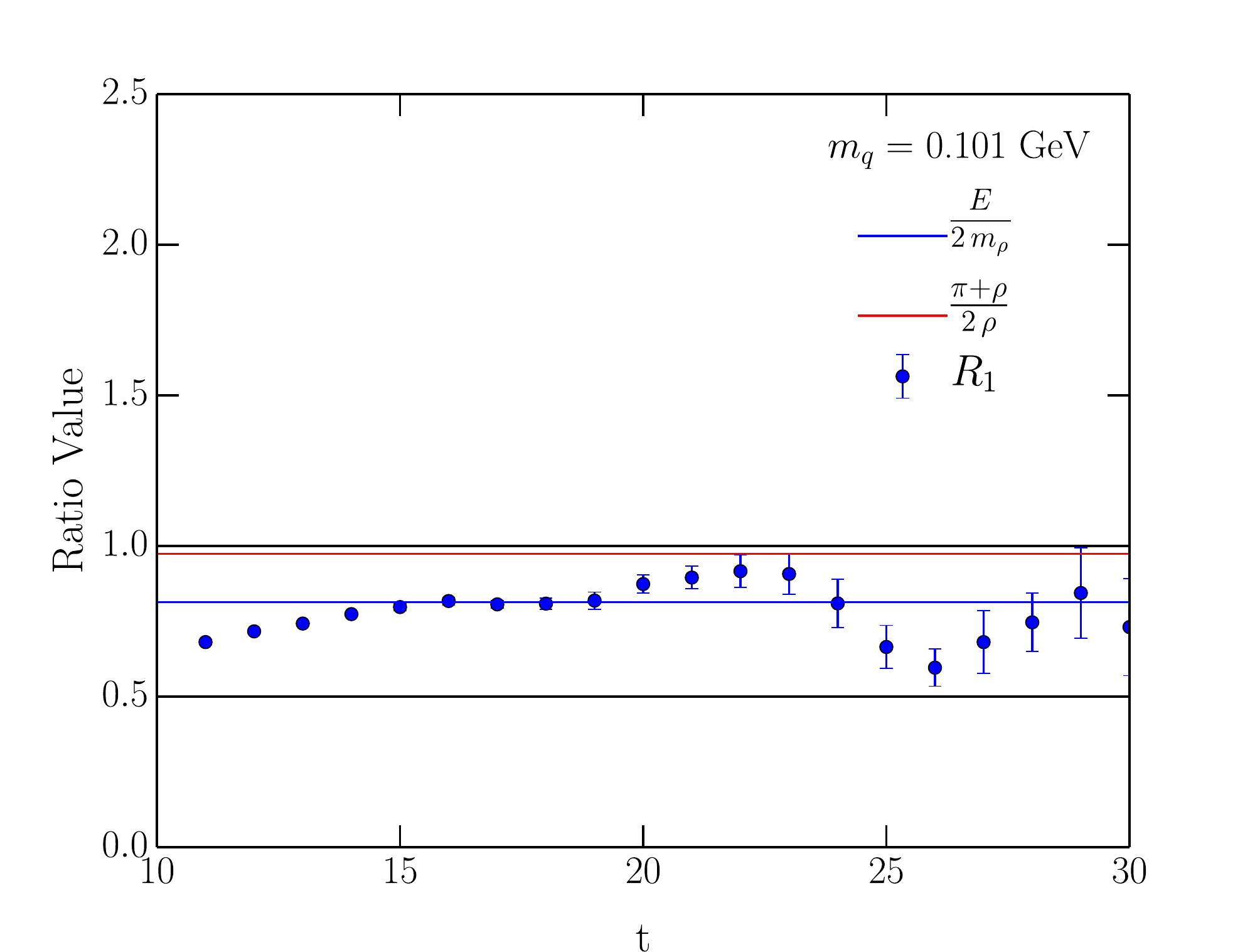}\includegraphics[width=0.5\columnwidth]{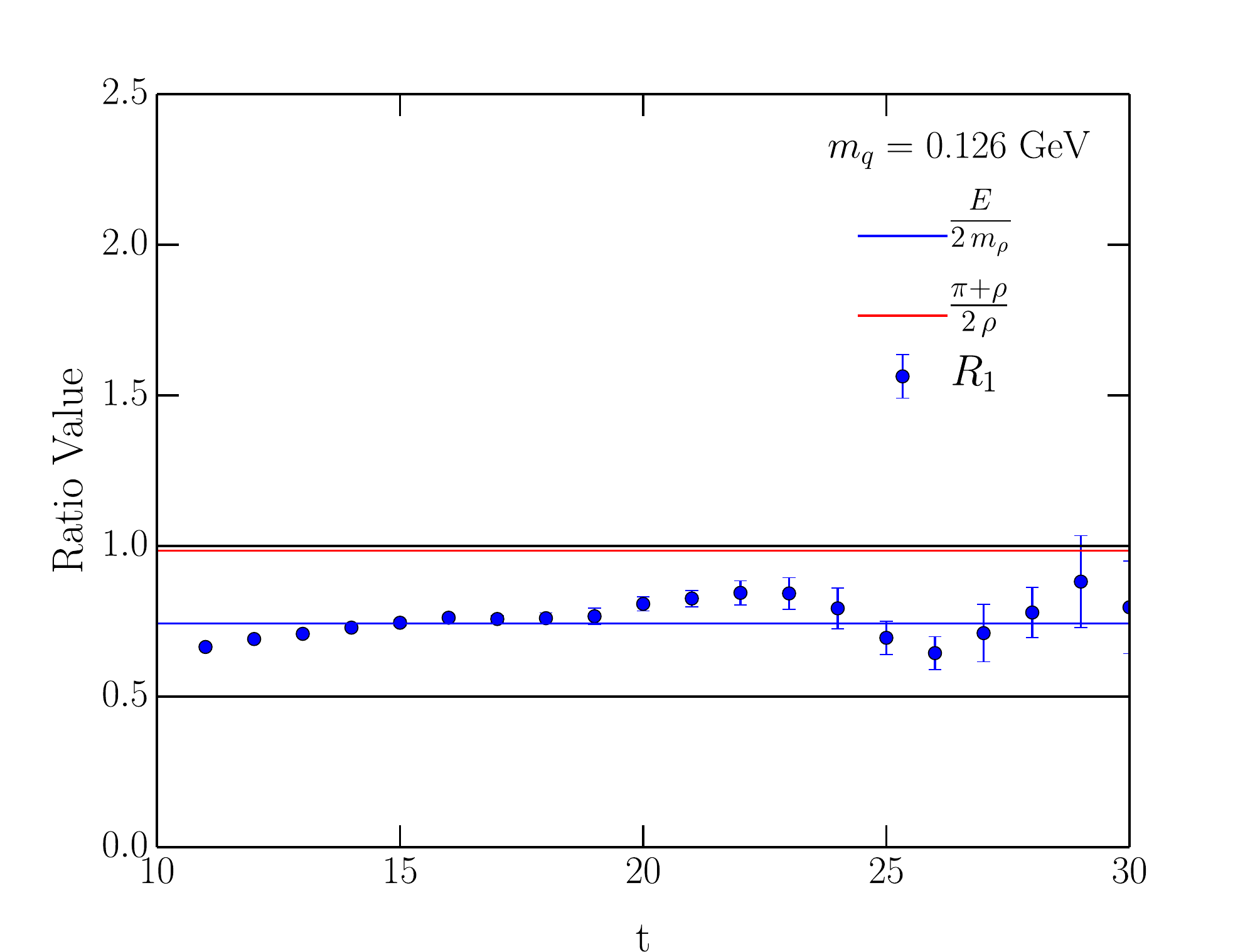}\\
  \includegraphics[width=0.5\columnwidth]{Plots/04000VRR1.pdf}\includegraphics[width=0.5\columnwidth]{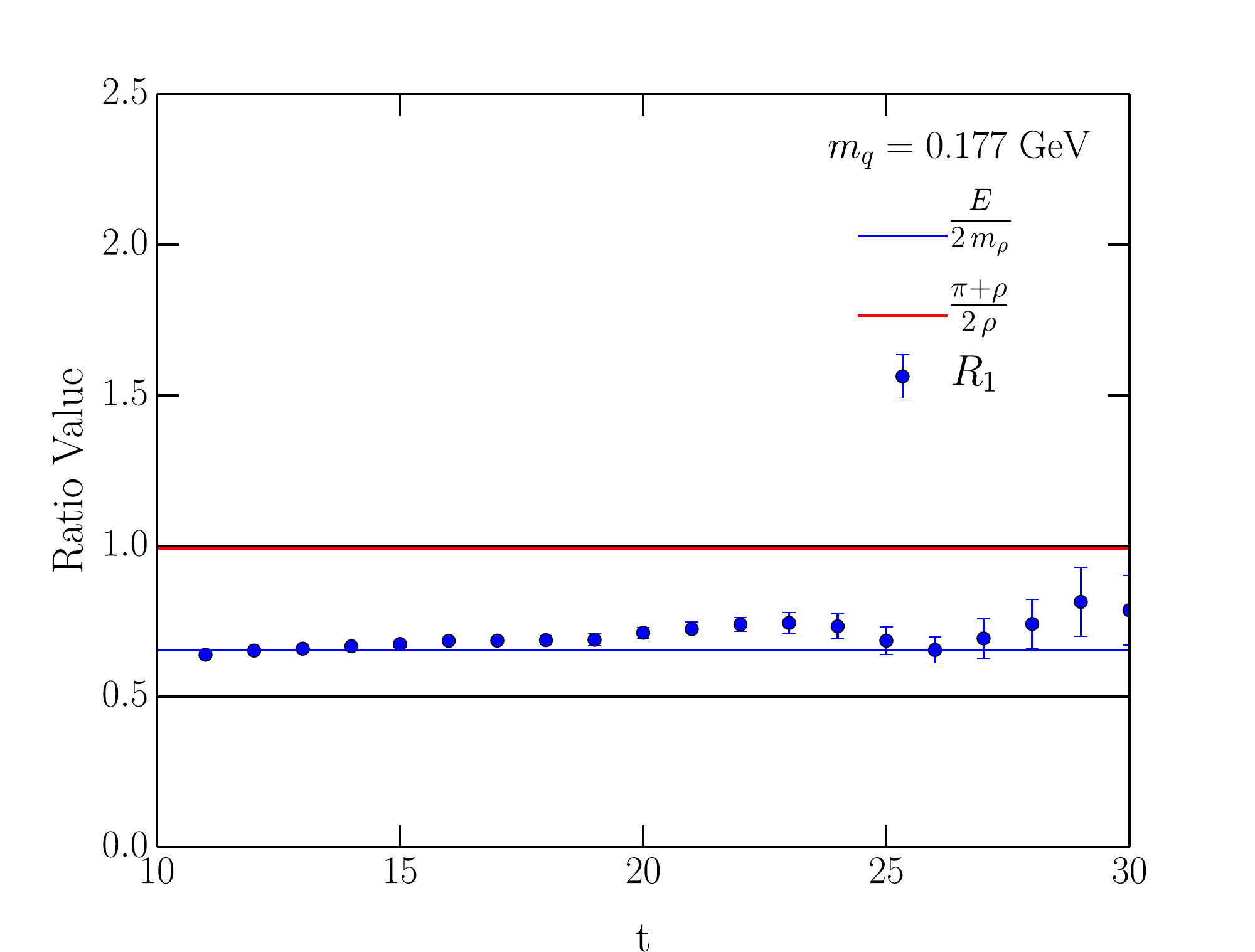}
  \caption{The ratio $R_{1}$ for the $a_{1}$ meson on the
    vortex-removed ensemble in the constituent regime, at heavy bare
    quark masses with $m_q=101,\ 126,\ 151,\ 177$ MeV increasing from left to right then top to bottom. Horizontal lines are drawn at $\frac{1}{2}$ ($\SU{2}_{L} \times \SU{2}_{R}$ symmetry), $\frac{(m_{\pi} + m_{\rho})}{2\,m_{\rho}}$, $1$, and $\frac{E}{2m_\rho}$ (two-quark state) to guide the eye.}
  \label{Fig:VRR1high}
\end{figure*}

%the left column of
We show the ratio $R_{1}$ for the $a_{1}$ meson at the light quark masses in Fig.~\ref{Fig:VRR1low}. At all four masses, $R_{1}$ hovers around the line $\frac{m_{\pi} + m_{\rho}}{2\,m_{\rho}}$ at early Euclidean times; this corresponds to the $a_{1}$ correlator being dominated by a $\rho$-$\eta'$ excited state. At the lightest two masses, the signal is too poor to provide evidence of any other state. However, at $m_{q} = 38$ MeV and $m_{q} = 50$ MeV, after time slice 25 another stable plateau is seen at $R_{1} = \frac{1}{2}$. This indicates the degeneracy of the $a_{1}$ with the rho meson, evidence of the restoration of the $\SU{2}_{L} \times \SU{2}_{R}$ symmetry. This reveals that at $m_{q} = 50$ MeV chiral symmetry breaking from the quark mass is still sufficiently small that the symmetry holds. This concurs with the results for the $\pi$ and $\rho$ mesons, which are not yet degenerate with $\frac{m_{\pi} + m_{\rho}}{2\,m_{\rho}} \simeq 0.91$ indicating a small but significant splitting at this quark mass. 

%% \begin{figure*}[thpb]
%% \subfigure[]{
%% \label{03200VRR1}
%% \includegraphics[width=0.5\columnwidth]{Plots/03200VRR1.pdf}}
%% \subfigure[]{
%% \label{04000VRR1}
%% \includegraphics[width=0.5\columnwidth]{Plots/04000VRR1.pdf}}
%% \subfigure[]{
%% \label{04800VRR1}
%% \includegraphics[width=0.5\columnwidth]{Plots/04800VRR1.pdf}}
%% \subfigure[]{
%% \label{05600VRR1}
%% \includegraphics[width=0.5\columnwidth]{Plots/05600VRR1.pdf}}
%% \caption{The ratio $R_{1}$ for the $a_{1}$ meson on the vortex removed ensemble, at bare quark masses of $100$ \subref{03200VRR1}, $126$ \subref{04000VRR1}, $151$ \subref{04800VRR1}, and $177$ MeV \subref{05600VRR1}. Horizontal lines are drawn at $\frac{1}{2}$ ($\SU{2}_{L} \times \SU{2}_{R}$ chiral regime), $\frac{m_{\pi} + m_{\rho}}{2\,m_{\rho}}$, $1$, and $\frac{2\,E_{q}}{4\,\cq}$ (two-quark state) to guide the eye.}
%% \label{Fig:VRR1high}
%% \end{figure*}
%the right column of 

The ratio $R_{1}$ for the $a_{1}$ meson at the four heavy quark masses is shown in Fig.~\ref{Fig:VRR1high}. The $\pi$ and $\rho$ are approximately degenerate here, such that the value of $\nicefrac{E}{2m_\rho}$ is less than the value of $\frac{m_{\pi} + m_{\rho}}{2\,m_{\rho}}$, \emph{i.e.}  the two quark state is lighter than the multi-particle $\rho$-$\eta'$ state at all four heavy masses. Indeed, we see that at high quark masses the ratio $R_{1}$ lies along the line at $\nicefrac{E}{2m_\rho},$ showing almost perfect agreement up to $t \simeq 20,$ with some small fluctuations at later times as the signal degrades. The two quark state behaviour mirrors that of $R_{0},$ indicating an onset of degeneracy between the $a_{0}$ and $a_{1}$ at high quark masses. While this is a feature seen also in the untouched ensemble, remarkably the masses of both are given within error bars by the value $E$ predicted by our simple model in Eq.~(\ref{eq:EQdef}). At the four heavy quark masses, the weakly-interacting constituent-quark like model is remarkably successful; all six hadrons considered are in agreement with the predictions at all of these masses.

\section{Summary}
\label{sec:hadspecsumm}

We have presented a novel examination of % for the first time studied 
the influence of centre vortices on the low-lying hadron spectrum over a wide range of bare quark masses using the chirally sensitive overlap operator. This has allowed us to use the behaviour of the low-lying hadron spectrum as a probe of the role of centre vortices in dynamical chiral symmetry breaking.

After a small amount of cooling, the vortex-only backgrounds are capable of recreating all the salient features of the low-lying hadron spectrum. While the ground state masses are slightly lower due to the use of smoothing~\cite{Thomas:2014tda}, the qualitative features of the spectrum are intact, in agreement with results seen for the quark propagator~\cite{Trewartha:2015nna, Trewartha:2015ida}. In particular, the pion remains much lighter than the other mesons. Its behaviour as a pseudo-Goldstone boson is a clear signal of the presence of dynamical chiral symmetry breaking on the vortex-only ensemble. Furthermore, there is a significant splitting in the masses of chiral partners. 

%There is also no sign of chiral symmetry restoration, with a clear separation of
%There is also no sign of chiral symmetry restoration, with a clear separation of the masses of chiral partners maintained.

On the vortex-removed ensemble, we have observed
%revealed a number of signals of 
the loss of dynamical chiral symmetry breaking. At low quark masses, there is strong evidence of the restoration of $\SU{2}_{L} \times \SU{2}_{R}$ chiral symmetry. The nucleon and $\Delta$ baryons become degenerate, as do the $a_{1}$ and $\rho$ mesons. The evidence for the restoration of the $\mathrm{U}(1)_{A}$ symmetry at our lowest quark mass is clear; at this mass the $a_{0}$ shows a degeneracy with the pion. 
% However, the signal is less satisfactory at the other light masses. It may be that this issue would be resolved with more statistics, or 
We have also observed that the $\mathrm{U}(1)_{A}$ symmetry is more sensitive to explicit chiral symmetry breaking from the bare quark mass.

At high quark masses, the vortex-removed hadron spectrum is consistent with the behaviour of weakly-interacting dressed constituent quarks in an otherwise near-trivial background, as seen in previous studies using a Wilson action~\cite{OMalley:2011aa}. In accord with the quark-propagator results \cite{Trewartha:2015nna, Trewartha:2015ida}, there is some residual dynamical mass generation on the vortex-removed ensemble.
%, and this is replicated here. 
The $\pi$ and the $\rho$ are approximately degenerate, as are the nucleon and Delta, such that the implied constituent quark mass from all four hadrons are in agreement. A constituent quark mass higher than the bare quark mass indicates a nontrivial dressing of the quarks from the gauge field fluctuations surviving vortex removal. Using the constituent quark mass extracted, the $a_{0}$ and $a_{1}$ mesons can be described successfully as a two quark state excited with the minimal non-trivial lattice momentum.

%The vortex-removed gauge fields behave as an almost trivial background. Unlike in the quark propagator, a previous study of the hadron spectrum~\cite{OMalley:2011aa} using Wilson-like fermions was able to correctly show this at high quark masses where the explicit chiral symmetry breaking present in the action .
%The vortex removed-ensemble deviates from a trivial background only in that there is a small residue of dynamical mass generation, as evinced by a constituent-quark mass larger than the bare quark mass.

Here, for the first time, using a vortex-removed gauge field ensemble we have been able to produce the hadronic degeneracies associated with the restoration of chiral symmetry. The use of the overlap fermion action, which respects chiral symmetry, is vital to revealing this property. Remarkably, we are able to reproduce all the salient features of QCD in the low-lying hadron spectrum from smoothed vortex-only backgrounds, while upon vortex removal we see a loss of dynamical chiral symmetry breaking. These results provide a further contribution to the already significant body of evidence~\cite{Kamleh:2017lij} that centre vortices are the fundamental mechanism underlying dynamical chiral symmetry breaking in $\SU{3}$ gauge theory.

\section*{Acknowledgements}

This research was undertaken with the assistance of resources at the NCI National Facility in
Canberra, the iVEC facilities at the Pawsey Centre and the Phoenix GPU cluster at the University of
Adelaide, Australia.  These resources were provided through the National Computational Merit
Allocation Scheme, supported by the Australian Government, and the University of Adelaide through
their support of the NCI Partner Share and the Phoenix GPU cluster.
This research is supported by the Australian Research Council through Grants No.\ DP150103164,
DP120104627 and LE120100181.

\vspace{24pt}

\bibliographystyle{iopart-num}
\bibliography{vortexHadSpec}

\end{document}